\documentclass[journal]{IEEEtran}

\usepackage{cite}
\usepackage{amsmath,amssymb,amsfonts}
\usepackage{algorithmic}
\usepackage{graphicx}
\usepackage{textcomp}
\usepackage{rotating}
\usepackage{subfigure}
\usepackage{url}
\usepackage{diagbox}
\usepackage{breqn}

\ifCLASSINFOpdf
\else
\fi

\begin{document}
%
\title{SAMIPS: A Synthesised Asynchronous Processor}
%
%
%

\author{Qianyi~Zhang,
        Georgios~Theodoropoulos
\thanks{Qianyi Zhang is currenlty with Cadence Design Systems, Shanghai, China (e-mail: qianyiz@gmail.com).}
\thanks{Georgios Theodoropoulos is currently with the Southern University of Science and Technology, Shenzhen, China (e-mail: theogeorgios@gmail.com).}
\thanks{Corresponding author: Georgios Theodoropoulos.}
\thanks{The design and development of SAMIPS took place at the University of Birmingham, UK, between 2002 and 2006. It was supported in part by RCUK Grant GR/S11091/01.}
}

%
%

\markboth
{Zhang and Theodoropoulos: SAMIPS: A Synthesised Asynchronous Processor}
{Zhang and Theodoropoulos: SAMIPS: A Synthesised Asynchronous Processor}
%



\maketitle

\begin{abstract}
Miniaturisation and ever increasing clock speeds pose significant challenges to synchronous VLSI design  with clock distribution becoming an increasingly costly and complicated issue and power consumption rapidly emerging as a major concern. Asynchronous logic promises to alleviate these challenges however its development and adoption has been hindered by the lack of mature design tools. Balsa is a response to this gap, encompassing a CSP-based  asynchronous hardware description language and a framework for automatically synnthesising asynchronous circuits. 
This paper discusses SAMIPS, an asynchronous implementation of the MIPS microprocessor and the first full scale asynchronous microprocessor to be synthesised in Balsa. The objectives of the paper are twofold: first to provide a holistic description of SAMIPS and its components, the approach that it has been followed for the asynchronisation of MIPS and the innovative solutions that have been developed to address hazard challenges and a quantitative performance analysis of the system; secondly, to provide insights about the effectiveness of Balsa as a  hardware description language and synthesis system. \end{abstract}


%
\IEEEpeerreviewmaketitle

\section{Introduction}
\label{sec:introduction}
%
%
%
%
\IEEEPARstart{G}{lobal} sustainability concerns and an ever expanding digital life utilising mobile, IoT and edge devices has rendered lower power requirements at the centre of the electronics industry. Asynchronous design has been advocated as a viable approach to reduce IC power consumption while also promising to liberate VLSI design from  clock skew problems, offer the potential for high performance  and better electromagnetic interference (EMI) and encourage a modular design philosophy which makes incremental technological migration a much easier task \cite{4586393}. 
As a result, several asynchronous design techniques  \cite{7d254e00d5d440df8cfc732c8112cff4} and a number of asynchronous processors \cite{8660629} have been developed.


One of the main factors that have hindered the further development and adoption of asynchronous systems is the lack of appropriate mature Electronic Design and Automation (EDA) tools  \cite{10.1145/513918.514024}. Amongst the various formalisms and languages  that have been proposed for asynchronous design,  Communicating Sequential Processes (CSP) \cite{Hoare85d} has been extensively advocated as particularly suitable for describing the behaviour of asynchronous systems and a number of CSP-based environments have been proposed for this purpose \cite{theo07,https://doi.org/10.1002/cpe.587,THEODOROPOULOS2000741,THEODOROPOULOS2002622,Martin90,656281,Brunvand1994,Ebergen91,10.1007/BFb0039070,30a9e74269d548bd8e7012590d02c80a,270625}. A robust and complete CSP-based tool is Balsa \cite{Bard00}, developed at the University of Manchester. Balsa is both a framework for synthesising asynchronous hardware and a language for specifying such systems.   


Balsa has been demonstrated in synthesising a number of asynchronous systems including the DMA controller of Amulet3i \cite{Bard00,Furber00}, SPA, an ARM-based core for smartcard applications\cite{Plan02}, ASCARTS \cite{ASCARTS} and Asynchronous FPGA designs \cite{KOMATSU2013}. 
This paper discusses SAMIPS, the first full scale asynchronous microprocessor to be synthesised in Balsa. SAMIPS is a fully anynchronous implementation of the MIPS processor (\textbf{S}ynthesised \textbf{A}synchronous \textbf{MIPS}). 

Different aspects of SAMIPS have been previously reported in  \cite{Theo2004,zhang03,Zhang03a,Zhang04}. This paper provides a description of the entire SAMIPS processor, its design process and a quantitative evaluation of the system. In particular the  main contributions of this paper are the following:

\begin{enumerate}
    \item It presents for the first time a detailed description of the SAMIPS processor and its Balsa specification and an extensive quantitative evaluation of the system. SAMIPS has been the first fully automatically synthesised asynchronous MIPS processor based on the theory of handshake circuits. 
     \item It outlines a roadmap for converting a synchronous processor to an asynchronous one.
    \item It revisits data hazards in asynchronous pipelines and presents alternative implementations and a detailed quantitative evaluation of an asynchronous data hazards approach the authors first outlined in \cite{Zhang03a}. 
    \item It revisits  control hazards in asynchronous pipelines, and presents alternative implementations and a detailed quantitative evaluation of an asynchronous data hazards approach the authors first outlined in  \cite{Theo2004}, also extending it to address interrupts.
    \item It presents a framework for the critical path analysis of the processor and demonstrates an approach for model-level design optimisations of the system. 
 
\end{enumerate}

The rest of the paper is structured as follows: Section \ref{Balsa} provides an overview of the Balsa language and environment. Section \ref{Asynchronising MIPS} outlines the requirements and approach taken to asynchronise the basic  data path and control of synchronous MIPS. Sections \ref{Data hazards problems} and  \ref{Control Hazards Problems} delve deeper into the data and control hazard challenges respectively. Section \ref{SAMIPS Organisation and Balsa Implementation} presents a detailed description of the SAMIPS components and their Balsa specification. Sections \ref{Evaluation and Refinement}, \ref{Evaluation of the Asynchronous Forwarding Mechanism} and \ref{Evaluation of the Multi-Colour Algorithm} present respectively a quantitative evaluation of the entire processor and the mechanisms that have been developed for data and  control hazards. Section \ref{Speed Based Refinement} presents a critical path analysis and  model-level design optimisations for SAMIPS and a detailed quantitative analysis of the resulting modifications. Finally, section \ref{Summary and Conclusions} concludes the paper.

\section{Balsa}
\label{Balsa}

\begin{figure}
\begin{center}
\includegraphics[width=0.45\textwidth]{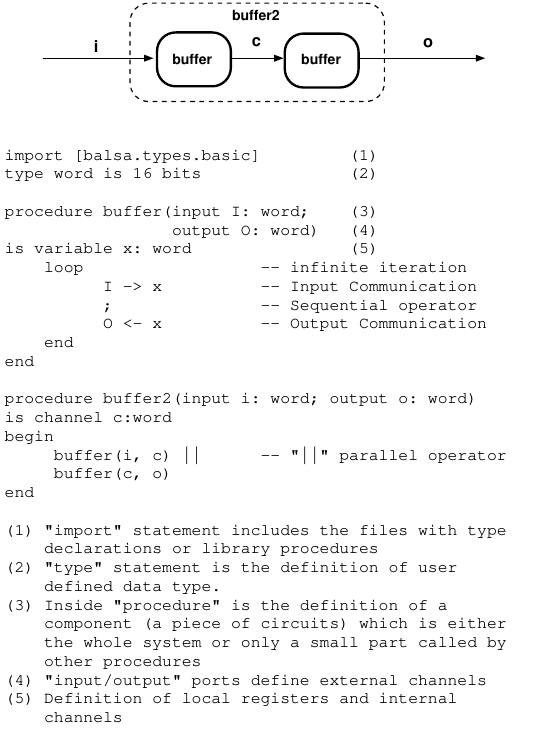}
\end{center}
\caption{Two-place buffer}
\label{Two-place buffer}
\end{figure}

Balsa uses CSP-based constructs to express Register Transfer Level design descriptions in terms of channel communications and fine grain concurrent and sequential process decomposition. Figure \ref{Two-place buffer} provides an illustrative Balsa example of a two a two-place buffer constructed from two single-place buffers; a list of comments are provided to explain each instruction. 
One notable and often used command of Balsa, which has not been illustrated in the example is \textsf{arbitrate} command for the arbiter construction. 

\begin{figure}
\begin{center}
\includegraphics[width=0.45\textwidth]{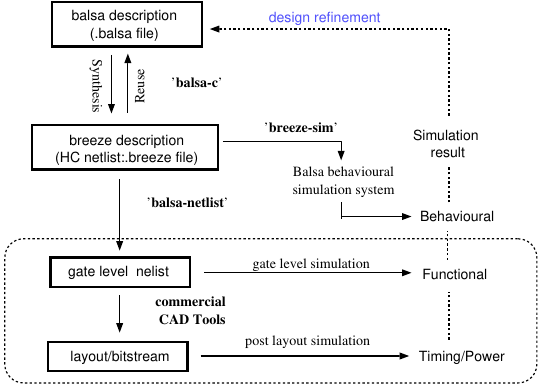}
\end{center}
\caption{Balsa System}
\label{Balsa System}
\end{figure}

The Balsa system provides a top-to-bottom design flow, shown in Figure \ref{Balsa System}. Descriptions of designs (\textit{.balsa} file) are then translated (\textit{\textbf{balsa-c}}) into implementations in a syntax directed-fashion with language constructs being mapped into networks of parameterised instances of "\textit{handshake components}" (\textit{.breeze} file) each of which has a concrete gate level implementation \cite{Berkel93}.  A handshake circuit is a network of handshake components connected via a number of handshake channels. Handshake circuits have been shown to be delay insensitive \cite{Berkel88}. For the same Balsa descriptions, there may be more than one handshake circuit implementations and the one selected depends on the particular compiler used. Balsa has about 48 handshake components in total and 30 among them are very frequently used. Each component has a unique name, symbol, definition and several implementations (based on different technologies). 

\begin{figure}[t]
\begin{center}
\includegraphics[width=0.4\textwidth]{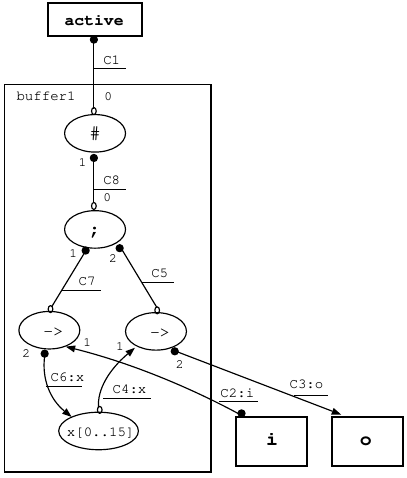}
\end{center}
\caption{Handshake Circuits for buffer of Figure \ref{Two-place buffer}}
\label{handshake circuits for buf1}
\end{figure}

 The automatically compiled Balsa handshake circuit for the single-place buffer example of Figure  \ref{Two-place buffer} is shown in Figure \ref{handshake circuits for buf1}. Each circle presents a handshake component. The \textit{activation port} (the active rectangle) starts the operation of the \textit{Repeater} (marked with \#) 
which initiates a handshake with the \textit{Sequencer} (marked with ;). 
\textit{Sequencer} first issues a handshake to left-hand \textit{Fetch} 
component, then the variable \textit{x} and finally right-hand \textit{Fetch} 
component. Once these operations finish, \textit{Sequencer} handshake with 
\textit{Repeater} to start the cycle again.

A number of tools are available to process the breeze handshake files. \textit{\textbf{balsa-netlist}} 

\begin{itemize}
\item \textbf{\textit{balsa-netlist}}: automatically generates CAD native netlist files, which can then be fed into the commercial CAD tools that further synthesize the netlist to the fabricable layout.

\item \textbf{\textit{breeze-cost}}: gives guideline area cost figures depending on a particular back-end implementation.  The cost units presented are linear microns of standard cells based on an 1$\mu$m library with a cell pitch of 37.5$\mu$m and a typical density of about 2/3 cells, 1/3 routing.

\item \textbf{\textit{breeze2ps}}: produces a postscript file with the handshake circuits graph.

\item \textbf{\textit{breeze-sim}}: performs the behaviour level simulation of the Balsa design. The simulation result can then be analysed through a waveform viewer, e.g. Gtkwave and SimVision.

\item \textbf{\textit{breeze-sim-control}}: is a graphical interface for simulation and visualisation.

\item \textbf{\textit{balsa-net-tran-cost}}: calculates the actual number of transistors based  on different technologies used after the gate level netlists are generated.

\item \textbf{\textit{balsa-mgr}}: a graphical interface of the Balsa design environment with project management facilities.

\end{itemize}

\begin{figure}
\begin{center}
\includegraphics[width=0.45\textwidth]{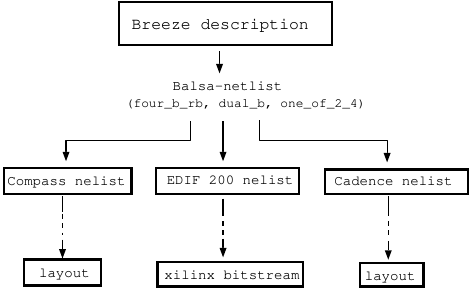}
\end{center}
\caption{Balsa Backend}
\label{Balsa Backend}
\end{figure}

Three commercial CAD systems are supported by Balsa: Compass Design Automation tools from Avant, the Xilinx Alliance FPGA design 
tools and Cadence Design Framework II (Figure  \ref{Balsa Backend}). During the target CAD netlist generation process, three data encoding technologies are provided for different design
requirements:
\begin{itemize}
\item {\em four\_b\_rb}: a bundled-data encoding with 4-phase broad/reduced broad protocol;
\item {\em dual\_b}: a delay-insensitive dual-rail encoding with broad synchronisation channel;
\item {\em one\_of\_2\_4}: a delay-insensitive 1-of-4 encoding. 
\end{itemize}

For SAMIPS the Cadence toolkit has been utilised. The Balsa description is translated into Cadence Verilog gate level netlists and then further synthesised to the final layout using the SGST HCMOS8D process in 0.18$\mu$m 6 layer metal technology which runs the core cells at 1.8v and the pad ring at 3.3v. 

Three levels of simulation are supported, namely behavioural, gate (functional) and physical (time/power). Behavioural simulation is conducted by \textbf{\textit{breeze-sim}}, while memory related behavioural 
simulation relies on the LARD toolkit \cite{Ende98}; the \textit{breeze2lard} tool helps to translate the .breeze file 
to a LARD simulation model. The native simulators of the supported commercial CAD tools carry out the other two low levels simulation (figure \ref{Balsa System}). There is also a parallel version of  \textit{breeze-sim} to conduct distributed simulation \cite{10.1007/978-3-642-45037-2_29,1612862}, while the balsa system also incorporates formal verification capabilities \cite{Wang04,Wang05}.

\section{Asynchronising MIPS}
\label{Asynchronising MIPS}

The MIPS  (\textit{Microprocessor without Interlocking Pipeline Stages}) family of microprocessors have two basic designs, namely the 32-bit MIPS32 and 64-bit MIPS64. The 32-bit series processors are based on a five-stage pipeline architecture while The 64-bit series include both superpipeline machines (such as R4000) and superscalar machines (such as R10000). 

For the development of SAMIPS,  MIPS32 as exemplified by the R3000 processor \cite{Kane92} was utilised as a base synchronous architecture for two main reasons: (a) one of the research goals has been to investigate the suitability of the Balsa synthesis system for a complex system design; there are  other asynchronous implementations of MIPS32 which can provide a basis for comparison, and (b) MIPS32 is a comparatively simple, clear and well documented architecture, yet presenting challenging problems for its asynchronous implementation that call for novel solutions. 

This section discusses the  approach that has been taken for asynchronising MIPS and the architectural features of MIPS that have been inherited by SAMIPS. For a more detailed description of MIPS and its ISA the reader is referred to \cite{Kane92,MIPSISA}. 
\subsection{MIPS R3000 Functional Blocks}
\begin{figure}
\begin{center}
\includegraphics[width=0.45\textwidth]{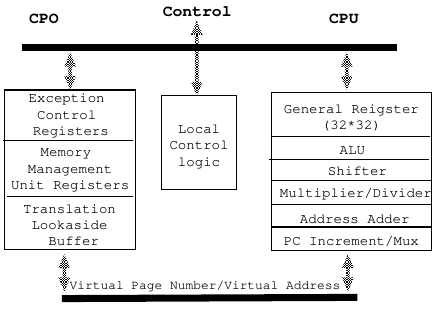}
\end{center}
\caption{R3000 Function Block}
\label{R3000 Function Block}
\end{figure}

MIPS R3000 consists of two tightly-coupled processors, namely a full 32-bit RISC CPU and a system control coprocessor CP0, as illustrated in Figure \ref{R3000 Function Block}. These are usually implemented on a single chip and can be extended with three off chip coprocessors. The CPU has 32 32-bit general-purpose registers, two 32-bit registers for multiply and divide operations (HI and LO), the Program Counter (PC), an ALU (Arithmetic Logic Unit), a Shifter, a Multiplier/Divider and an Address Adder. 

CP0 includes the exception and control units and a memory management system which provides hardware support for address translation in the form of an on-chip 64 entry 
 Translation Lookaside Buffer (TLB).

SAMIPS focuses on the design of the processor core. A simplified CP0 has also been implemented for test purposes.

\subsection{Instruction Set}

In MIPS R3000, there are three types of instructions, namely \textit{I-TYPE (immediate)}, \textit{J-TYPE (jump)} and \textit{R-TYPE (register)}, as illustrated in Table \ref{SAMIPS Instruction Type}.

\begin{table}
\caption{SAMIPS Instruction Type}
\begin{center}
\renewcommand{\baselinestretch}{1}
\begin{tabular}{l|l|l|l|l|l|l|}  \cline{2-7} 
 I-TYPE: &  Opcode & rs & rt & \multicolumn{3}{c|}{immediate} \\ \cline{2-7}
 J-TYPE: &  Opcode & \multicolumn{5}{c|}{target} \\ \cline{2-7} 
 R-TYPE: &  Opcode & rs & rt & rd & sa & funct  \\ \cline{2-7}
\end{tabular}
\renewcommand{\baselinestretch}{1.5}
\end{center}
{\small \textit{Opcode} is the operand code, \textit{rs}, \textit{rt}, \textit{rd} are register numbers, \textit{immediate} is either an immediate operand or
offset for branch/memory address calculation, \textit{target} is for jump target offset, \textit{funct} field is an supplement to Opcode and \textit{sa} is a shift 
amount.}
\label{SAMIPS Instruction Type}
\end{table}

\begin{figure}
\begin{center}
\includegraphics[width=0.45\textwidth]{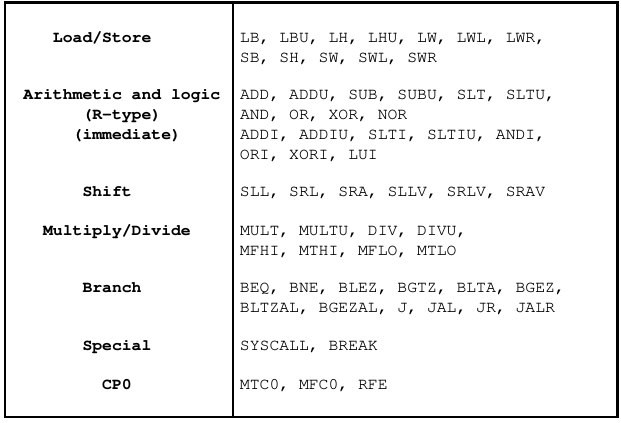}
\end{center}
\caption{SAMIPS Instruction Set}
\label{SAMIPS Instruction Set}
\end{figure}

SAMIPS ISA is fully compatible with MIPS ISA, implementing all the CPU instructions of R3000 and three CP0 instructions for exception handling.  Figure \ref{SAMIPS Instruction Set} shows the instruction set implemented by SAMIPS, divided into 6 groups according to different functionalities. 

\begin{itemize}
\item \textbf{Load/Store Instructions}. Load/store instructions execute memory access operations, which transfer data between the register bank and the memory. They are all I-TYPE instructions and the memory virtual address is calculated by adding the value in the base register with the 16-bit, signed immediate offset. This virtual address is translated later by the TLB in CP0 to the physical memory address. The data transferred can be a 32-bit word, a 16-bit half word or a 8-bit byte.

\item \textbf{Arithmetic and Logic Instructions}. This group of instructions carry out all the basic computational operations of the processor except for multiplications and divisions. They are either R-TYPE (both operands are read from CPU registers) or I- TYPE (one operand is a 16-bit immediate value) instructions.

\item \textbf{Shift Instructions}. There are two types of shift operations in MIPS R3000, namely shift with a constant distance specified in the \textit{sa} field and shift with a variable distance based on the low order five bits of the general register \textit{rs}. All shift instructions are R-TYPE.

\item \textbf{Multiply/Divide Instructions}. Results of the multiplication of two 32-bit numbers may be larger than 32-bit, but are definitely within 64-bit. Therefore, multiplication and division results are first calculated and stored in two special purpose registers, namely HI and LO; then sent to CPU general registers by other instructions. Instructions in this group are also R-TYPE.

\item \textbf{Branch Instructions}. Branch operation may potentially change the control flow of a program. MIPS supports two types of branch operations: (a) an unconditional jump is absolute, and the target address is specified in the 26-bit \textit{target} field (J-type) or in a general register \textit{rs} (R-type); (b) a conditional branch is based on the comparison result of two registers, and the target address is the current PC added by the 16-bit offset, shifted left two bits and sign-extended (I-TYPE).

\item \textbf{Special Instructions}. Instructions in this group perform special system tasks. There are only two special instructions in MIPS R3000, namely system calls and breakpoint, and both of them are R-TYPE.

\item \textbf{CP0 Instructions}. The main function of CP0 is to perform the memory management and handle exceptions. CP0 operations include maintenance and modification of the coprocessor registers and data transfers between them and the processor core. During the exception handling, three (out of seven in total) operations are used and therefore implemented in SAMIPS, namely {\small MTCO} (move to CP0), {\small MFCO} (move from CP0) and {\small RFE} (restore from exception).

\end{itemize}

\subsection{The Datapath}
\label{The Datapath}
\begin{figure}
\begin{center}
\includegraphics[width=0.5\textwidth]{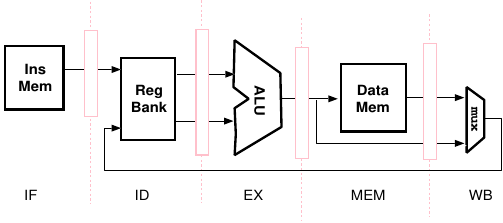}
\end{center}
\caption{MIPS R3000 Five Stage Pipeline}
\label{Stage}
\end{figure}

The MIPS R3000 processor is based on a pipeline with five stages referred to Instruction Fetch (IF),
Decode/Register File Read (ID), Execution or Address Calculation (EX), Memory Access (MEM) and Register Write-back (WB), shown in Figure \ref{Stage}. It is a Harvard architecture, with two memory interfaces referred to as \textsf{InsMem} and \textsf{DataMem} in the figure, one for instruction fetch and one for data access respectively. The separate memory interfaces can be exploited by either swapping memory access between two ports through an arbitrator or using two caches, namely the instruction cache and the data cache. MIPS has adopted the latter approach with different caches for instructions and data in the IF and MEM stages respectively.

\begin{figure}
\begin{center}
\includegraphics[width=0.5\textwidth]{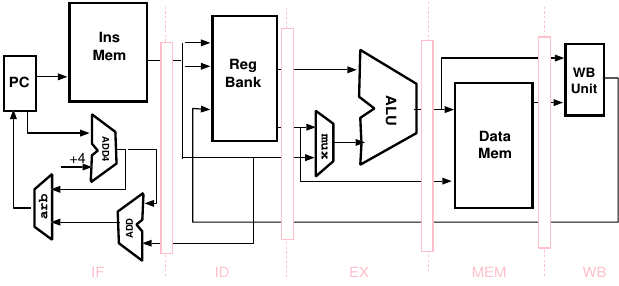}
\end{center}
\caption{The Datapath of SAMIPS}
\label{The Datapath of SAMIPS}
\end{figure}

\begin{figure}
\begin{center}
\includegraphics[width=0.5\textwidth]{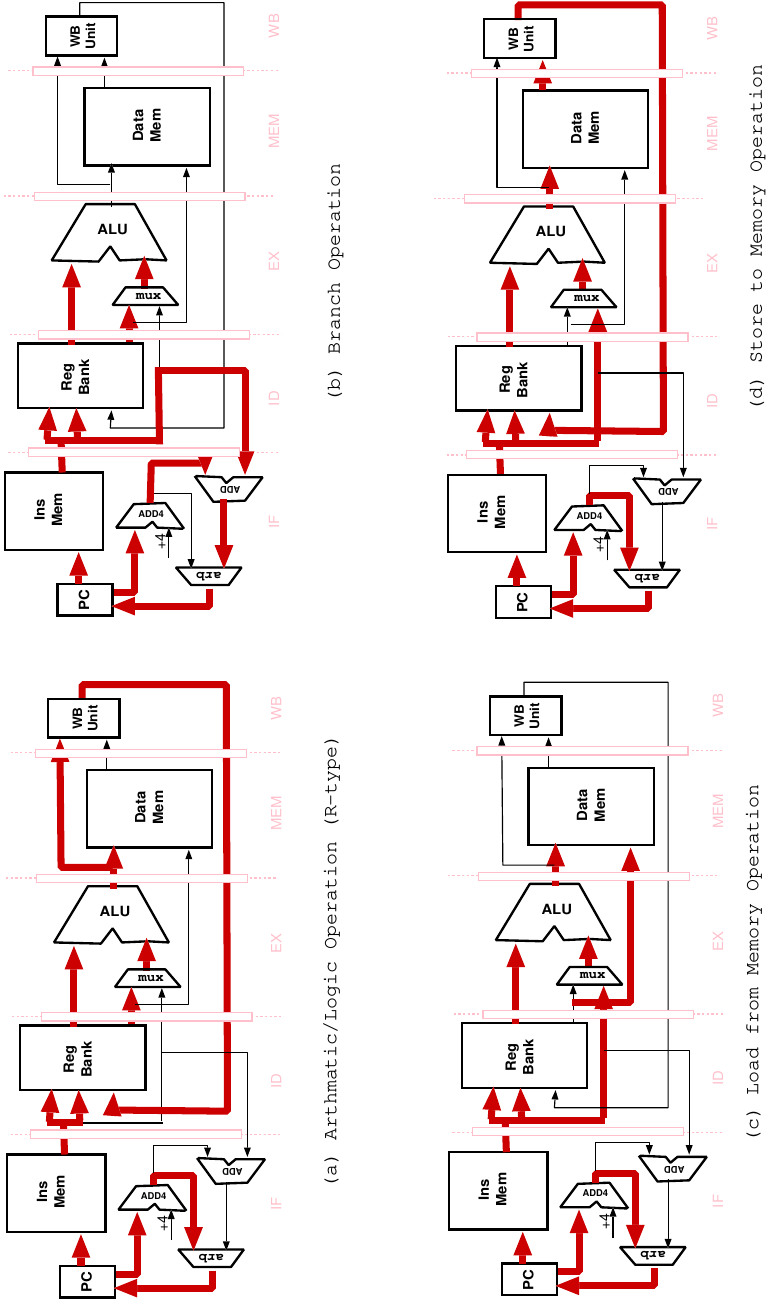}
\end{center}
\caption{The Datapath Activities in SAMIPS}
\label{Datapath Activities}
\end{figure}

An important decision for the design of SAMIPS was to adhere to the five-stage pipeline of the synchronous MIPS. Two main reasons lie behind this decision. Firstly, a five-stage pipeline design can provide a basis for comparison with the synchronous MIPS R3000 implementation. Secondly, the five-stage pipeline introduces challenging hazard-related problems that call for innovative asynchronous solutions. Therefore, the datapath of SAMIPS remains very similar to that of its synchronous counterpart, as illustrated in Figure \ref{The Datapath of SAMIPS}. IF generates the address of the next instruction to be fetched. The address is normally the contents of the PC incremented by four, or in the case of a branch, the branch target address (offset) which arrives from the ID stage. In synchronous MIPS, a multiplexer is used to choose between the two. In SAMIPS, this has been replaced by an arbiter, since the arrival of branch target offsets is nondeterministic. The new instruction is fetched from the \textsf{InsMem} and passed to the ID stage, where it is decoded into different parts and the operands are read from the \textsf{RegBank}. In the EX stage the instruction is executed in the \textsf{ALU}, which calculates add, logical, shift, etc., results and decides whether a branch is taken or not. If the ALU result is a memory address, it is passed to the \textsf{DataMem} for an memory operation, otherwise it is directly sent to the \textsf{WBUnit} in the WB stage to be written back to the \textsf{RegBank}. 


Figure \ref{Datapath Activities} shows the datapath activities for different types of operations, including arithmetic/logic operations (R-TYPE), branch operations and load/store operations. The thick arrows indicate the data flow. For R-TYPE arithmetic/logic operations, both operands are read from the \textsf{RegBank}, and the execution result is finally written back to the \textsf{RegBank}. For branch operations, the branch decision is calculated in the \textsf{ALU} and will be sent back to the IF stage through a control signal. Both load and store operations need to calculate the memory address in the \textsf{ALU} and access the main memory or the data cache through the \textsf{DataMem} interface. 
  
\subsection{The Control}
\begin{figure}
\begin{center}
\includegraphics[width=0.5\textwidth]{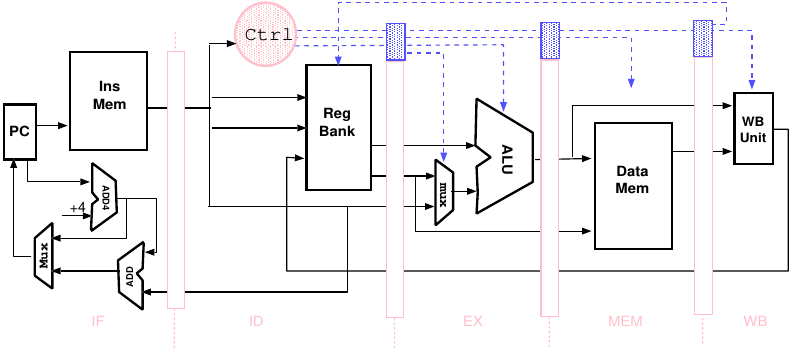}
\end{center}
\caption{The Control of Synchronous MIPS R3000}
\label{Control of Synchronous MIPS R3000}
\end{figure}

\begin{figure}
\begin{center}
\includegraphics[width=0.5\textwidth]{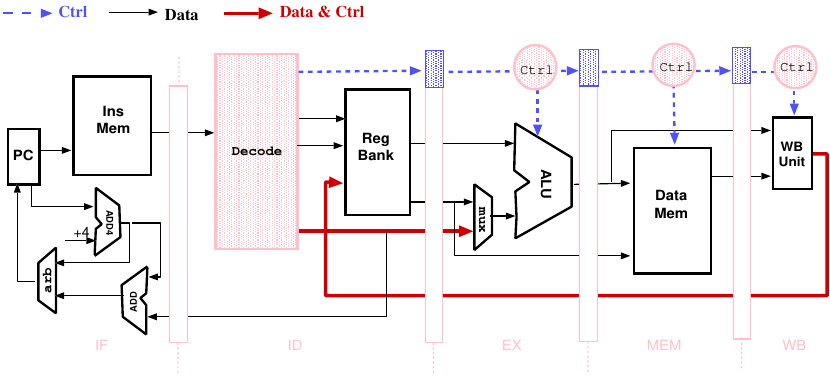}
\end{center}
\caption{The Control of SAMIPS}
\label{Control of SAMIPS}
\end{figure} 

The synchronous  MIPS R3000 processor utilises a centralised control unit in the ID stage to produce the necessary control signals, which propagate through the pipeline together with the data to drive circuits in the different stages of the datapath, as shown in Figure \ref{Control of Synchronous MIPS R3000}. 

An asynchronous system is a data-driven system, whereby each part of the circuits is activated when there are requests in its inputs. Thus the scheme used in synchronous MIPS provides a natural basis to generate and distribute the control information in SAMIPS too. A Decode unit is placed in the ID stage to perform the instruction decoding and generate the control signals required for different stages; these signals will thereafter follow the data through the pipeline, as shown in Figure \ref{Control of SAMIPS}. Certain control information is bundled with the data passed between stages, i.e. the control for the \textsf{mux} in the EX stage and the write back control for the \textsf{RegBank} in the WB stage, which are shown in the thick arrows in the figure.
\subsection{Exceptions}
\label{Exceptions}

\begin{figure}[t]
\begin{center}
\includegraphics[width=0.4\textwidth]{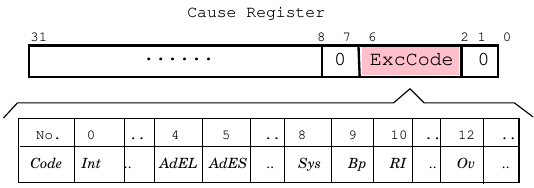}
\end{center}
\caption{ExcCode Field of the Cause Register}
\label{ExcCode Field of the Cause Register}
\end{figure}

In addition to the branch operations, there are two other types of events, which may change the control flow of a program, namely exceptions and interrupts. An exception is an event from within the processor, e.g. an arithmetic/address error or a system call. An interrupt is from the outside world, e.g. an I/O device request. Interrupts are considered as a special type of exception in MIPS architecture. An exception will enforce the processor to terminate the current job, save the current state (in \textit{Status} register), the current address (in \textit{EPC}, the Exception Program Counter register), and cause of the exception (in \textit{Cause} register ExcCode field, see Figure \ref{ExcCode Field of the Cause Register}) in CP0 and start to execute the exception handling program. After an exception, \textit{EPC} points at the current instruction which caused this exception, unless it is in a ``branch delay slot" (see Section \ref{Solution in Synchronous R3000}), in which case the \textit{EPC} points to the preceding branch instruction. SAMIPS implements six types of exceptions in total as explained below. 

\begin{itemize}
\item \textbf{Arithmetic Overflow Exceptions}. An arithmetic overflow exception occurs when the calculation result of an arithmetic instruction, like ADD, ADDI or SUB, is 2's-complement overflowed.  Common exception vector (starting from base address Ox80000000) is used for this exception and the \textit{Ov} code is set in the \textit{Cause} register. This exception is generated in the EX stage.
\item \textbf{Reserved Instruction Exceptions}. A reserved instruction exception occurs when the processor is going to execute an undefined instruction. Common exception vector is used for this exception and the \textit{RI} code is set in the \textit{Cause} register. This exception is generated in the ID stage.
\item \textbf{Address Error Exceptions}. An address error exception occurs when (a) load/store a word not aligned on a word boundary or a halfword not on a halfword boundary; (b) access a kernel address space (address from 0x80000000 to 0xffffffff) from a user mode (address from 0x00000000 to 0x7fffffff). Common exception vector is used for this exception and the \textit{AdEL} (for load operation) or \textit{AdES} (for save operation) code is set in the \textit{Cause} register. This exception is generated in the MEM stage. 
\item \textbf{Breakpoint Exceptions}. A breakpoint exception occurs when the processor executes a BREAK instruction. Common exception vector is used for this exception and the \textit{BP} code is set in the \textit{Cause} register. This exception is generated in the ID stage.
\item \textbf{System Call Exceptions}. A System Call exception occurs when the processor is going to execute a SYSCALL instruction. Common exception vector is used for this exception and the \textit{Sys} code is set in \textit{Cause} register. This exception is generated in the ID stage.
\item \textbf{External Interrupts}. An interrupt occurs when an external event happens, which generates a request to CP0. CP0 receives the request and sets one of the eight interrupt conditions in the \textit{Cause} register according to the exception enabled and priority conditions. Common exception vector is used for this exception and the \textit{Int} code is set in the \textit{Cause} register.
\end{itemize}

\section{Data Hazards}
\label{Data hazards problems}

In a pipelined system, the execution of instructions is overlapped. However, there are situations in which the execution of next instruction has to be delayed, because it depends on the result of a previous instruction still in the pipeline. Such a situation is referred to as a \textit{data hazard} \cite{Patt97}. The cause of a data hazard is the inherent dependency between instruction operands and results. Due to the overlapped execution of instructions, it is possible that an incorrect register reading or write operation will occur and finally affect the ALU calculation result. In other words, the sequence of the reading (fetching operands) and writing (storing results) operations of the same register is not interchangeable.  Dependencies are typically classified as RAW (also known as true dependencies), WAW (also known as output dependencies) and WAR (also known as anti dependencies). The problem of data hazards is exacerbated in asynchronous pipelines due to the lack of global synchronisations. The fundamental problem is that in a distributed, nondeterministic system such as an asynchronous architecture, global snapshots of the state are not easily or efficiently obtained. 

Different solutions to the data hazard problem have been proposed for both synchronous and asynchronous systems, each taking advantage of the particular processor architectural characteristics. Possible synchronous solutions to the RAW dependencies include: enforcing the compiler to reorder the instructions to remove the dependency \cite{Patt96}; locking the register pending to be written \cite{Bayko}; and getting the missing item earlier from the subsequent stages in the pipeline, which is referred to as \textit{result forwarding} \cite{Patt97}. Synchronous solutions for WAW dependencies include pipeline stalling \cite{Gilbert97}, register renaming \cite{Yeager96,Toma67}, scoreboarding \cite{Patt96} and the use of a reorder buffer \cite{SmP188}.  Possible synchronous solutions for WAR dependencies include: pipeline stalling \cite{Ibbe82} and  register renaming. Asynchronous solutions to the data hazard problem have utilised  register locking \cite{588033,Christ98,Amde03,5563539},  forwarding \cite{Furber97,Sproull94,ElstonS95,5563539,7314387}, reorder buffer \cite{Gilbert97, Furber00a, Martin97, kol98}, pipeline stalling \cite{Rich96} and register renaming  \cite{kol98}.

\subsection{Result Forwarding in Synchronous R3000}
\begin{figure}
\begin{center}
\includegraphics[width=0.5\textwidth]{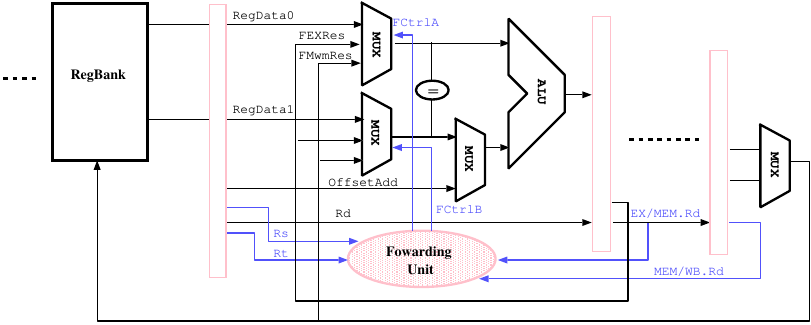}
\end{center}
\caption{Register Bank with Result Forwarding \cite{Patt97}}
\label{Register Bank with Result Forwarding}
\end{figure} 

In synchronous MIPS R3000 only RAW dependencies occur, because neither out-of-order completion nor out-of-order issuing is supported. 
MIPS uses forwarding to handle the data hazard problem, as illustrated in Figure \ref{Register Bank with Result Forwarding}. Each operand of the ALU has three sources, namely RegBank, a forwarded result from the EX stage or a forwarded result from the MEM stage. The decision as to which source the ALU should use at any particular moment is taken by a centralised control unit, which drives the multiplexers at the ALU input.

Referring to the example shown in Figure \ref{Data hazards: An Example}, the following behaviour will be exhibited: 

\begin{figure}
\begin{center}
\includegraphics[width=0.5\textwidth]{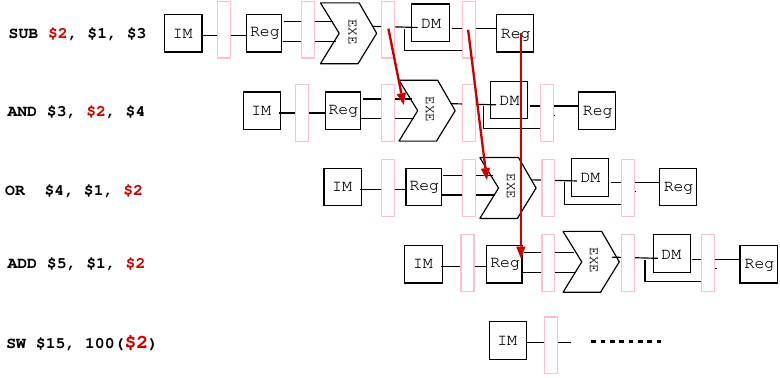}
\end{center}
\caption{Data hazards: An Example \cite{Patt97}}
\label{Data hazards: An Example}
\end{figure} 

\begin{itemize}
\item For instruction {\small AND}, operand (\$2) is the result of instruction SUB and can be forwarded from the register between the EX and the MEM stage; 
\item For instruction {\small OR}, operand (\$2) is the result of instruction SUB and can be forwarded from the MEM and the WB stage; 
\item For instruction {\small ADD}, the ID stage is separated into two substages: first decoding and writing register files, then reading from register files
\item For instruction {\small SW}, a data hazard will never happen, since the execution of instruction {\small SUB} has already finished.
\end{itemize}

\begin{figure}[t] 
\begin{center}
\includegraphics[width=0.5\textwidth]{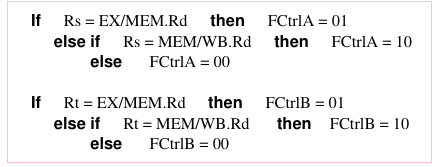}
\end{center}
\caption{Comparison Algorithm inside the \textsf{Forwarding Unit}}
\label{syncFA}
\end{figure} 

For each instruction entering the EX stage, the \textsf{Forwarding Unit} (Figure \ref{Register Bank with Result Forwarding}) needs to compare the source register addresses of the  instruction with the destination addresses of the previous ones which have already been passed to the following pipeline stages. As illustrated in the figure, the comparison is between \textit{Rs}, \textit{Rt} (source addresses) and \textit{EX/MEM.Rd} (the destination address passed to the MEM stage), \textit{MEM/WB.Rd} (the destination address passed to the WB stage). Based on this, control signals to drive the multiplexers are generated, namely \textit{FCtrlA} and \textit{FCtrlB}. The comparison algorithm inside the \textsf{Forwarding Unit} is illustrated in Figure \ref{syncFA}.

\subsection{Asynchronous Result Forwarding in SAMIPS}
\label{Asynchronous Forwarding in SAMIPS}

\begin{figure}
\begin{center}
\includegraphics[width=0.5\textwidth]{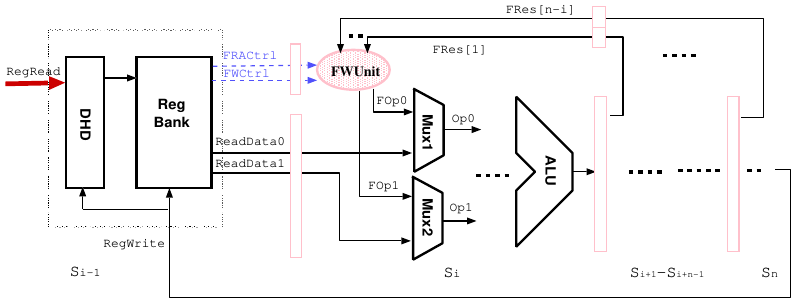}
\end{center}
\caption{Asynchronous Forwarding}
\label{Asynchronous Forwarding}
\end{figure} 




Since in MIPS R3000 only RAW dependencies occur, asynchronous solutions for the WAW and WAR are not directly related. However some of them can be inspiring, such as the forwarding approach adopted in AMULET2 and the idea of the \textsf{look up} unit used in AMULET3's reorder buffer mechanism. 

For the SAMIPS design, the objective is to develop a mechanism that would allow the centralised Forwarding Unit to be removed and have the control signals that drive the multiplexers at the ALU somehow sent down the pipeline along with the corresponding data. The control signals should specify which stage will potentially forward a result  and whether the result will be forwarded or whether it is needed. The solution that has been devised does not depend on global current knowledge but rather on knowledge of the past. It is based on two observations for the system: 

\begin{enumerate}
\item  the sequence of instructions will be preserved during processing (after all, even highly concurrent and non-deterministic, this is still a von-Neumann system);
\item a data hazard can be detected when doing register reading at the ID stage if some history information is recorded.
\end{enumerate}

Figure \ref{Asynchronous Forwarding} depicts the proposed forwarding mechanism in the context of a generalised datapath with n-stages ($S_1$ to $S_n$); the execution stage (EX) is the \textit{i}th stage ($S_i$). \textit{DHD}, a data hazard detection unit, is
first added into the \textsf{RegBank}. The two control signals, namely \textit{FRACtrl} and \textit{FWCtrl}, are generated by \textit{DHD} and sent to the \textsf{FWunit} in the 
stage $S_i$, which is also the final destination of the forwarded results:
\textit{FRes[1]} to \textit{FRes[n-i]} from stage $S_{i+1}$ to stage $S_n$ respectively (\textit{FEXRes} and \textit{FMEMRes} in SAMIPS). Based on the information on \textit{FRACtrl} and \textit{FWCtrl}, these forwarded results are either only acknowledged or used as CPU operands. Two multiplexers, referred as \textsf{Mux1/2} in Figure \ref{Asynchronous Forwarding}, are used to merge the data from the register file and the 
\textsf{FWunit}.


For the above mechanism to work, the control signals sent to the \textsf{FWunit} must carry the following information:

\begin{enumerate}
\item The \textit{FRACtrl} (Forwarded Result Acknowledgment
Control) need to specify which following stages will potentially generate a result in each cycle. This is because the following stages do not know whether their results are going to be used immediately before the succeeding instructions detect a data hazard, and thus could only forward back all the newly generated results.  In SAMIPS, the result may come from two places, namely the ALU output in the MEM stage and the memory result in the WB stage.

\item The \textit{FWCtrl} (Forwarding Control) need to specify whether the forwarded results are going to be used by the following instructions. This control signal is generated during the detection of data hazards.
\end{enumerate}

\subsubsection{Data Hazard Detection Table and Queue}
\label{Data Hazard Detection Table/Queue}

\begin{figure}[t]
\begin{center}
\includegraphics[width=0.5\textwidth]{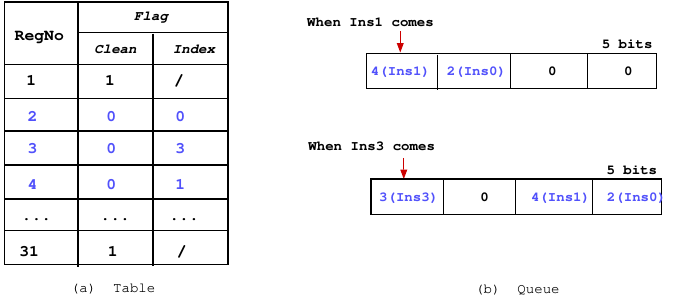}
\end{center}
\caption{Data Hazard Detection} 
\label{Data Hazard Detection Table} 
\end{figure}

To detect data hazards, a novel approach has been devised as part of the SAMIPS design. The approach utilises a \textit{DHD} unit inthe \textsf{RegBank} which keeps a record of all instructions pending to modify the register file and are still in the pipeline. There are different ways to implement the \textit{DHD}. 

One implementation is the DHDT (Data Hazard Detection Table), based on a table structure, as illustrated in Figure \ref{Data Hazard Detection Table}(a). The algorithm makes use of a variable \textit{CurIndex} which is incremented by one (modulo $n-i+2$, where \textit{n} is the number of  stages and \textit{i} is the stage where forwarding takes place  - modulo 4 in SAMIPS) every time a new instruction enters the \textsf{RegBank}.  The DHDT table contains additional information for each register referred to as the register \textit{Flag}. The \textit{Flag} consists of one bit (\textit{Clean}) indicating whether the register is pending to be written (1 sets the register to ``clean" and 0 to ``dirty"), and $\log_2(n-i+2)$ bits (\textit{Index}) essentially indicating which instruction will rewrite the register (or, in other words, which stage will forward the result). For SAMIPS where $n=5$ and $i=3$, at most four instructions can be in the pipeline at any time from the ID to the WB stage, and therefore two bits are enough for the \textit{Index} field.  Every time a register write instruction marks the register as dirty, the value of the current \textit{CurIndex} is copied into the \textit{Index} field for that register.  When a register read instruction arrives, the difference between the Index value of the register from the current value of the \textit{CurIntex} indicates how many instructions in the past the register was marked as dirty and therefore which instruction will rewrite the register. The storage cost introduced by this scheme inside the \textsf{RegBank} is the cost of the \textit{Flag} plus the cost of the \textit{CurIndex}. The total additional bits would be $(\log_2(n-i+2)+1)\times 31 + \log_2(n-i+2)$. 

An alternative implementation is the DHTQ (Data Hazard Detection Queue), based on a queue structure, as illustrated in Figure \ref{Data Hazard Detection Table}(b). Unlike the table, it only records the 5-bit address of each register, pending to be written by the instructions still in the pipeline. Size of the queue is $n-i+2$, and hence is 4 in SAMIPS. Each instruction has an entry inside the queue, before it gets removed from the pipeline, either a 5-bit write back address or 0 if it does not need to modify the \textsf{RegBank} (since register 0 always return 0 and is not allowed to be modified in MIPS, we are able to use it to indicate non write back operations). When a new instruction arrives, the queue pushes it into the tail and pops out one item from the top. Thus, DHDQ provides a snapshot of the pipeline at any particular moment. By looking at the DHDQ, a register read instruction can decide which preceding instruction will write back the register. When a register-writing request arrives, it will rewrite that register and reset the first matched entry it finds in the queue back to 0. The total additional bits introduced for a queue scheme would be $(n-i+2)\times 5$. 

A DHDT is easier to understand, but from the formula, it is obvious that the storage cost of it is much higher than that of a DHDQ. Table \ref{Storage Cost of DHDT and DHDQ} shows the storage cost of DHDT and DHDQ with $i=3$ and $n \in \{5..9\}$.  

\begin{table}
\caption{Storage Cost of DHDT and DHDQ}
\begin{center}
\begin{tabular}{|l|c|c|c|c|c|} \hline
           & $n=5$   & $n=6$   & $n=7$   & $n=8$   & $n=9$  \\  \hline \hline
 DHDT      & 95      & 127     & 127     & 127     & 127    \\\hline
 DHDQ      & 20      & 25      & 30      & 35      & 40     \\ \hline
\end{tabular}
\label{Storage Cost of DHDT and DHDQ}
\end{center}
\end{table}

With the information provided by the DHDT/DHDQ, the \textit{FWCtrl} is generated. It is also possible to extract the \textit{FRACtrl} from them, but for simplicity in operation a 1-bit queue - FRAQ (Forward Result Acknowledgement Queue) is maintained. This bit simply indicates whether there is going to be a forwarded result in each cycle. The size of the queue is $(n-i)$, which implies the total number of stages that will potentially forward results. In SAMIPS, the size is 2. 
\subsubsection{The Forwarding Algorithm}
\label{The Forwarding Algorithm} 
To implement the forwarding mechanism, input \textit{RegRead} 
should contain three addresses, two for reading and one for writing, while \textit{RegWrite} should consist of the writing address and data (see figure \ref{Asynchronous Forwarding}). The proposed forwarding algorithm includes three parts, namely initialisation, register read and register write. 

\begin{figure}[t]
\begin{center}
\includegraphics[width=0.45\textwidth]{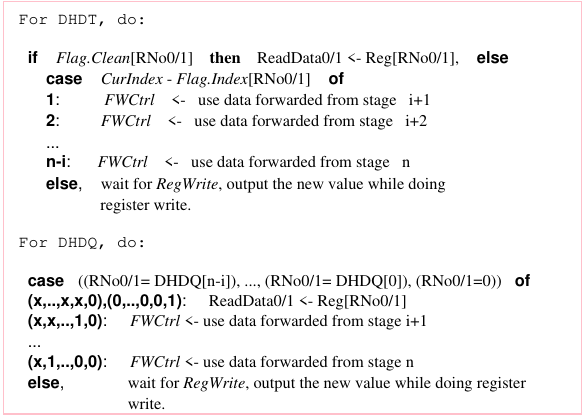}
\end{center}
\caption{Data Hazard Detection} 
\label{Data Hazard Detection} 
\end{figure}

\textbf{Initialisation}. During this period, both FRAQ and DHDT/DHDQ need to be emptied.  All items in FRAQ are set to zero. For DHDT, set all \textit{Flag.Clean} to 1 (it means clean) and \textit{CurIndex} to 0; for DHDQ, set all items in the queue to 0 (it means no register pending to be written). 

\textbf{Register Read.} \textit{RegRead} and \textit{RegWrite} are received through an arbiter since it is not guaranteed these two channels will not fire simultaneously. The innovative characteristic of the register read is that the data hazard is detected simultaneously with register reading based on the history information recorded in DHDT/DHDQ. The corresponding algorithm is illustrated in Figure \ref{Data Hazard Detection}. For DHDT, the algorithm first checks \textit{Flag.clean} and then compares \textit{CurIndex} with \textit{Flag.Index} of the registers to be read; for DHDQ, the algorithm compares the two register reading addresses in \textit{RegRead} with those saved in the queue. Based on the comparison result, the data hazard is detected. Besides the data hazard checking and register reading operations, the \textsf{RegBank} also needs to send out the present FRAQ as \textit{FRACtrl} and then update it according to the writing address contained in the current \textit{RegRead}.
\begin{table*}[t]
\caption{Data Hazards in SAMIS: Example} 
\begin{small}
 \begin{center}
 \begin{tabular}{|c|c|c|c|p{5cm}|}  \hline
 Instruction          &  FRAQ  & DHDT       & DHDQ   & Operation                   \\ 
                            &        &            &        &                                    \\ \hline \hline
           /                & (0, 0) & (1, 1, /)  & (0, 0, & Initialisation                     \\
	                    &        & (2, 1, /)  &  0, 0) &                                    \\
			    &        & ( ..... )  &	   &			                \\ \hline   
 SUB \textbf{\$2}, \$1, \$3 & (1, 0) & (1, 1, /)  & (2, 0, & \$3, \$1 are both read             \\  
                            &        & (2, 0, 0)  &  0, 0) & from RegBank                       \\ 
			    &        & (3, 1, /)  &	   &                                    \\ \hline  
 AND \$3, \textbf{\$2}, \$4 & (1, 1) & (2, 0, 0)  & (3, 2, & \$2 not valid in DHDT (appears     \\
                            &        & (3, 0, 1)  &  0, 0) & in DHDQ), using the result         \\
			    &        & (4, 1, /)  &        & forwarded from the MEM stage           \\ \hline
 OR  \$4, \$1, \textbf{\$2} & (1, 1) & (2, 0, 0)  & (4, 3, & \$2 not valid,                     \\
                            &        & (3, 0, 1)  &  2, 0) & using the forwarded result         \\
			    &        & (4, 0, 2)  &        & from the WB stage                      \\ \hline
 ADD \$5, \$1, \textbf{\$2} & (1, 1) & (2, 0, 0)  & (5, 4, & \$2 still not valid, no forwarding \\
                            &        & (3, 0, 1)  &  3, 2) & opportunity, waiting for a result   \\
			    &        & (4, 0, 2)  &        & coming back                        \\
			    &        & (5, 0, 3)  &        &                                    \\ \hline
 SW  \$5, 100(\textbf{\$2}) & (1, 1) & (2, 1, /)  & (5, 5, & \$2 should be valid again,         \\
                            &        & (3, 0, 1)  &  4, 3) & otherwise waiting for a result     \\
			    &        & (4, 0, 2)  &        & coming back                        \\ 
			    &        & (5, 0, 0)  &        &                                    \\ \hline
\end{tabular}
\end{center}
\end{small}

\label{DH example} 
\end{table*}

\textbf{Register Write.} 
\begin{figure}
\begin{center}
\includegraphics[width=0.5\textwidth]{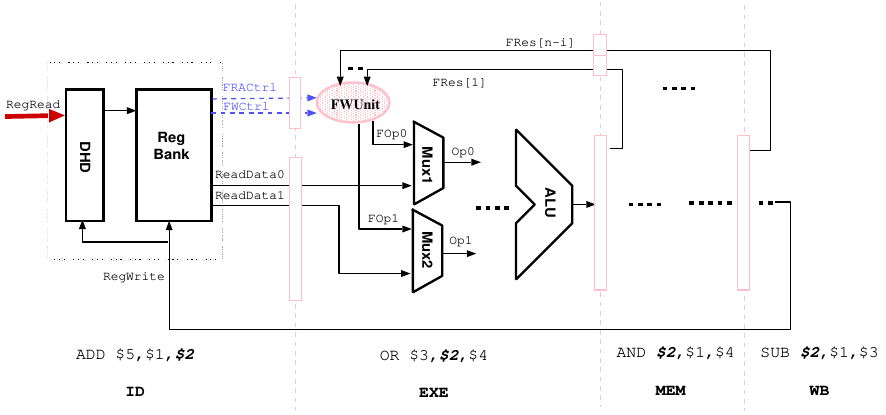}
\end{center}
\caption{Consecutive Instructions Modify the Same Register}
\label{ForwardE}
\end{figure} 
The new feature of the register write operation is that after updating the value inside the register files, the DHDT/DHDQ need to be adjusted accordingly as well. During updating DHDT, a problem may occur if two consecutive
instructions modify the same register. Considering a instruction sequence in SAMIPS datapath as shown in Figure \ref{ForwardE}, where both instructions {\small SUB} and {\small AND} will modify register \$2. In this case, the first \textit{RegWrite} signal will
mark the register ``clean'', which is of course wrong since the register is still to be modified by the next instruction. This problem can be
solved if \textit{RegWrite} carries with it a 2-bit snapshot of the value of the \textit{CurIndex} as it was when the instruction
passed through the \textsf{RegBank}. This value is then compared with the corresponding \textit{Flag.Index} inside the DHDT. The mismatch means there is at
least one other instruction waiting to rewrite this register in the
pipeline. 
\subsubsection{FWunit}
The \textsf{FWunit} is the final destination of forwarded results. Inside the
\textsf{FWunit} these forwarded results are acknowledged (based on the content of \textit{FRACtrl}) and sent to the ALU through two multiplexers (\textsf{Mux1} and \textsf{Mux2}) if necessary (based on the content of \textit{FWCtrl}). 

\subsubsection{Example}
As an illustrative example of the proposed forwarding mechanism, Table \ref{DH example}  presents a series of snapshots of the state of the DHDT/DHDQ and forwarding algorithm inside the \textsf{Regbank} during the execution of the instruction sequence of Figure \ref{Data hazards: An Example} in SAMIPS.

\section{Control Hazards}
\label{Control Hazards Problems} 
In a pipelined system, there are situations that prefetched instructions are invalid and should be removed although they have already entered the pipeline. Such a situation that execution of one instruction is cancelled due to the change of the flow of control is referred to as a \textit{control hazard} \cite{Patt97}. 
There are two situations when a control hazard may occur: branches and exceptions including interrupts (see section \ref{Exceptions}). 


A branch instruction is always declared explicitly in a program, and thus its occurrence is predictable. Since the following instructions are aware if there is a pending branch operation ahead, they are possible to be delayed a while until the decision is made and the new target address is generated. The delayed period is called the ``branch delay slot" \cite{Patt97}. Alternative approaches include splitting branches \cite{SH-5} and branch predictions \cite{Yeager96}. 



Exception dependencies on the other hand are unpredictable, and the processor must be able to flush the pipeline correctly. A processor is said to have \textit{precise exceptions} if it can be stopped so that the instructions just before the faulting instruction are completed and those after it can be restarted from scratch, otherwise it is said to have \textit{imprecise exceptions} \cite{Patt96}. The developed solutions include the Pending Consistent State (PCS) \cite{Hw87},  the history buffer \cite{SmP188}, the future file \cite{John91}, and the instruction windows \cite{Akk03}.  

\subsection{Control Hazards in Asynchronous Pipelines}
In synchronous pipelined systems, the depth of prefetching, or the number of instructions that have entered the processor and must be deleted in the event of a control hazard, is determined by the clock cycles and is thus deterministic. In an asynchronous microprocessor however the exact amount of prefetched instructions is nondeterministic and unpredictable. In this instance, the prefetching depth depends on the precise moment at which the prefetching is interrupted by a branch or an exception. The processor must be able to distinguish between instructions originating from the branch or exception target, which may be processed, and prefetched instructions at the time of the hazard, which must be discarded.

\subsubsection{Asynchronous Solutions to Branches}

In an asynchronous pipeline the maximum branch delay slot is still a fixed value, limited by the total number of buffers between pipeline stages. Thus, regardless of the inefficiency, if ``bubbles" are inserted according to the maximum branch delay slot, this problem can be solved. TTITAC-2 \cite{Takamura97} and Amde's asynchronous DLX Microprocessor \cite{Amde03}, both based on a five-stage pipeline, have two branch delay slots instead of one to prevent the processor from fetching any invalid instructions before a branch is determined. NSR \cite{Brunvand93} avoids the problem by pushing control decisions to the IF stage based on conditions set up in advance by the execution unit, so there is no prefetching as such. FRED \cite{Rich96} separates the branch operation to two phases, namely address generating and sequence change, which also avoid undoing prefetched instructions. Sun's counterflow processor \cite{Sproull94} takes advantage of the two asynchronous pipelines running in opposite directions, so that the hazard information flows backwards, invalidating prefetched instructions on its way \cite{Theo2004}.

\subsubsection{Asynchronous Solutions to Exceptions}

Solutions for branch related control hazards cannot necessarily solve hazards caused by exceptions, though the reverse is generally true. 
One approach is based on the idea of ``tag", a label attached to instructions. In Caltech's MiniMIPS \cite{Martin97}, the `tag" is called the ``valid-again" bit. When an exception occurs, the WB will keep sending a ``kill" message to the register unit to cancel results until it receives a request from the DECODE with the ``valid-again" bit, which is generated by the PC and sent along with fetched instructions to the DECODE unit. In AMULET1 \cite{588033}, the ``tag"  is a single bit used to ``colour"  the state of the processor at any particular moment. When an exception occurs the color changes and instructions whose color does not match that of the state of the processor are rejected. 
Another approach is the asynchronous reorder buffer developed in AMULET3, which is also an asynchronous solution for WAR data hazard problem. The execution results are reordered in the buffer and cannot be written to the register bank before the previous ones finish their operations. Therefore, the processor is able to draw a line between the results before and after an exception in the buffer. Similarly, the Instruction Window (IW) has been adopted to deal with exceptions in FRED.  There have also been some recent attempts to deal with interrupts in asynchronous pipelines \cite{9951785}.

\subsection{Control Hazards in Synchronous R3000}
\label{Solution in Synchronous R3000}
\begin{figure}
\begin{center}
\includegraphics[width=0.45\textwidth]{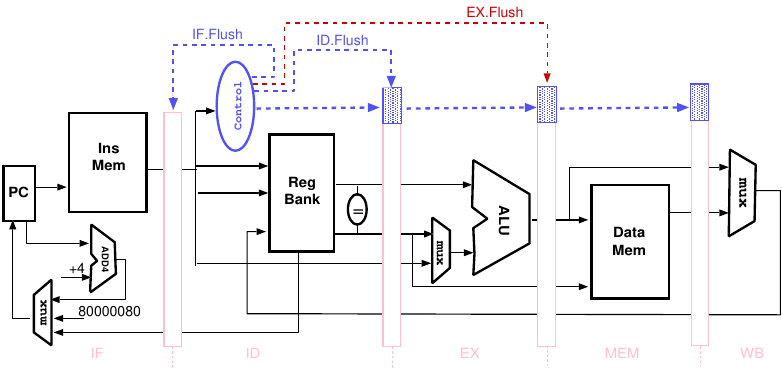}
\end{center}
\caption{Branch and Exception Handling in MIPS R3000}
\label{Exception Handling in MIPS R3000}
\end{figure}

In MIPS R3000, it takes a fixed time to execute an instruction, e.g. the execution of a logical instruction takes 1/clock rate $\times$ 5 cycles. If a branch decision is made in the EX stage, exactly two consecutive instructions are affected which indicates the branch delay slot is two. The control hazard caused by a branch can be solved by just inserting two ``bubbles" (NOP instructions in MIPS) after every branch instruction. R3000 adopts this approach, and to reduce the effect to the performance, it pushes the decision of a branch instruction back to the ID stage by adding a comparison logic on two outputs from the RegBank, so one NOP instruction is enough (Figure \ref{Exception Handling in MIPS R3000}). The efficiency is further improved by replacing the NOP with a useful instruction logically preceding the branch. MIPS R3000 supports precise exceptions. When an exception occurs, the \textsf{Control} unit of R3000 will send flushing pipeline control signals to the IF, ID and EX stages, as shown in Figure \ref{Exception Handling in MIPS R3000}, e.g. when it is an arithmetic overflow occurs in the EX, flush signals are sent to the IF and ID stages (although not shown in Figure \ref{Exception Handling in MIPS R3000}, the ALU overflow signal is an input to the \textsf{Control} unit). The execution of all affected instructions is thus cancelled. 


\subsection{Addressing Control Hazards in SAMIPS  }
\label{A Distributed Multi-Colour algorithm}
The fundamental problem is dealing with control hazards in asynchronous system without global synchronisation is that global snapshots of the state of the system at any particular moment are not easily obtainable. Therefore it is not obvious where the information about the change of the control flow is obtained and kept.  
The solution that has been devised to address this problem in the context of the design of SAMIPS is a  Distributed Multi-Colour algorithm which is based on the fundamental observation that stages that are deeper in the pipeline have higher priority than stages before them; a control transfer event invalidates any other events that may occur in pipeline stages earlier in the pipeline, even if the latter precede the former in time. Based on this, a colour vector with priority has been defined to represent the state of the processor at any particular moment. The algorithm was first introduced and described in detail in \cite{Theo2004} and is outlined below for the sake of completeness. Illustrative examples of the algorithm maybe found in \cite{Theo2004} while the algorithm has been formally verified using CSP \cite{Wang04, Wang05}.


\begin{figure}
\begin{center}
\includegraphics[width=0.45\textwidth]{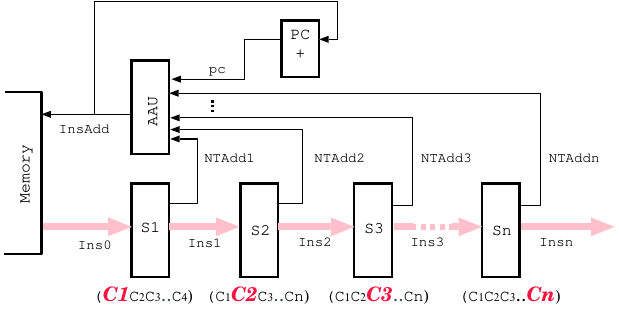}
\end{center}
\caption{Pipeline Stages Colour State Vectors}
\label{Stages}
\end{figure}

 
%
The algorithm defines the state of the system as a "Colour Vector with Priority":
\begin{equation}
C = (C_1, C_2..., C_n) \in C^n  
\end{equation}
{\em \small where $C$ is a set of colours $C = \{0,1\}$, $n$ is the number of stages in the pipeline and $C_i$ is the colour of the stage $i$. Priority of $C_i >$ Priority of $C_j$,  $i>j$.} \\

\begin{figure}
\begin{center}
\includegraphics[width=0.3\textwidth]{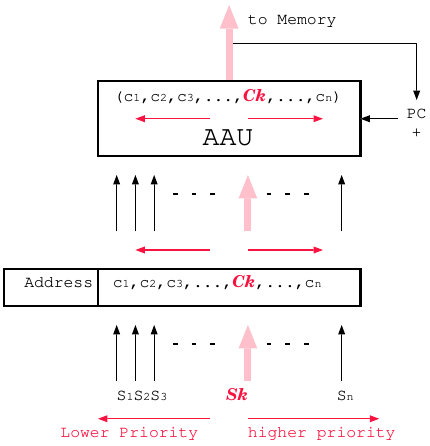}
\end{center}
\caption{The Address Arbitration Unit}
\label{AAU}
\end{figure}

The basic function of the algorithm is illustrated in Figure \ref{Stages}. Each pipeline stage maintains a copy of the colour vector state $C = (C_1, C_2, C_k,... C_n)$ which is sent together with the control transfer address to a prefetching unit, referred to as Address Arbitration Unit (AAU).   For each new instruction that arrives for execution at a stage, its colour state vector is compared against the state vector of the stage. If the stage's own colour bit is different than the corresponding bit in the instruction vector then, this instruction is one of the instructions following an instruction that has already caused a control hazard in the stage, and therefore the instruction is rejected.  Otherwise, the instruction is executed and the state vector of the instruction becomes its own. If any higher priority colour bit in the instruction is different than the corresponding colour bit of the stage, the instruction is the first address of a control transfer that has taken place deeper in the pipeline, then the instruction is executed and the state vector of the instruction becomes the state of the stage. 
Arbitration is required since target addresses may be generated at any time by any pipeline stage. The AAU is illustrated in figure \ref{AAU}. The role of the AAU is to let through to memory  instruction addresses either from the Programme Counter or those that are the result of high priority control hazards, while blocking any subsequent lower priority target addresses from reaching memory. The AAU keeps a record of the colour state of the processor (vector C), which it updates based on the colour vectors of the instruction addresses arriving to it. 






\subsection{Multi-Colour Algorithm in SAMIPS}
\label{Multi-Colours Algorithm in SAMIPS}

For the utilization of the multi-colour algorithm in SAMIPS, a small modification of the algorithm is required, which is due to one function of the MIPS compiler. Since synchronous MIPS uses one ``branch delay slot" to solve the control hazard problem (see section \ref{Solution in Synchronous R3000}), most MIPS compilers automatically replace the NOP instruction in the ``branch delay" slot with a useful instruction.



This means that the immediate instruction followed by a branch is a valid one and should be executed normally, rather than discarded. This conflicts with the  proposed multi-colour algorithm. Therefore, to be compatible with such MIPS compilers, the colour change inside the pipeline stage $S_i$ should also be artificially delayed one cycle until the next instruction has arrived. Section \ref{SAMIPS Organisation and Balsa Implementation} provides further explanation as to how to delay the colour flip and keep the ``branch happened" information inside $S_i$. Of course, there is no need of such modification if the ``automatic swap" function of the adopted MIPS compiler could be switched off.    

\subsection{Interrupts in SAMIPS}


Interrupts are slightly different from ordinary exceptions:

\begin{itemize}
\item most interrupts are external events and independent of the CPU's normal instruction stream while exceptions are internal; 
\item exceptions are dealt with whenever they are produced, but interrupts cannot be always handled immediately and the response time usually depends on the current status of CPU, e.g. if CPU is handling an exception or a higher priority interrupt occurs at the same time; 
\item in a pipelined architecture, there are always some instructions still in the pipeline when the processor core receives a valid interrupt signal, whether to execute those instructions before processing an interrupt needs to be decided.  
\end{itemize}
In Synchronous MIPS R3000, interrupts are treated as a  special type of exceptions and there are six external interrupts driven by CPU input pins \cite{Sweetman99}. 
An active level on any of the interrupt pins is sensed in each cycle and it will cause an interrupt exception if it is enabled in the CP0 \textit{Status Register} (SR). The number of the enabled interrupts is recorded in the \textit{Interrupt Pending} (IP) field. Interrupt processing begins after the CPU receives an exception and finds out it is an interrupt in CP0 \textit{Cause Register} (the \textit{ExcCode} field of \textit{Cause} records the type of exception occurring). During the processing of an exception, all interrupts are disabled. In MIPS, interrupts have by default equal priorities in the hardware. The priority can be implemented in software if this is required.

One feature of the interrupt processing in MIPS is the definition of interrupt victims, namely the instructions that have been fetched but cannot be executed immediately because of the interrupt. In the synchronous MIPS, ``the last instruction that needs to be completed before interrupt processing starts will be the one that just finished its MEM stage when the interrupt is detected. The exception victim will be the one that has just finished its ALU stage" \cite{Sweetman99}. As mentioned above, the processor core needs to deal with those instructions still in the pipeline when it receives a valid interrupt request, namely either to execute or flush them. It is costly to flush prefetched instructions, but the problem is what if one of the current instructions in the pipeline generates an exception or causes a branch. Since no exceptions occur in the WB stage in MIPS, all the instructions that have not finished their MEM stages are flushed. 

\begin{figure}
\begin{center}
\includegraphics[width=0.5\textwidth]{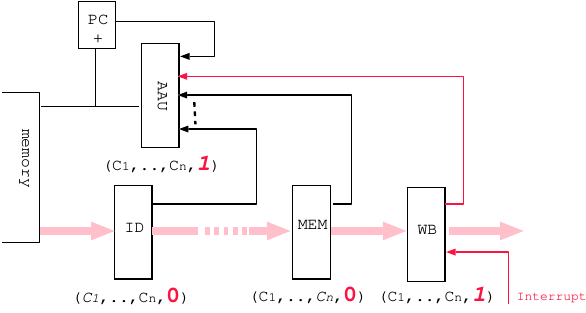}
\end{center}
\caption{Dealing With Interrupts}
\label{Interrupt1}
\end{figure}

\subsubsection{Multi-Colour Algorithm for Interrupts}
 The solution for interrupts in SAMIPS has inherited all features from synchronous MIPS described above, except the definition of the interrupt victim. In SAMIPS, there is no guarantee that there is one instruction which has just finished its MEM stage and is beginning the WB stage when an interrupt arrives. But it is still possible to stop all instructions, which have not reached their WB stages utilizing the multi-colour algorithm. Hence, the last instruction that needs to be completed before the interrupt processing starts will be the one which has already begun its WB stage.  The solution is to assign one colour bit to the interrupt signal which is maintained inside the WB stage. Following the multi-colour algorithm, this scheme gives interrupts the highest priority. When an interrupt signal arrives, it is sent to the WB stage first, as shown in Figure \ref{Interrupt1}. WB acknowledges it after finishing the current operation; it then sends a control hazard request to AAU and changes its colour accordingly. Since an interrupt has the highest priority, \textsf{AAU} will accept it and stop other control hazard requests. WB will also stop all the prefetched instructions by checking their colour bits. This introduces very little control overhead, except for calculating the back point after interrupt processing, To achieve that, WB must maintain a copy of the current instruction address. 

\begin{figure}
\begin{center}
\includegraphics[width=0.4\textwidth]{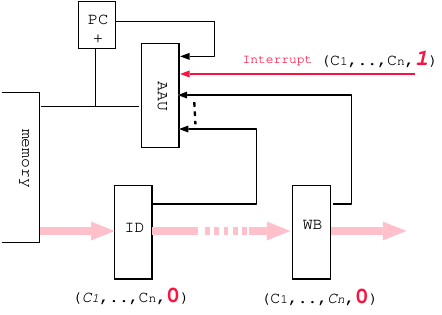}
\end{center}
\caption{Dealing With Interrupts: An alternative solution}
\label{Interrupt2}
\end{figure}

\subsubsection{An Alternative Algorithm}
 
\begin{figure}
\begin{center}
\includegraphics[width=0.5\textwidth]{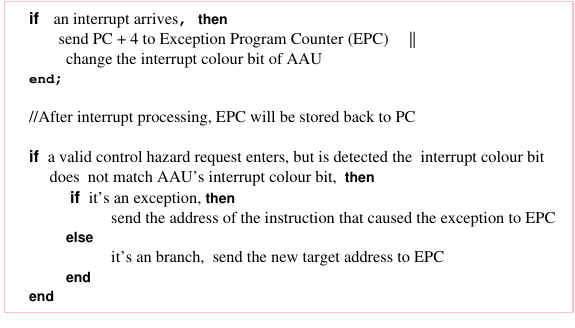}
\end{center}
\caption{Calculating the Back Point}
\label{Calculating the Back Point}
\end{figure}

Synchronous MIPS chooses to execute only one but flush all the remaining instructions in the pipeline. Although this approach appears to avoid executing unnecessary branches or exceptions from the flushed instructions, flushing the pipeline is expensive in time and energy. An alternative, more efficient solution would be to execute these prefetched instructions normally, but stop exceptions or branches in a safe way. With the extended use of the multi-colour algorithm, this is achievable. The basic idea is to send the interrupt signal to the \textsf{AAU} directly instead of to the WB stage.  When \textsf{AAU} receives an interrupt signal, the corresponding colour bit will change, as illustrated in Figure \ref{Interrupt2}. Hence, \textsf{AAU} is able to distinguish the instructions between interrupt from the new instruction stream of the interrupt handler. The algorithm inside the \textsf{AAU} to calculate the back point after returning from interrupt processing is shown in Figure \ref{Calculating the Back Point}. 
In this alternative algorithm the interrupt victim is redefined to be:
\begin{itemize}
\item either the next instruction which has not been prefetched yet before the interrupt signal arrives, if no instruction in the pipeline generates a control hazard; 
\item or the instruction which generates a valid exception, if one or more exceptions happens;
\item or the instruction from the target address, if a valid branch/jump occurs. \end{itemize}
 

\section{SAMIPS Organisation and Balsa Implementation}
\label{SAMIPS Organisation and Balsa Implementation} 
\begin{figure*}
\begin{center}
\includegraphics[width=0.5\textwidth,angle=-90]{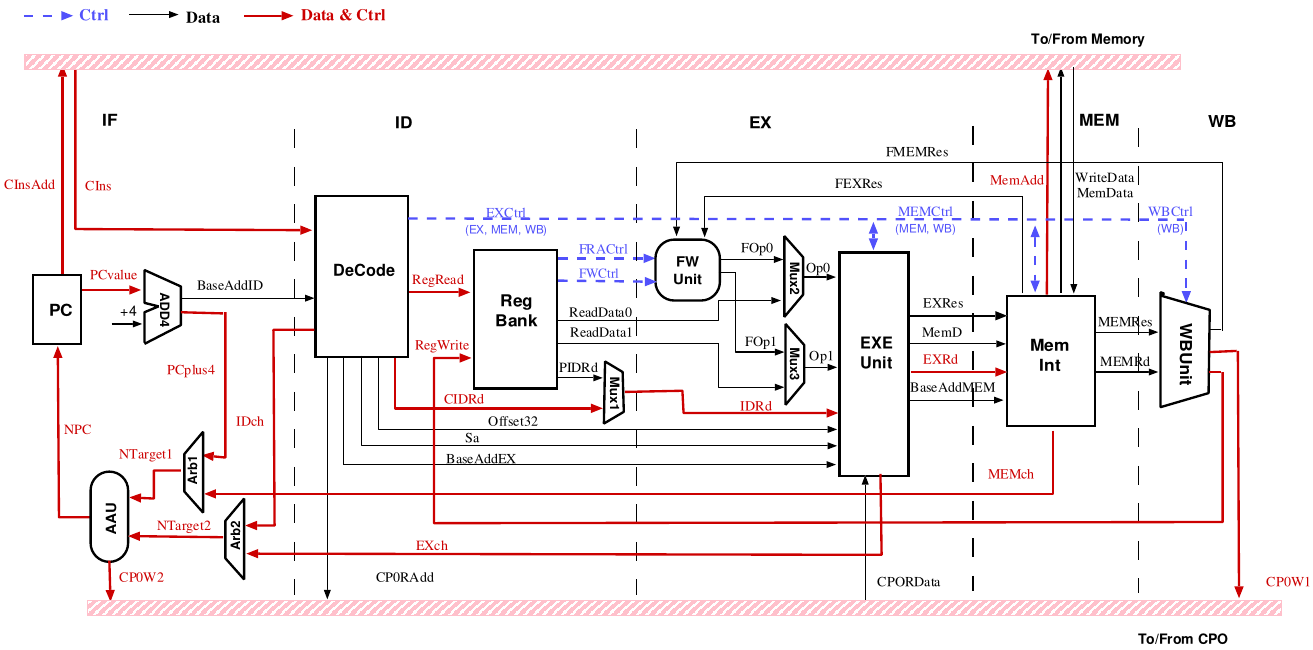}
\end{center}
\caption{SAMIPS RTL Design}
\label{SAMIPS RTL Design}
\end{figure*}

The development of SAMIPS has followed Balsa's design flow introduced in section \ref{Balsa}. A behavioral model of SAMIPS was specified which was then automatically compiled into a gate level netlist. Therefore, the design of SAMIPS is heavily influenced by the syntax and semantics of Balsa. Figure \ref{SAMIPS RTL Design} shows the organisation of SAMIPS, consisting of five pipeline stages as illustrated in section \ref{The Datapath}. Each functional block in each stage has been specified as a Balsa process while the interconnections (shown in solid or dotted arrows) have been mapped to Balsa handshake channels. There are three different types of interconnections between functional blocks, namely the data channels which only contain data (the solid black arrows), the control channels which only carry the control signals (the dotted grey arrows) and the combined channels which send data and control together (the solid grey arrows). The definition of each channel is provided in Tables \ref{Data Channels}, \ref{Control Channels} and \ref{Combined Data and Control Channels}.   
The detailed implementation and the interfaces of different functional blocks of each stage are discussed in the following sections. 

\begin{table*}[t]
\caption{Data Channels}
\begin{center}
\begin{tabular}{|l||c|p{10cm}|} \hline   
Channel  & Length & Description \\\hline\hline
BaseAddID     & 32 bits   & current instruction address plus 4 sent to the ID stage \\ \hline
BaseAddEX     & $\cdots$  & current instruction address plus 4 sent to the EX stage \\\hline
BaseAddMEM    & $\cdots$  & current instruction address (or plus4) sent to the MEM, for branch delay slot  \\\hline
CIDRd         & 5  bits  & CP0 register writing adresses \\\hline
CP0RAdd       & $\cdots$ & CP0 register reading address \\\hline
EXRes         & 32 bits  & EX stage results \\\hline
FOp0/1        & $\cdots$ & forwarded results passed to the \textsf{EXEUnit} as operands  \\\hline
FEXRes        & $\cdots$ & forwarded EX stage results \\\hline
FMEMRes       & $\cdots$ & forwarded MEM stage results \\\hline
MemD          & $\cdots$ & data pending to write to main memory \\\hline
MemData       & $\cdots$ & data read from main memory \\\hline
MEMRd         & 5 bits   & CPU register writing address   \\\hline
MEMRes        & 32 bits  & MEM stage results \\\hline
Offset32      & $\cdots$ & signextended offset address or immediate data \\\hline
Op0/1         & $\cdots$ & \textsf{EXEunit} operands \\\hline
ReadData0/1   & $\cdots$ & register data read from \textsf{RegBank} \\\hline
Sa            & 5  bits  & shift amount \\\hline
WriteData     & 32 bits  & data sent to main memory \\\hline

\end{tabular}
\label{Data Channels}
\end{center}
\end{table*}
\begin{table*}[t]
\caption{Control Channels}
\begin{center}
\begin{tabular}{|l||c|p{10cm}|} \hline   
Channel  & Length & Description \\\hline\hline
EXCtrl    & \{\{1,2\}, \{2, 3\}, 6\} bits & EXCtrl contains all the control information for EX, MEM and WB stages\\ \hline
MEMCtrl   & \{\{1,2\}, \{2, 3\}\} bits    & MEMCtrl contains the control information for both MEM and WB stages \\\hline
WBCtrl    & \{1, 2\} bits                 & control information for the WB stage  \\\hline
FRACtrl   & \{1, 1\} bits                 & a 2-bit control that indicates whether FEXRes and FMEMRes will fire in the current cycle \\\hline
FWCtrl    & \{2, 2\} bits                 & FWCtrl tells whether FEXRes and FMEMRes will be forwarded to \textsf{EXEunit} as operands \\\hline
\end{tabular}
\label{Control Channels}
\end{center}
\end{table*}
\begin{table*}[t]
\caption{Combined Data \& Control Channels}
\begin{center}
\begin{tabular}{|l||c|p{10cm}|} \hline   
Channel  & Length & Description \\\hline\hline
CIns       & \{\{1,1,1\}, 32\} bits       & instructions read from main memory, carrying ``colour" information\\ \hline
CInsAdd    & $\cdots$                           & instructions address sent to memory with ``colour" \\\hline
CP0W1/2    & \{32, 5, 1\} bits            & CP0 register writing channel, consisting of data, address and writing command \\\hline
EXRd       & \{5, 32\} bits               & register writing address passed from the EX stage, with original register data attched   \\\hline
IDRd/PIDRd & $\cdots$                     & register writing address passed from ID stage, with original register data attached   \\\hline
EXch       & \{\{1,1,1\}, 2, 1, 32\} bits & control hazard report channel from the EX stage  \\\hline
IDch       & $\cdots$                     & control hazard report channel from the ID stage \\\hline
MEMch      & \{\{1, 1, 1\}, 1\} bits      & control hazard report channel from the MEM stage \\  \hline
MEMAdd     & \{1, 3, 32\} bits            & data address sent to memory, with access and data type command \\\hline
NTarget1/2 & \{\{1,1,1\}, 2, 1, 32\} bits & control hazard report channels to \textsf{AAU} \\\hline
NPC        & \{\{1,1,1\}, 32\} bits       & new PC value with ``colour"  \\\hline
PCvalue    & $\cdots$                     & current PC value with ``colour" \\\hline
PCplus4    & $\cdots$                     & current PC value plus 4 with ``colour" \\\hline
RegRead    & \{5, 5, 5\} bits             & \textsf{RegBank} read command  \\\hline
RegWrite   & \{32, 5, 1\} bits            & \textsf{RegBank} write command   \\ \hline
\end{tabular}
\label{Combined Data and Control Channels}
\end{center}
\end{table*}

\subsection{The IF Stage} 

\begin{figure}
\begin{center}
\includegraphics[width=0.35\textwidth]{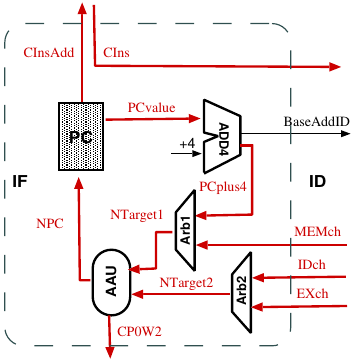}
\caption{IF Stage}
\label{IF Stage}
\end{center}
\end{figure}

IF is the first stage of the pipeline and is illustrated in Figure \ref{IF Stage}. This is essentially the 
instruction-prefetching unit of the processor which incorporates the program counter (PC) circuitry. The \textsf{AAU} unit, described in section \ref{Control Hazards Problems} for the arbitration of control hazard requests, is also part of this stage.    In this stage, the physical address, which is either the current PC plus 4 or a new 
target address from the datapath if a control hazard occurs, is calculated and then sent back to the \textsf{PC} through the \textsf{AAU} and finally to the memory. Control hazards may happen at any of the three stages in SAMIPS, namely ID, EX and MEM. The target address attached with the colour vector, stage number and the control hazard type, will be sent to the \textsf{AAU} for arbitration through channel \textit{IDch}, \textit{EXch} or \textit{MEMch}. 

A notable feature of the IF stage is that all the internal and external channels
except \textit{BaseAddID} (the current PC value used as a base
address for the new branch/jump target calculation) carry colour
information. 

\subsubsection{AAU} 
The Balsa description of the \textsf{AAU} in SAMIPS is presented in Figure \ref{AAUcode}. Two arbiters (\textsf{Arb1} and \textsf{Arb2}) together with an arbiter inside the \textsf{AAU} (specified in ``\textit{arbitrate}" command in Balsa) are used to construct a binary arbiter tree for the four input requests, namely \textit{Pcplus4}, \textit{IDch}, \textit{EXch} and \textit{MEMch}, which is finally stored in a Balsa register \textit{ch\_R}. As illustrated in section \ref{A Distributed Multi-Colour algorithm}, \textsf{AAU} has to do the colour checking on \textit{ch\_R} according to the multi-colour algorithm and only the address with the matched colour is passed through. If an exception request is valid, the \textsf{PC} is loaded with the common exception vector 0x80000080 and the return address after the exception handling is sent to the CP0's EPC register through channel \textit{CP0W2}. If a branch request is valid, the \textsf{PC} is loaded with the new target address.


\begin{figure*}[t]
\renewcommand{\baselinestretch}{1}
{\footnotesize 
\line(1, 0){430}
\begin{verbatim}
loop
  arbitrate NTarget1 then ch_R := NTarget1
  |         NTarget2 then ch_R := NTarget2
  end;

  -- check the color
  if ((ch_R.St = PC) and (ch_R.C = AAUC)) then 
    NPC  <- {ch_R.a, ch_R.C}
  else 
    if (((ch_R.St=ID) and (ch_R.C.EX=AAUC.EX) and (ch_R.C.MEM=AAUC.MEM)) or
        ((ch_R.St=EX) and (ch_R.C.MEM=AAUC.MEM)) or 
         (ch_R.St=MEM)) then 
         if (ch_R.eNj as bit) then  -- an exception
              CP0W2 <- {r, 13, ch_R.a} ||   -- send current address to CP0's EPC 
              NPC   <- {0x80000080, ch_R.C} -- load the exception vector
         else                       -- a branch
              NPC  <- {ch_R.a, ch_R.C}      -- send the new target address
         end ||                                to the memory
         AAUC := ch_R.C             -- change the colour
     end
   end
end
\end{verbatim}
\line(1, 0){430}
}
\caption{Balsa Model of \textsf{AAU}}
\label{AAUcode}
\end{figure*}


\renewcommand{\baselinestretch}{1.5}
 
\subsection{The ID Stage} 
\label{ID Stage}

\begin{figure}[t]
\begin{center}
\includegraphics[width=0.5\textwidth]{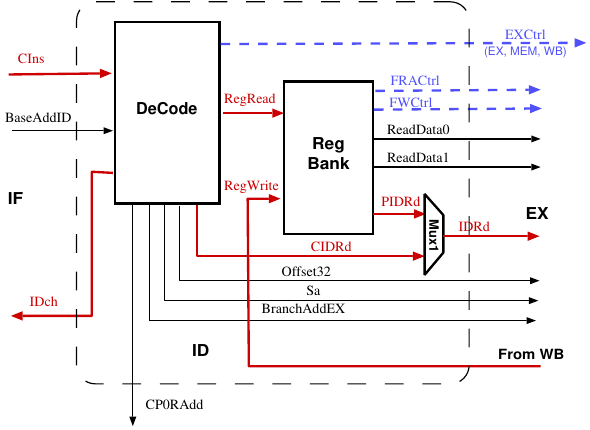}
\caption{ID Stage}
\label{ID stage} 
\end{center}
\end{figure}

The ID stage is illustrated in Figure \ref{ID stage}. The main instruction decoder \textsf{DeCode} of the system lives at this 
stage. \textsf{DeCode} checks the first 6 bits (\textit{Opcode} field) and the last 5 bits (\textit{funct} field in R-type) of the instruction. Based on this, the operands are extracted and control information for the later stages is generated. Register file reading and writing
is carried out at this stage as well. The innovative feature of the \textsf{RegBank} is the implementation of an asynchronous forwarding mechanism
as described in section \ref{Data hazards problems}. 

\subsubsection{DeCode}

\begin{table*}
\caption{Control signal encoding for the EX stage} 
\begin{small}
\begin{center}
\begin{tabular}{|c||c|c|c|c|c|c|c|c|}    \hline 
 \diagbox{bits 2..0}{bits 5..3}  & 000   & 001  & 010  & 011  & 100  & 101    & 110  & 111    \\ \hline \hline 
 000    & BEQ   & BNE  & BGTZ & BLEZ & BLTZ & BLTZAL & BGEZ & BGEZAL  \\ \hline
 001    & *     & JR   & JALR & JAL  & *    & *      & *    & *       \\ \hline
 010    & ADD   & SUB  & ADDU & SUBU & AND  & OR     & XOR  & NOR     \\ \hline
 011    & EXC   & EXCS & MA   & COR  & SLTU & SLT  & *      & *       \\ \hline
 100    & SLLV  & SRLV & SRAV & NOP  & SLL  & SRL  & SRA    & *       \\ \hline
 101    & *     & *    & *    & *    & *    & *      & *    & *	      \\ \hline
 110    & MULTU & MULT & DIVU & DIV  & MTHI & MTLO   & MFHI & MFLO    \\ \hline
 111    & *     & *    & *    & *    & *    & *      & *    &         \\ \hline 
\end{tabular}
\label{ContrlEX} 
\end{center}
\end{small}
\end{table*}

\begin{table*}
\caption{Control signal encoding for the MEM stage} 
\begin{small}
\begin{center}
\begin{tabular}{|c||c|c|c|} \hline
& \multicolumn{1}{|c|}{AccType}       & \multicolumn{2}{|c|}{DataType (bit 0)} \\ \cline{3-4}
  bits  1..0 / 2..1   &     & 0           &  1           \\ \hline \hline
00      & memory READ                         & *           &  Word        \\ \hline
01      & memory WRITE                        & Word Left   &  Word Right  \\ \hline
10      & IMMediate operand is transferred by 
          offset32 but non memory operation   & Half Signed & Half Unsigned \\ \hline
11      & NON memory operation                & Byte Signed & Byte Unsigned \\ \hline
\end{tabular} 
\label{ContrlMEM}
\end{center}
\end{small}
\end{table*}

\begin{table*}
\caption{Control signal encoding for the WB stage} 
\begin{small}
\begin{center}

\begin{tabular}{|c||c|c|c|} \hline
bit 0/1      & cNp            & \multicolumn{2}{|c|}{wNe (bit 0)} \\ \cline{3-4}
        &               & \multicolumn{1}{|c|}{0} &  \multicolumn{1}{|c|}{1}           \\ \hline \hline
0      & CP0 operation  & NUN operation    &  Exception operation, need also write to cp0    \\ \hline
1      & CPU operation & register Write   &  register Reset  \\ \hline
\end{tabular} 
\label{ContrlWB}
\end{center}
\end{small}
\end{table*}

The \textsf{DeCode} unit has two input channels and eight output ones. The 
data on channels \textit{Offset32} (signextended 32-bit from the 16-bit
\textit{immediate} field), \textit{Sa} (5-bit shift amount from the \textit{sa} field), \textit{CIDRd} (5-bit CP0 register write address from the \textit{rd} field) and \textit{CP0R} (5-bit CP0 register read address from the \textit{rd} field) are directly extracted from the instruction. During the exception handling, data is stored to and fetched from the coprocessor. \textit{CIDRd} is the channel carrying the coprocessor register write address, which is extracted directly from original instructions during decoding and passed to the EX stage after merging with the CPU register write address (\textit{PIDRd}) from the \textsf{RegBank} through a multiplexer (\textsf{Mux1}).

The remaining three are either control or combined data and control channels generated by the \textsf{DeCode} unit:
\begin{itemize}

\item \textit{EXCtrl}. Since all the control information is first generated by 
the \textsf{DeCode}
and then propagated down the pipeline
(Figure \ref{ID stage}), \textit{EXCtrl} carries all the control
information required by next three stages, namely \textit{EX}, \textit{MEM} 
and \textit{WB}. It is then left to the local 
circuits of each of the following stages to select the related control 
information. Tables \ref{ContrlEX} to \ref{ContrlWB} illustrate the encoding of control signals for EX, MEM and WB stages respectively. In Table \ref{ContrlEX}, two exception operation encodings, namely \textit{EXC} and \textit{EXCS}, are used to distinguish between those exceptions occurring in a ``branch delay slot" from the others. \textit{MA} is a memory access operation and \textit{COR} represents a CP0 read operation, while the rest correspond to MIPS instructions. The control information for the MEM stage is shown in Table \ref{ContrlMEM}. It consists of two parts, namely a 3-bit memory access type (\textit{AccType}) and a 2-bit data fetch type (\textit{DataType}). 

\item \textit{RegRead}. This channel carries the three register addresses. In synchronous MIPS, only register file reading addresses are sent to \textsf{RegBank} in the ID stage. The register writing address is passed through the entire pipeline and eventually sent back to \textsf{RegBank} at the WB stage. To implement our proposed asynchronous forwarding mechanism, the writing address is also required at the ID stage, to enable \textsf{RegBank} to keep a record of all the instructions pending to modify register data. 

\item \textit{IDch}. The \textit{IDch} channel in the ID stage consists of a 32-bit new target address, a 3-bit colour vector, a 2-bit stage number and a 1-bit control hazard type (branch or exception). 

\end{itemize}

\begin{figure*}
\renewcommand{\baselinestretch}{1}
{\footnotesize 
\line(1, 0){415}
\begin{verbatim}
...
CIns, BaseAddID -> then Ins_R := CIns.Ins || C_R := CIns.C || ...  end
                                          
-- check colour information
if ((C_R = IDC) or (C_R.EX /= IDC.EX) or (C_R.MEM /= IDC.MEM)) then 
  /*---------------------------------------------------------------------*/
  -- when a branch happens, the colour is changed when the next instruction  
     arrives and use another register S to indicate the "branch delay slot"
  if ( R and (C_R.C = IDC)) then IDC.ID := not IDC.ID ||
                                 S      := 1
                            else IDC    := C_R   end;
  R := 0 ;  -- reset R 
  /*--------------------------------------------------------------------*/	
  ...
  case ((#(Ins_R as J_TYPE).Op)[2..5] as 4 bits) of 
     ...
     | 10, 11  then 
         SignExtend() || 
         RegRead <- {(Ins_R as I_TYPE).rs, (Ins_R as I_TYPE).rt, 0} ||
         case (Ins_R as J_TYPE).Op of
           SW    then EXCtrl <- {MA, {write,W},   {non,p}}
         | SH    then EXCtrl <- {MA, {write,H},   {non,p}}
     | SB    then EXCtrl <- {MA, {write,B},   {non,p}}
         | SWL   then EXCtrl <- {MA, {write,WL},  {non,p}}  	 
         | SWR   then EXCtrl <- {MA, {write, WR}, {non,p}}
         else 
	    -- for those undefined instrution, an exception is generated. 
            if S then EXCtrl <- {EXCS, {imm,N}, {e,c}}
                 else EXCtrl <- {EXC,  {imm,N}, {e,c}}  end ||
            IDch   <- {1024, e, ID, C_R}  || 
	    -- exception doesn't have any "delay slot", colour changes immediately
            IDC.ID := not IDC.ID
         end 
     ...
     | 0 then 
         case (Ins_R as J_TYPE).Op of   
         ...     
         | J  then  
            RegRead <- {0, 0,  0} ||
            EXCtrl  <- {NOP, {non,N}, {non,p}} ||
            IDch    <- {(LShiftJ {(0 as 2 bits), (Ins_R as J_TYPE).target, 
                      (#(BaseAdd_R)[31..28] as 4 bits)} as word), j, ID, C_R} ||
           /*--------------------------------------------------------*/
            -- keep the "branch happened" information in a register R, 
               but not change the current colour      
            R       := 1  
           /*--------------------------------------------------------*/
         ...
...;
S := 0  -- reset S
\end{verbatim}
\line(1, 0){415}
}
\caption{Balsa Model of \textsf{DeCode}}
\label{DeCodecode}
\end{figure*}
\renewcommand{\baselinestretch}{1.5}

Figure \ref{DeCodecode} gives a sample of the Balsa specification of the \textsf{DeCode}. A ``\textit{case}" command is used to decode instructions. Inside the ``\textit{if}" construct is the colour checking. By comparing the colour vector of the incoming instruction with that of the current stage, \textsf{DeCode} passes only valid instruction streams to the following stages. As discussed in section \ref{Multi-Colours Algorithm in SAMIPS}, a slight modification in the multi-colour algorithm is required in order to execute the machine code generated from an ordinary MIPS compiler. Figure \ref{DeCodecode} illustrates the modified parts. The basic idea is to employ a 1-bit Balsa register \textit{R} to hold the ``branch happened" information without changing the current colour vector of the stage. This register \textit{R} has to be checked in addition to the colour vector each time a new instruction arrives. Hence the colour vector is changed in the following cycle after a branch occurs. To distinguish an exception occurring in a ``branch delay slot" from a normal one, another 1-bit register \textit{S} is employed. \textit{S} is set to 1 when \textit{R} equals 1, which is only one cycle after a branch. It is important to reset both \textit{R} and \textit{S} back to zero.    
\subsubsection{RegBank}

\begin{figure*}
\renewcommand{\baselinestretch}{1}
{\footnotesize 
\line(1, 0){430}
\begin{verbatim}
...
variable R               : array 32 of word
variable DHDQ            : DHDQ
variable FWcase0,FWcase1 : FWcase
variable FRAQue          : FRACtrl
...
arbitrate RegRead then 
   FRACtrl  <- FRAQue ;                                     -- send out FRACtrl
   FRAQue  :=   {(RegRead.WNo /= 0), FRAQue.FR1A} ||        -- update FRAQue
   begin 
      if (RegRead.WNo /= 0) then PIDRd <- {R[RegRead.WNo], RegRead.WNo} end ||  
      /*---------- data hazard detection-----------------*/              
      clean0 <- {(RegRead.RNo0 = 0),      (RegRead.RNo0 = DHDQ.W0),
                 (RegRead.RNo0 = DHDQ.W1),(RegRead.RNo0 = DHDQ.W2)} ||  
      clean1 <- ... ||        
      clean0 -> then 
               case (clean0 as 4 bits) of 	       
                   0bxx10 then  FWcase0 := EXER	                             
               |   0bx100 then  FWcase0 := MEMR 
               |   0b1000 then  FWcase0 := WBR
               else ReadData0 <- R[RegRead.RNo0] ||  FWcase0 := non  end 
      end  ||
      clean1 -> then ... ;
      /*-------------------------------------------------*/
      FWCtrl <- {FWcase0, FWcase1} ||         -- send out FWCtrl  
      begin                                   
         DHDQ := {RegRead.WNo, DHDQ.W0, DHDQ.W1, DHDQ.W2};   -- update DHDQ
         if (FWcase0 = WBR) or (FWcase1 = WBR) then  
	     ...  -- wait for a RegWrite signal come
	     if FWcase0 = WBR then ReadData0 <- RegWriteI.RegData end ||
	     if FWcase1 = WBR then ReadData1 <- RegWriteI.RegData end	   
         end   
      end
   end
| RegWrite then       
   if (RegWrite.rNw as bit) then R[RegWrite.Rd] := RegWrite.RegData end||

   /*-------------- update DHDQ, after register writing -----------------*/
   begin    cleanW := { (RegWrite.Rd = DHDQ.W0), (RegWrite.Rd = DHDQ.W1), 
                        (RegWrite.Rd = DHDQ.W2), (RegWrite.Rd = DHDQ.W3)}; 
            case (cleanW as 4 bits) of
               0b0001  then DHDQ.W0 := 0
            |  0b001x  then DHDQ.W1 := 0 
            |  0b01xx  then DHDQ.W2 := 0 
                       else DHDQ.W3 := 0 
            end   end 
   /*--------------------------------------------------------------------*/
end
\end{verbatim}
\line(1, 0){430}
}
\caption{Balsa Model of \textsf{RegBank} DHDQ Implemenation}
\label{RegBankcode}
\end{figure*}
\renewcommand{\baselinestretch}{1.5}

\begin{figure*}[t]
\renewcommand{\baselinestretch}{1}
{\footnotesize 
\line(1, 0){430}
\begin{verbatim}
...
variable Flag            : array 32 of Flag
variable CurIndex        : 2 bits
channel  Dist0, Dist1    : 2 bits
...
arbitrate RegRead then 
    ...
    /*--------------------- data hazard detection --------------------------*/              
    if Flag[RegRead.RNo0].Clean then  ReadData0 <- R[RegRead.RNo0] ||
		                      FWcase0   := non
    else Dist0 <- (CurIndex + not(Flag[RegRead.RNo0].Index) +1 as 2 bits) || 
         Dist0 -> then   case Dist0 of 
                           1 then FWcase0 := EXER
                         | 2 then FWcase0 := MEMR
                         | 3 then FWcase0 := WBR   end   
	          end 
    end || 
    /*---------------------------------------------------------------------*/              
    ...;                                  
    if (RegRead.WNo /= 0) then PIDRd <- {R[RegRead.WNo], CurIndex, RegRead.WNo} ||
                               Flag[RegRead.WNo].Clean := 0 || 
                               Flag[RegRead.WNo].Index := CurIndex
                          end ;
    ...
| RegWrite then 
    if (RegWrite.rNw as bit) then  R[RegWrite.Rd] := RegWrite.RegData  end||

    /*------ update DHDT, after register writing -------*/
    if (RegWrite.Index = Flag[RegWrite.Rd].Index) then 
        Flag[RegWrite.Rd].Clean := 1  end 
   /*---------------------------------------------------*/
end
\end{verbatim}
\line(1, 0){430}
}
\caption{Balsa Model of \textsf{RegBank} DHDT Implemenation}
\label{DHDTcode}
\end{figure*}
\renewcommand{\baselinestretch}{1.5}

\textsf{RegBank} contains thirty-two 32-bit general purpose registers and 
has two input channels (\textit{RegRead} and
\textit{RegWrite}) and two output channels (\textit{ReadData0} and
\textit{ReadData1}), as depicted in Figure \ref{ID stage}. 

As discussed in section \ref{Data hazards problems} where the asynchronous forwarding mechanism was presented, the \textsf{RegBank} implements the forwarding algorithm (see section \ref{Asynchronous Forwarding in SAMIPS}). Figure \ref{RegBankcode} shows the Balsa description of the \textsf{RegBank} DHDQ implementation in SAMIPS. 

When a new \textit{RegRead} signal is arrived, \textit{FRACtrl} is sent out first and FRAQ is updated by popping one bit out on the top and pushing another (corresponding to the current instruction) into the tail. This bit will be either 1, if the instruction will do register write, or 0 if not. The DHDQ is checked for data hazards and then \textit{FWCtrl} and \textit{ReadData0/1} are sent out (see Figure \ref{Asynchronous Forwarding}). Finally DHDQ has to be updated according to the writing address in \textit{RegRead}.


In the case of a register write instruction, its passing through the
\textsf{Regbank} will mark the register as pending to be written. A mechanism
is required to unmark the register and enable following read instructions to
get the correct operand, in the case when the write instruction is rejected
in the execution (EX) stage because of a colour mismatch (see also section \ref{The EX Stage}). The solution that has been devised is as follows: the register write instruction carries with it the value of the register (on channel \textit{PIDRd}). If the instruction is rejected, that value is (a) written back to the register bank to unmark the register (on channel \textit{RegWrite} through \textit{EXRes}) and (b) passed on to the \textsf{FWunit} to be used as a forwarded result for read instructions that have found the register in the \textsf{RegBank} marked as dirty.

When a new \textit{RegWrite} signal arrives, the new value is written to the corresponding register \textit{R[i]}.  The DHDQ is parsed and the first matched register address is reset to 0. This is specified in a ``\textit{case}" command in Balsa.

The \textsf{RegBank} DHDT implementation varies from DHDQ implementation in the data hazard detection part, as shown in Figure \ref{DHDTcode}. 

Due to its efficiency, DHDQ has been chosen for the implementation of SAMIPS and the evaluation performed in section \ref{Evaluation and Refinement}.

\subsection{The EX Stage} 
\label{The EX Stage}
\begin{figure}[t]
\begin{center}
\includegraphics[width=0.5\textwidth]{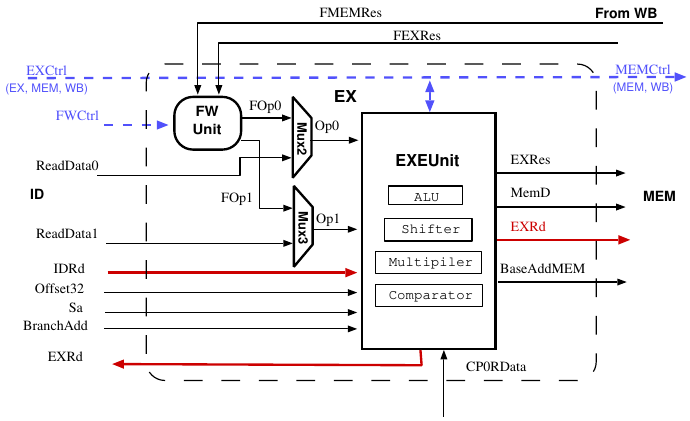}
\caption{EX Stage}
\label{EX stage}
\end{center}
\end{figure}

This stage is the one in which instructions are finally executed (Figure \ref{EX stage}). Since the conditional branch and arithmetic overflow exception occur at this stage, the \textsf{EXEunit} is also responsible for checking the ``colour" and discarding invalid instructions. The EX stage also hosts the \textsf{FWunit} unit, which is one of the main parts beside the \textsf{RegBank}, to implement the asynchronous forwarding mechanism.

\subsubsection{EXEunit}
The \textsf{EXEunit} consists of an ALU for arithmetic and logic operations, a shifter for shift operations, a Multiplier for multiplication and division operations and a comparator for conditional branch operations. It has seven input channels and four output channels. 

Among the input channels of \textsf{EXEunit}, \textit{Op0} and \textit{Op1} are the ALU operands, which are either read from the \textsf{RegBank} or forwarded from the MEM and WB stages (see also section \ref{ID Stage}); \textit{CP0RData} carries the data fetched from the coprocessor registers; all the remaining inputs are generated from the \textsf{DeCode} and have been described in the previous section. In each cycle, not every input channel will fire, and thus the \textsf{EXEunit} is based on the control information inside the \textit{EXCtrl} channel to acknowledge its input requests. This is different from a synchronous system where the communication is driven by time not data. 

The output channels of \textsf{EXEunit} include:
\begin{itemize}
\item \textit{MEMCtrl}: a bundled channel of the control information for the MEM and WB stages.
\item \textit{EXRes}: a 32-bit instruction execution result. 
\item \textit{MemD}: a 32-bit channel carrying data to be written into the main memory.
\item \textit{EXRd}: a 37-bit channel containing register writing address passed from the EX stage, with the original register data attached\end{itemize}

\begin{figure*}[t]
\renewcommand{\baselinestretch}{1}
{\footnotesize 
\line(1, 0){410}
\begin{verbatim}
...
-- data and control signal receiving
EXCtrl -> then EXCtrl_R := EXCtrl ||                       
               if (#(EXCtrl.EX)[2..5] as 4 bits) = 9 then sa -> sa_R end ||
               if EXCtrl.M.AccType /= non then Offset32 -> Offset32_R  end 
end;
if EXCtrl_R.EX = COR then CP0RegData -> Offset32_R end ||
if #(EXCtrl_R.WB.wNe)[1] then IDRd   -> IDRd_R     
...
if (CIns_R.C = EXC or (CIns_R.C.MEM /= EXC.MEM)) then
case (#(EXCtrl_R.EX)[4..5] as 2 bits) of
      
   0 then                  -- Compare operation for Branch decision-making 
         if (((EXCtrl_R.EX = BEQ) and (Op0_R = Op1_R)) or 
            ((EXCtrl_R.EX = BNE)  and (Op0_R /= Op1_R)) or 
            (((EXCtrl_R.EX = BGEZ) or (EXCtrl_R.EX = BGEZAL)) ...
             and  (#Op0_R[31] or (Op0_R = 0))) or #(EXCtrl_R.EX)[3] ) then 
                  EXch <- {Op0_R, j, EX, C} || R := 1 || ...
         end

 | 1 then                        - ALU operation
         case EXCtrl_R.EX of  
              ADD, ADDU then  EXEunitRes <- (Op0_R + Op1_R as word)
          |   SUB, SUBU then  EXEunitRes <- (Op0_R - Op1_R as word)
          |   AND       then  EXEunitRes <- (Op0_R and Op1_R as word) 
          -- ID stage exceptions operation
          |   EXC       then  EXEunitRes <- BaseAdd_R
          |   EXCS      then  EXEunitRes <- (BaseAdd_R-4 as word)    
          ...
end 
\end{verbatim}
\line(1, 0){410}
}
\caption{Balsa Model of \textsf{EXEunit}}
\label{EXEunitcode}
\end{figure*}
\renewcommand{\baselinestretch}{1.5}

Figure \ref{EXEunitcode} is the Balsa sample code of the \textsf{EXEunit}. Input data are received first; followed by the colour information check and instruction execution. Since the conditional branch happens at the EX stage, the modification of the multi-colour algorithm for the ``branch delay slot" is necessary. For the exception operation, if the control signal encoding for the EX stage inside \textit{EXCtrl} is \textit{EXC}, the current program counter (saved in \textit{BaseAdd\_R}) is sent out to the co-processor; otherwise if it is \textit{EXCS}, which means the exception occurs in a ``branch delay slot" (see the description of the exception handling in section \ref{Exceptions}), it is the PC minus 4 (the address of the preceding branch instruction) sent to the co-processor.  

\subsubsection{FWunit}
\begin{figure*}
\renewcommand{\baselinestretch}{1}
{\footnotesize 
\line(1, 0){430}
\begin{verbatim}
FRACtrl -> FRACtrl_R;
if FRACtrl_R.EXA then FEXRes->FEXRes_R end ||    -- results come from the MEM stage
if FRACtrl_R.MEMA then FMEMRes->FMEMRes_R end || -- results come from the WB stage

FWCtrl -> FWCtrl_R;
case FWCtrl_R.FWcase0 of                         -- forward result to EXEunit
    EXR then  FOp0 <- FEXRes_R 
|   MEMR then  FOp0  <- FMEMRes_R
end ||
case FWCtrl_R.FWcase1 of ... end 
\end{verbatim}
\line(1, 0){430}
}
\caption{Balsa Model of \textsf{FWunit}}
\label{FWunitcode}
\end{figure*}
\renewcommand{\baselinestretch}{1.5}

The Balsa implementation of the \textsf{FWunit} is shown in Figure \ref{FWunitcode}.  \textsf{FWunit} is the final destination of the forwarded results from the MEM and WB stages(\textit{FEXRes} and \textit{FMEMRes}). Inside the \textsf{FWunit} these two forwarded results are acknowledged and sent to the \textsf{EXEunit} through two multiplexers (\textsf{Mux1} and \textsf{Mux2}) if necessary, based on the content of \textit{FRACtrl} and \textit{FWCtrl}. 

\subsection{MEM stage} 
\begin{figure*}
\renewcommand{\baselinestretch}{1}
{\footnotesize 
\line(1, 0){410}
\begin{verbatim}
... 
 if  (#(MEMCtrl_R.WB.wNe)[1]) then EXRd -> Rd_R end || 
 if ((#(MEMCtrl_R.WB.wNe)[1]) and ((MEMCtrl_R.WB as 3 bits) /= 2)) then 
    FEXRes <- EXRes_R end ||  -- forward result back to the EX stage

 if  (MEMC.MEM = MEMCtrl_R.C.MEM) then  -- check colour vector
    ...
    if #(MEMCtrl_R.WB.wNe)[1] then MEMRd <- Rd_R.Rd end || 
    if #(MEMCtrl_R.M.AccType)[1] then  -- need not to access the memory
        MEMRes <- EXRes_R ||  WBCtrl <- MEMCtrl_R.WB
    else  ...
        MemAdd <- {EXRes_R, MEMCtrl_R.M.DataType,
                   (#(MEMCtrl_R.M.AccType)[0] as rNw) } ||
        if #(MEMCtrl_R.M.AccType)[0] then    -- fetch from the memory              
              WriteData <-  MemD_R                       
        else  WriteData <- Rd_R.RegData  ||  -- store to the memory
              MemData -> then MEMRes <- MemData end ||
              WBCtrl <- {w, p}                              
        end
...
\end{verbatim}
\line(1, 0){410}
}
\caption{Balsa Model of \textsf{MemInt}}
\label{MemIntcode}
\end{figure*}
\renewcommand{\baselinestretch}{1.5}

MIPS is a Harvard architecture with separate ports for the instruction and data in the main memory. Figure \ref{MemIntcode} is the Balsa description of the \textsf{MemInt}. The \textsf{MemInt} interacts with the memory via three external channels: \textit{MemAdd} (data address), \textit{MemData} (memory read data) and \textit{WriteData} (memory write data). The execution result is forwarded back to the EX stage via the \textit{FEXRes} channel. The register writing address passed from the EX stage on channel \textit{EXRd} is acknowledged and stored in a local register and then passed to the next stage on channel \textit{MemRd} to finally arrive, through the WB stage, back to the \textsf{RegBank}. 

\subsection{WB stage} 

\begin{figure*}
\renewcommand{\baselinestretch}{1}
{\footnotesize 
\line(1, 0){410}
\begin{verbatim}
...
     select WBCtrl then WBCtrl_R := WBCtrl end;
     ...
     -- CPU register writing
     if (#(WBCtrl_R.wNe)[1] and ((WBCtrl_R as 3 bits) /= 2)) then 
         RegWrite <-{((not #(WBCtrl_R.wNe)[0]) as rNw), Rd_R, MEMRes_R} ||
         FMEMRes <- MEMRes_R    -- forward result back to the EX stage
     end ||
     -- CP0 register writing during the exception handling
     if not (WBCtrl_R.cNp as bit) then 
         ...
         CP0W1 <- {(#(WBCtrl_R.wNe)[1] as rNw), Rd_R, MEMRes_R}
     end
...
\end{verbatim}
\line(1, 0){410}
}
\caption{Balsa Model of \textsf{WBUnit}}
\label{WBUnitcode}
\end{figure*}
\renewcommand{\baselinestretch}{1.5}

WB is the last stage of the MIPS pipeline. The function of this stage is very simple compared to the previous four and contains only one functional block \textsf{WBUnit}. Figure \ref{WBUnitcode} is the Balsa description of the \textsf{WBUnit}. It generates a \textit{RegWrite} signal when the current instruction requires a write operation to \textsf{RegBank}, and forwards the result from the MEM stage back to the EX stage on channel \textit{FMEMRes}. If an exception occurs, the return address is sent to CP0 through \textit{CP0W1}.

\section{Evaluation of SAMIPS}
\label{Evaluation and Refinement} 
This section presents a quantitive analysis of SAMIPS in terms of area cost, speed and power consumption. 









\subsection{BALSAMIPS} 
\label{BALSAMIPS}


\begin{figure*}
\begin{center}
\includegraphics[width=0.9\textwidth]{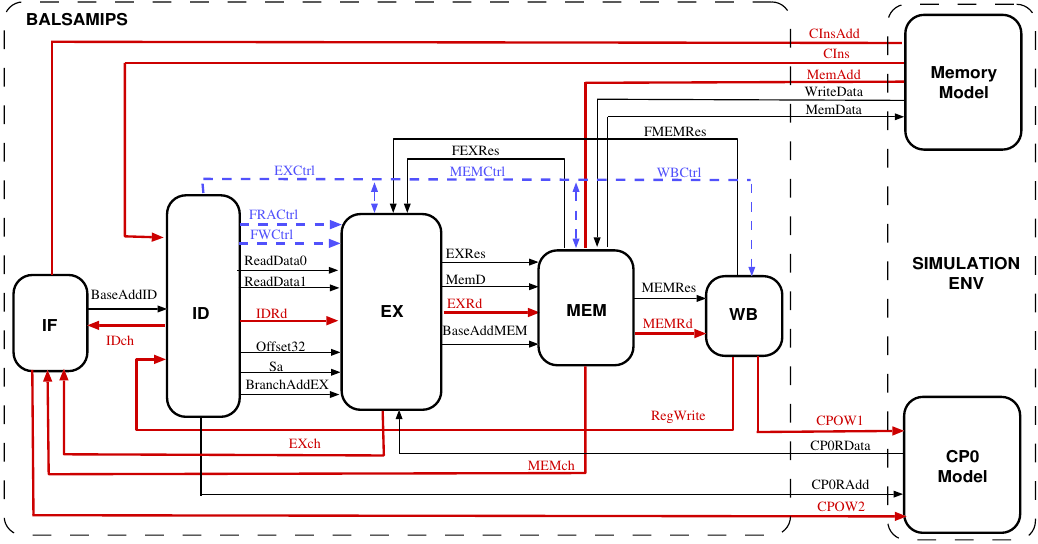}
\caption{BALSAMIPS Top Level Process Graph} 
\label{BALSAMIPS Top Level Process}
\end{center}
\end{figure*}

BALSAMIPS is the Balsa model of SAMIPS. It is structured as a hierarchy of concurrent Balsa processes and consists of approx. 900 lines of Balsa code\footnote{The source code of SAMIPS can be found at: \url{https://samips.github.io/}}. It (meta-)executes MIPS machine code produced by a standard MIPS cross-compiler (the GNU C MIPS cross compiler installed in the linux-i386 (intel) platform has been utilised in this case). Figure \ref{BALSAMIPS Top Level Process} shows the top level process structure of BALSAMIPS. It consists of five concurrent Balsa processes, one for each pipeline stage. Some of these processes are themselves structured as a network of lower-level concurrent Balsa processes as shown in Figure \ref{SAMIPS RTL Design}, while others may not have any sub-processes due to simplicity, e.g. MEM and WB. 
The Coprocessor (CP0) and memory model have also been modelled. They communicate with BALSAMIPS separately through four and five handshake channels respectively.  

\subsection{CP0 model} 
\label{CP0}
A simplified CP0 has been modelled in Balsa, which mainly implements coprocessor register reading and writing operations.  It has thirty-two 32-bit registers including the \textit{Status} register (\$12), \textit{Cause} register(\$13) and \textit{EPC} register (\$14). It has two read ports, namely \textit{CP0RAdd} for address and \textit{CP0RData} for data, and two write ports, \textit{CPW1} and \textit{CPW2} for the writing request from the WB and IF stages respectively.   
\subsection{Memory model} 
\label{Memory}

To enable simulation at different levels, three memory models have been developed, namely a LARD memory model (for LARD simulator), a Balsa memory model (for Balsa behavioural simulator) and a Verilog memory model (for Verilog simulator and NanoSim VCS co-simulator). The communication to BALSAMIPS is through five external channels, namely \textit{CInsAdd}, \textit{CIns} for instruction interface and \textit{MemAdd}, \textit{WriteData}, \textit{MemData} for data interface.  

\subsection{The Simulation Environment}
\label{Simulation Environment}
Following the design flow of the Balsa system as shown in Figure \ref{Balsa System}, three levels of simulation have been performed, namely behavioural, gate and physical level simulation. 

At the behavioural level, BALSAMIPS has been compiled into a network of handshake components, which can be simulated using both  LARD and the Balsa {\em breeze-sim} behavioural simulator. At this level, the functional correctness of SAMIPS is validated through (meta)executing a number of benchmarks. The LARD and Balsa memory models are used. At the same level, Balsa provides area cost estimations for a particular back-end implementation through the {\em breeze-cost} utility as mentioned in section \ref{Balsa}. 

\begin{figure}[t]
\begin{center}
\includegraphics[width= 8cm]{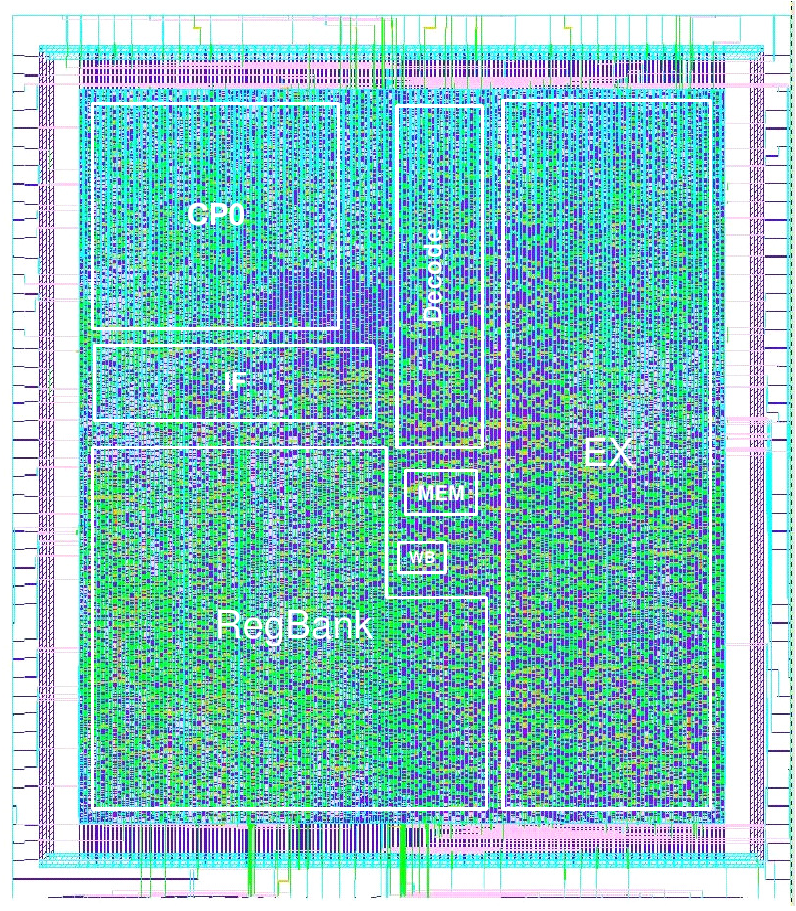}
\end{center}
\caption{SAMIPS Physical Layout}
\label{SAMIPS Layout}
\end{figure}

The {\em balsa-netlist} utility of Balsa maps the BALSAMIPS Breeze description into its technology-based implementation and produces a set of Verilog gate level netlists. The cell library used for SAMIPS is the the AMULET ST 0.18$\mu$m library and three encoding technologies have been considered, namely {\em four\_b\_rb}, {\em dual\_b} and {\em one\_of\_2\_4} (see section \ref{Balsa}).  At this level, the actual number of transistors can be calculated from {\em balsa-net-tran-cost}.  The gate-level Verilog description has been simulated together with the Verilog memory model by the Verilog-XL simulator (Cadence Verilog-XL Simulator 05.10.003-s). Although the power estimation is not provided at gate level, the latency analysis is more accurate than the behavioural level, where latencies in handshake components are nominal. Area costs in terms of transistor counts, are also calculated based on different technologies, and are hence accurate. 


To obtain accurate performance and the power consumptions of SAMIPS, the Verilog model of SAMIPS is further synthesised into its final physical layout using the SGST HCMOS8D 0.18$\mu$m 6 layer metal process which runs the core cells at 1.8v and the pad ring at 3.3v, shown in Figure \ref{SAMIPS Layout}. The functional block allocation is only a guideline as the system is automatically synthesised in a flat form. The post layout simulation at the physical level uses the NanoSim-VCS mix-signal co-simulation, which enables the combination of gate-level performance with transistor-level accuracy.  
\subsection{Benchmarks}
\label{Benchmarks}
\begin{figure}
\begin{center}
\includegraphics[width=0.5\textwidth]{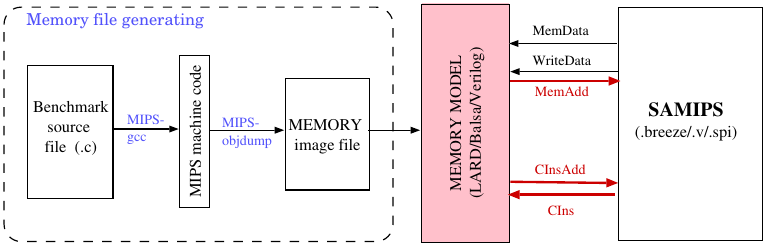}
\end{center}
\caption{Memory Mapping}
\label{Memory Mapping}
\end{figure}

Since SAMIPS's instruction set is fully compatible with MIPS ISA, it is able to execute the MIPS machine code generated from the standard compilers.  All the benchmarks adopted are originally written in C (except for ExcTest, which is partly written in the MIPS assembly language and partly in the MIPS machine code) and compiled into MIPS machine code by the MIPS cross compiler as illustrated in Figure \ref{Memory Mapping}. A memory image file is finally constructed from the MIPS machine code and fed into the memory model as a test bench. 


A set of four standard benchmarks have been used for the evaluation of SAMIPS:
\begin{table*}
\caption{Dynamic Instruction Usage}
\begin{center}
\begin{tabular}{|l||l|l|l|l|l|} \hline
   &  Dhrystone1.2 & Dhryston2.1 & Quicksort & Heapsort & ExcTest  \\ 
   &        &         &      &  &  \\ \hline \hline
 Memory     & 42.46\% & 43.80\% & 42.15\% & 36.39\% & 20.22\%   \\ \hline 
 Arithemtic & 23.03\% & 19.39\% & 18.23\% & 14.03\% & 9.33\%    \\ \hline 
 Logic      & 2.82\%  & 1.67\%  & 4.01\%  & 3.12\%  & 28.77\%   \\ \hline 
 Shift      & 1.90\%  & 1.09\%  & 5.72\%  & 7.84\%  & 3.73\%    \\ \hline 
 Nop        & 21.48\% & 25.33\% & 23.44\% & 27.90\% & 11.82 \%  \\ \hline 
 Multiply   & 0.28\%  & 0.17\%  & 0\%     & 0\%     & 0\%       \\ \hline 
 Branch     & 8.03\%  & 8.56\%  & 6.45\%  & 10.73\% & 9.95\%    \\ \hline 
 Special    & 0\%     & 0\%     & 0\%     & 0\%     & 5.60\%    \\ \hline 
 CP0        & 0\%     & 0\%     & 0\%     & 0\%     & 10.5\%    \\ \hline \hline 
 Total      & 1420    & 2290    & 3319    & 4298    & 643       \\ \hline 
\end{tabular}
\end{center}

\label{Dynamic Instructions Usage}
\end{table*}
 \textsf{Dhrystone 1.2}\cite{Weicker84} and \textsf{Dhrystone 2.1}\cite{York02} (both of these Dhrystone versions have been used for the evaluation of SAMIPS due to the different dynamic instruction usage);  \textsf{Quicksort}\cite{Hoare61} (the C program used here is applying quick sort on an array of 10 integers);  \textsf{Heapsort} \cite{Williams64} (exhibits higher usage of shift and branch operations,  an array of 10 integers is also used). In addition, since none of the standard programs will generate an exception, \textsf{ExcTest} is a short program for exceptions testing only, designed specifically for the evaluation of SAMIPS. \textsf{ExcTest} incorporates  4 arithmetic overflows, 4 memory address errors, 4 undefined instructions, 4 break instructions and 32 system calls. 

Table \ref{Dynamic Instructions Usage} illustrates the dynamic instruction usage of the five benchmarks, generated from the simulation trace file. 

\subsection{Analysis}
\label{Analysis}
\begin{table*}[t]
\caption{Behavioural Level Analysis}
\begin{center}
\begin{tabular}{|c|c|c|c|c|c|} \hline
        \multicolumn{5}{|c|}{Latency(units/per instruction)}   & Area Cost  \\ \cline{1-5}
        Dhrystone  & Dhrystone & Quicksort & Heapsort & ExcTest &  (microns)         \\ 
            1.2   &   2.1     &           &          &           &            \\	\hline \hline
     654.3  &   536.1   &  637.5    & 649.0    &  725.8    & 557,074     \\ \hline
\end{tabular}
\end{center}

\label{Behaviour Level Simulation Result}
\end{table*}
\begin{table*}[t]
\caption{Gate Level Analysis }
\begin{center}
\begin{tabular}{|l||c|c|c|c|c|c|} \hline
                     & \multicolumn{5}{|c|}{Latency(ns/per instruction)}                          & Transistor  \\ \cline{2-6}
                     & Dhrystone  & Dhrystone & Quicksort & Heapsort  & ExcTest & Count       \\
                     &      1.2   &   2.1     &           &           &           &             \\  \hline \hline
 {\em four\_b\_rb}   &    6.270      &    6.324 &     6.017    & 5.946     &  6.267    &  168,906        \\ \hline
 {\em dual\_b}       &    8.431      &    8.505 &  8.030    & 8.127     &  7.960    &  441,935        \\ \hline
 {\em one\_of\_2\_4} &    8.087      &    8.128 &  7.709    & 7.801     &  7.830    &  582,237        \\ \hline
\end{tabular}

\label{Gate Level Simulation Result}
\end{center}
\end{table*}

\begin{table*}[t]
\caption{Physical Level Analysis}
\begin{center}

\begin{tabular}{|l||c|c|c|c|c|c|c|c|c|c|c|c|}

\hline

& \multicolumn{10}{c|}{Speed (MIPS) and Power Efficiency (MIPS/W)} & Area Cost\\
\cline{2-11}

 & \multicolumn{2}{c|}{Drystone 1.2} & \multicolumn{2}{c|}{Drystone 2.1}& \multicolumn{2}{c|}{Quicksort}& \multicolumn{2}{c|}{Heapsort}& \multicolumn{2}{c|}{ExcTest} & (mm$^2$)\\
\cline{2-11}

 & Speed & Power & Speed & Power& Speed & Power& Speed & Power& Speed & Power& \\ \hline

{\em four\_b\_rb}  & 16.9  &  2197.8    &    16.6  & 2031.3   & 16.8    &  2120.7 &  17.8  & 2206.0  &    17.4 & 2132.1 &  0.9  \\ \hline

{\em dual\_b} &    6.8  &   543.3 &    6.7   & 560.4 &  6.8   & 586.1   &  7.0  & 541.7   &    7.6   & 649.6 & 2.1  \\ \hline

 {\em one\_of\_2\_4} &    7.7  &    902.7      &    7.6   &  873.6&   7.7  &   910.4    &  N/A & N/A   &   8.5  &   955.1 & 2.7 \\ \hline
\end{tabular}

\label{Physical Level Simulation Result}
\end{center}

\end{table*}



SAMIPS's performance has been evaluated and analysed through three levels of simulation, as described in section \ref{Simulation Environment}. 
The original simulation results are presented in Tables \ref{Behaviour Level Simulation Result}-\ref{Physical Level Simulation Result} and are further analysed in terms of area cost, power and speed in sections \ref{Analysis of Area Cost Distribution}-\ref{Analysis of Latency Distribution}. 

Table \ref{Behaviour Level Simulation Result} shows the Balsa behavioural level simulation results. 
At this level, data encoding and semiconductor technology is irrelevant and the simulation time is calculated based on the number of handshake events rather than the estimated gate latencies. Therefore, the execution latency per instruction shown in the table is only an approximation. For instance, the average execution time per instruction in Dhrystone1.2 is much higher than that of Dhrystone2.1 in Table \ref{Behaviour Level Simulation Result}, but they are similar in gate level simulation in Table \ref{Gate Level Simulation Result}. 
Balsa behavioural simulation is still useful for the analysis of the system in a technology independent way at the early stages of the design process. 

Table \ref{Gate Level Simulation Result} illustrates gate level simulation results. At this level, technology is introduced but the capacitance is not considered, and thus there is no power estimation. The shown latency is based on the simulation performed by Cadence's Verilog-XL on three different data encoding technologies, which is also calculated per instruction. As is clear in the table, the {\em four\_b\_rb} implementation of SAMIPS which is not delay insensitive, has the highest speed and smallest size. Comparing the two delay insensitive implementations, {\em one\_of\_2\_4} is a little higher in speed but larger in transistor count.

Table \ref{Physical Level Simulation Result} presents the physical level (post layout) simulation result. The estimated speed is presented in MIPS (Million Instructions per Second). {\em four\_b\_rb} implementation is 2.5 times faster than the {\em dual\_b} and 2.2 times than the {\em one\_of\_2\_4}. {\em one\_of\_2\_4} is the highest in area cost, about 3 times larger than {\em four\_b\_rb}. Power efficiency is reported in MIPS/W (Million Instruction per Second per Watt). Again {\em four\_b\_rb} is the most power efficient. 

\subsubsection{Area Cost}
\label{Analysis of Area Cost Distribution} 

\begin{figure} 
\begin{center}
\includegraphics[width=0.5\textwidth]{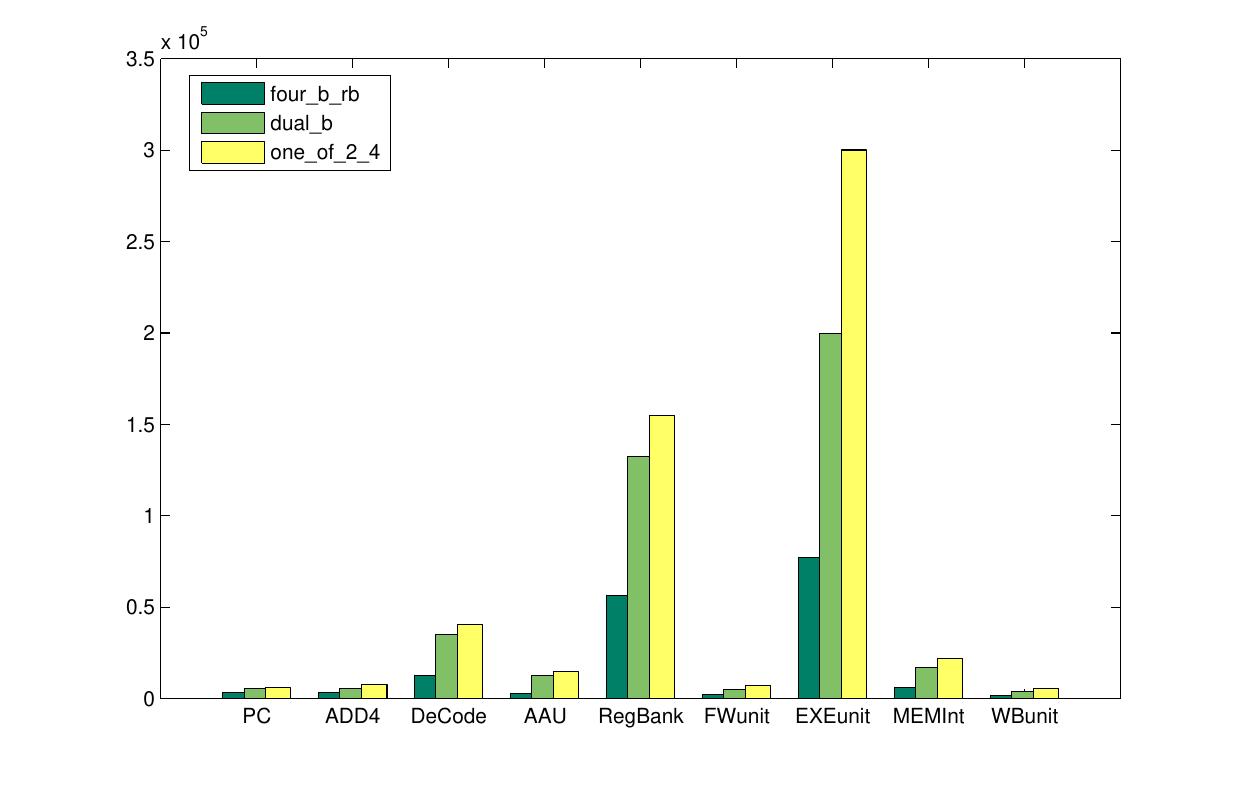}
\caption{SAMIPS Transistor Distribution per Functional Block }
\label{SAMIPS Transistor Distribution on Functional Block Analysis}
\vspace{1cm}

\includegraphics[width=0.5\textwidth]{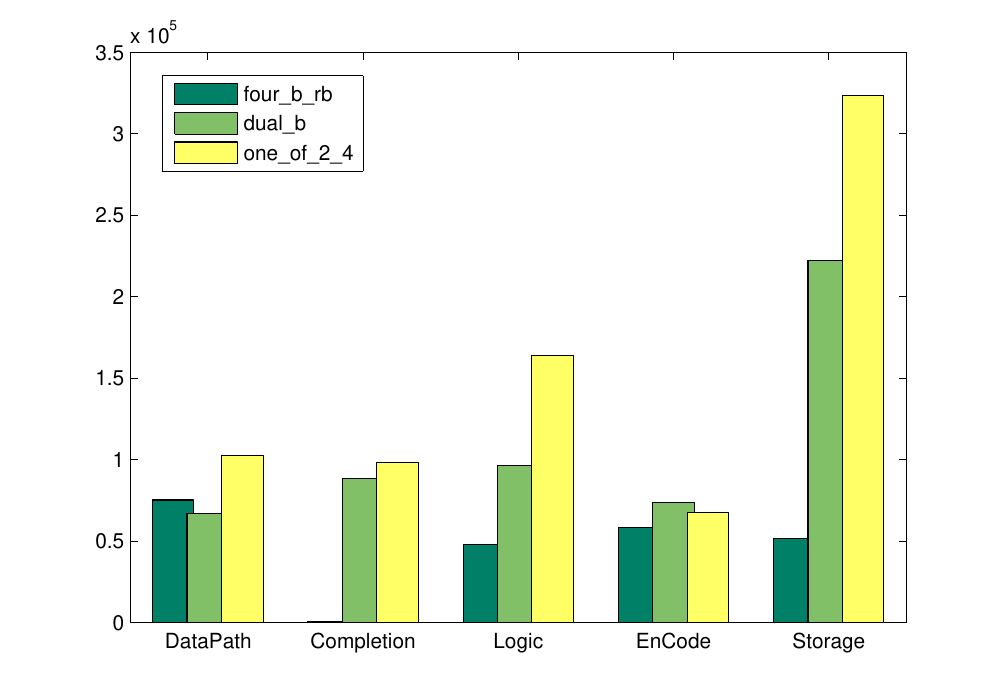}
\end{center}
\caption{SAMIPS Transistor Distribution per Implementation Logic \cite{Tom05}}
\label{SAMIPS Transistor Distribution Implementation Logic Analysis}
\end{figure}

Following the standard Balsa synthesis flow, area cost distribution analysis for SAMIPS has been undertaken at the gate level in terms of transistor count. Figure \ref{SAMIPS Transistor Distribution on Functional Block Analysis} illustrates the SAMIPS transistor distribution for nine major functional blocks (see also Figure \ref{SAMIPS RTL Design}). 
It is evident that the largest space is occupied by the \textsf{EXEunit} (47\%-54\%) and the \textsf{RegBank} (28\%-34\%). With regard to the latter the 32 general purpose registers occupy about 90\% space (see also section \ref{Evaluation of the Asynchronous Forwarding Mechanism}). 
Further analysis of the \textsf{EXEunit} at the behavioural level gives Table \ref{Area Cost Distribution of EXEunit}. The multiplier contributes the most in space and should be a good candidate to start with if area cost based optimisation is considered. On the other hand, modifications of the multiplier would not affect the overall speed much due to the very low usage of multiplication instructions (see Table \ref{Dynamic Instructions Usage}).

\begin{table}
\caption{Area Cost Distribution in \textsf{EXEunit}}
\begin{center}
\begin{tabular}{|c|c|c|c|c|} \hline
               shifter  & multiplier   & ALU   & comparator   & others  \\ 
 \hline \hline
 10.67\%    & 41.92\%   & 28.85\%   & 7.07\%   & 18.55\%  \\  \hline
\end{tabular}

\label{Area Cost Distribution of EXEunit}
\end{center}
\end{table}

An additional area cost distribution analysis of SAMIPS, based on the implementation logic of Balsa circuits, has been conducted.
and is illustrated in Figure \ref{SAMIPS Transistor Distribution Implementation Logic Analysis}. This analysis divides the Balsa handshake components of SAMIPS into five separate categories: \textsf{Datapath}, \textsf{Completion} \textsf{Logic} \textsf{EnCode} and \textsf{Storage}.
Both delay-insensitive ({\em dual\_b} and {\em one\_of\_2\_4}) implementations spend considerably more space on completion and storage than the speed-insensitive ({\em four\_b\_rb}) implementation. This confirms that the handshake control logic needed in the delay-insensitive encoding is much more complicated than that needed in the speed-insensitive one. 
It also suggests that the combined utilisation of technologies may result to a more efficient implementation, e.g. {\em one\_of\_2\_4} is a good alternative to {\em dual\_b} in \textsf{Logic} and \textsf{Storage}. 


\subsubsection{Power}
\label{Analysis of Power Distribution} 
\begin{figure} [t]
\begin{center}
\includegraphics[width=0.5\textwidth]{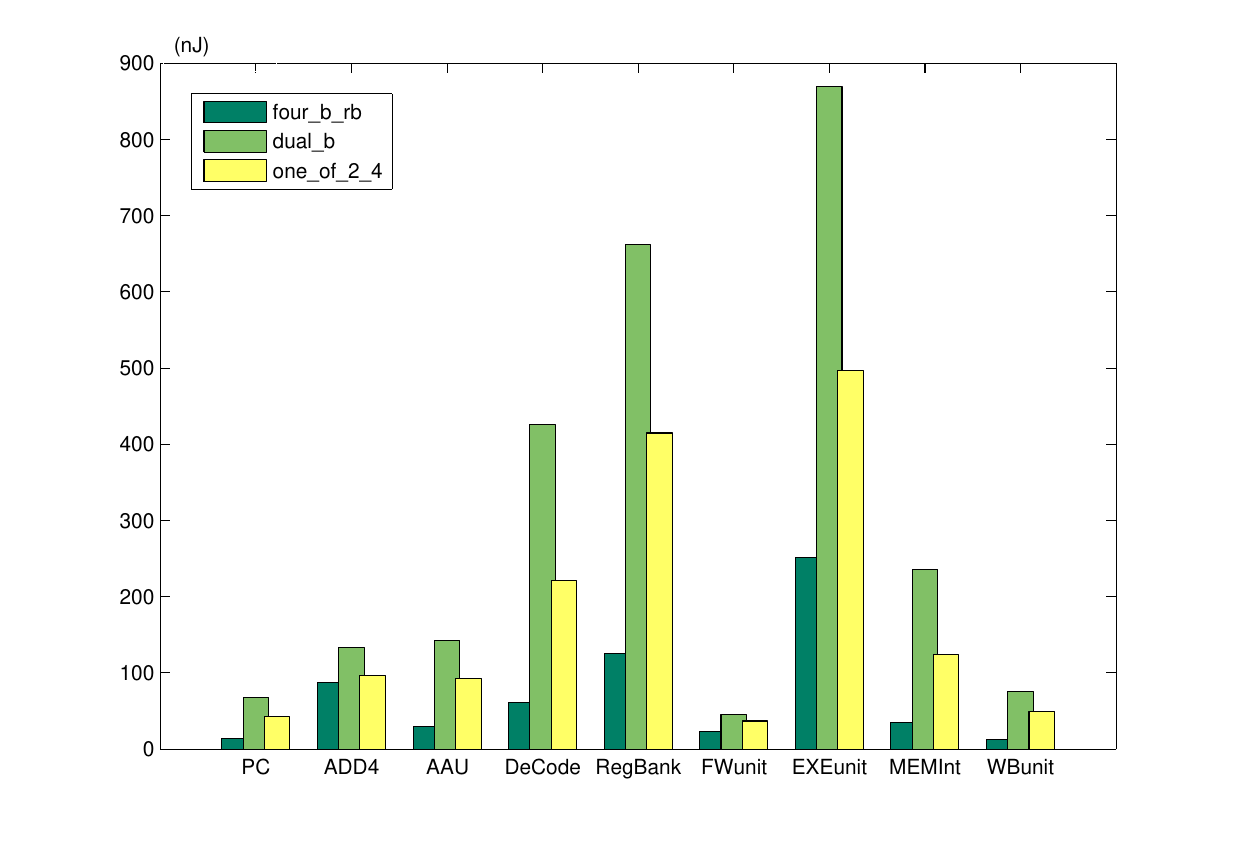}
\caption{SAMIPS Power Distribution per Functional Block }
\label{SAMIPS Power on Functional Block Analysis}

\includegraphics[width= 0.5\textwidth]{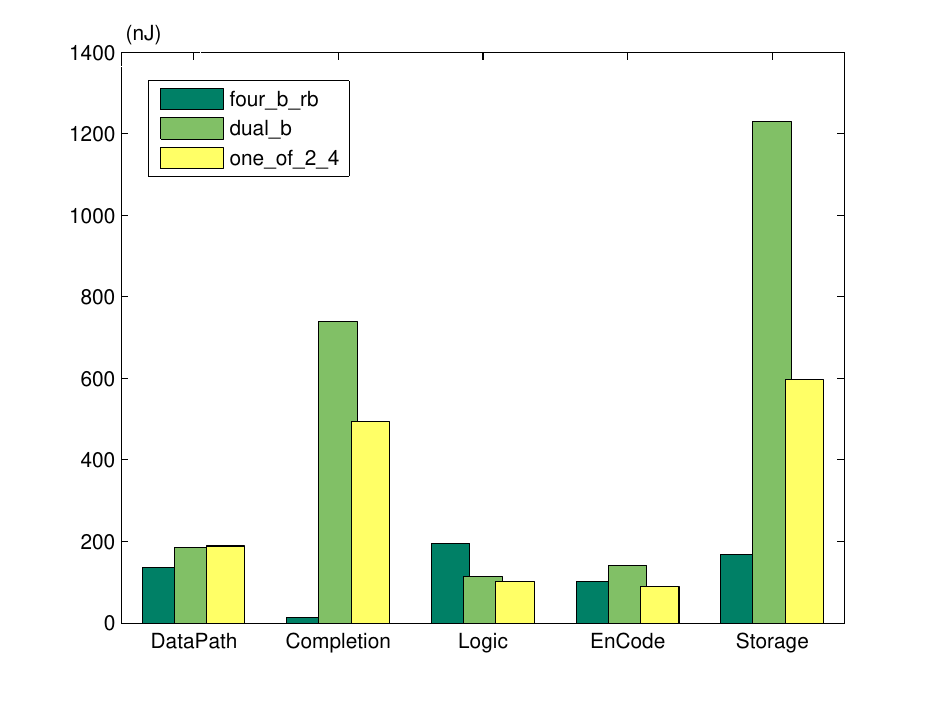}
\caption{SAMIPS Power Distribution per Implementation Logic. 
}
\label{SAMIPS Power Implementation Logic Analysis}
\end{center}
\end{figure}

Figure \ref{SAMIPS Power on Functional Block Analysis} illustrates the SAMIPS power distribution for nine major functional blocks of SAMIPS (data is taken from the post layout simulation of Dhrystone 1.2). The \textsf{EXEunit} (32\% - 39\%) and the \textsf{RegBank} (20\% - 26\%) consume most power. This is expected as both units are in the main datapath and hence very busy throughout the whole execution period and both have a high transister cost.  
The third largest contribution is from the \textsf{DeCode}, in that it is also activated in each execution cycle. Consequently, any attempt to increase the power efficiency of SAMIPS should consider \textsf{EXEunit}. For instance, it could be possible to terminate some instructions before the EX stage and remove them from the pipeline earlier, e.g. NOP. 

Figure \ref{SAMIPS Power Implementation Logic Analysis} shows the power distribution analysis on the implementation logic. 
Most power goes to completion and storage components in the {\em dual\_b} and {\em one\_of\_2\_4}. For the storage logic, the {\em one\_of\_2\_4} column is only about the half of the {\em dual\_b} one. If the combined use of the encoding technologies is supported, {\em one\_of\_2\_4} could save nearly half of the power compared to {\em dual\_b} with only 30\% more transistors. 

\subsubsection{Latency}
\label{Analysis of Latency Distribution}
\begin{figure}[t]
\begin{center}
\includegraphics[width= 0.45\textwidth]{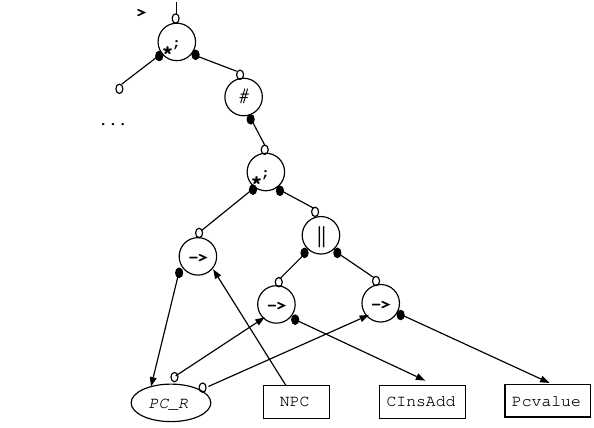} 
\caption{Handshake Circuit for the Program Counter (\textsf{PC})}
\label{Handshake Circuit for PC}
\end{center}
\end{figure}

Since each functional block of SAMIPS works at its own rate and may have different internal operation latencies, each functional block can be in one of the three following states, namely \textit{busy}, \textit{wait} and \textit{idle}. During the \textit{busy} time ($T_{busy}$), a functional block is processing its input requests and generating the output requests to the other blocks; during the \textit{wait} time ($T_{wait}$), a functional block has finished all the internal processing and is waiting for its output requests to be correctly acknowledged; during the \textit{idle} time ($T_{idle}$), a functional block is inactive. 

The latency analysis aims to measure the internal processing time $T_{busy}$, the external delay $T_{waiting}$ and the $T_{idle}$ of each functional block. Such analysis can identify bottlenecks in the system and has been performed 
using the gate level verilog-XL simulation results. 
Each functional block has a set of input and output handshake channels. The calculation of the $T_{waiting}$ is based on the activity analysis of the output channels. 

As an example, figure \ref{Handshake Circuit for PC} shows the handshake network for the \textsf{PC} unit of SAMIPS (see also Figure \ref{SAMIPS RTL Design}). It has two outputs, \textit{PCvalue} and \textit{CInsAdd}. After the initiation, the handshake 
circuit enables an endless loop to its right branch, which keeps reading in the new PC value from the Balsa register \textit{NPC}, 
storing it in a register \textit{PC\_R} and then passing the new PC to the two output channels. For the {\em four\_b\_rb} implementation, two waiting times can be distinguished: (a) from the time the last output request up ($r\uparrow$) is sent out to the time the first acknowledgement up ($a\uparrow$) is received; (b) from the time the first acknowledgement down ($a\downarrow$) is received  to the time the last acknowledgement up ($a\uparrow$) is received. Formula 2 shows the calculation of $T_{waiting}$ for the \textsf{PC}.

\begin{equation}
\begin{split}
T_{waiting} = \Big(MAX\{PCvalue\_0a\uparrow,  CInsAdd\_0a\uparrow\}\\
- MIN\{PCvalue\_0a\downarrow,
CInsAdd\_0a\downarrow\}\Big)\\
+ \Big(MIN\{PCvalue\_0a\uparrow, CInsAdd\_0a\uparrow\}\\
- MAX\{PCvalue\_0r\uparrow, CInsAdd\_0r\uparrow\}\Big)
\end{split}
\end{equation}

Since the \textsf{PC} only has one pull input \textit{NPC}, it is deactivated after sending out the request of \textit{NPC} until receiving the acknowledgement with the data from the \textsf{AAU}. The calculation of $T_{idle}$ is shown in Formula 3.

{\setlength\arraycolsep{2pt}
\begin{eqnarray}
T_{idle} & = & NPC\_0a\uparrow - NPC\_0r\uparrow  
\end{eqnarray}}

 $T_{busy}$ is the total processing time, and is simply calculated as the total simulation time minus $T_{waiting}$ and $T_{idle}$, shown in Formula 4.

{\setlength\arraycolsep{2pt}
\begin{eqnarray}
T_{busy} & = & T_{total} - T_{waiting} - T_{idle}
\end{eqnarray}}

\begin{figure}[t]
\begin{center}
\includegraphics[width=0.5\textwidth]{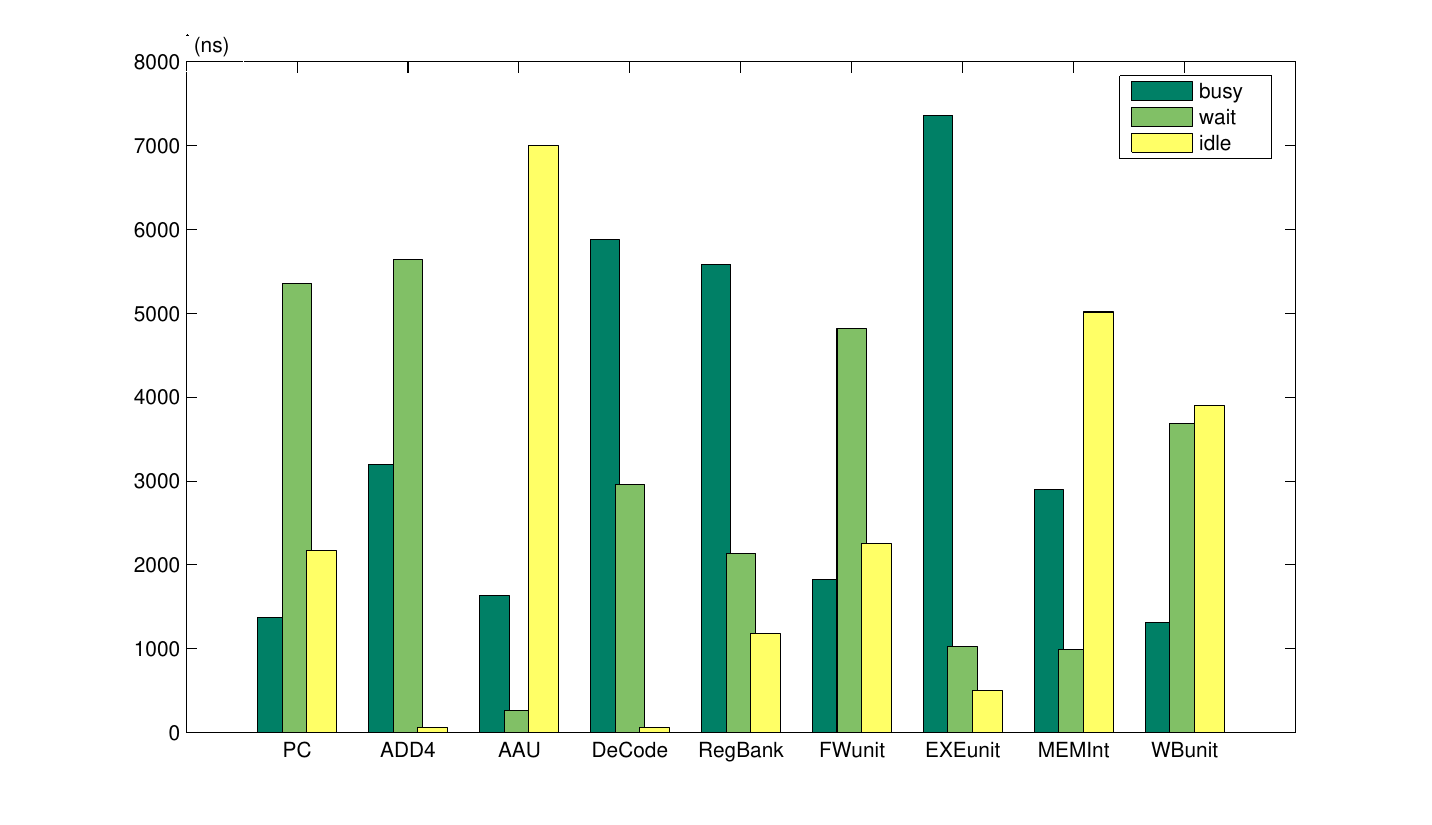} 
\caption{SAMIPS Latency Distribution Analysis ~(\textit{four\_r\_rb})}
\label{Functional Block Utilisation}
\end{center}
\end{figure}

Figure \ref{Functional Block Utilisation} illustrates the latency distribution results of the \textit{four\_r\_rb} implementation. It clearly shows that the \textsf{EXEunit} is the busiest one, which causes the other units to wait. The second and third busiest blocks are the \textsf{DeCode} and the \textsf{RegBank}. Some blocks have very short busy times but long waiting times, such as the \textsf{PC}, \textsf{ADD4}, \textsf{FWunit} and \textsf{WBunit}.   This is because they are connected to the component with a long processing latency, and hence their output requests cannot be acknowledged before the receiver finishes its own processing, e.g. the \textsf{ADD4} has to wait for the \textsf{DeCode} to acknowledge its output channel \textit{BaseAddID} (see also figure \ref{SAMIPS RTL Design}). 

Table \ref{Channel Activity} presents an analysis of  utilisation of major SAMIPS channels. 
Although channel based analysis does give a clear latency distribution figure, the calculation on those complex functional blocks with multiple inputs and outputs could be very complicated and time consuming. An alternative approach is the isolation layout at the physical level. The average processing latency ($L_{avg}$) of each functional block  could be measured if they are laid out and simulated separately. $T_{busy}$ is then calculated using Formula 5. 
\begin{eqnarray}
T_{busy} & = & L_{avg} \times No. \; of \; input \; request
\end{eqnarray}

\begin{table}
\caption{Channel Utilisation}
\begin{center}
\begin{tabular}{|p{1.7cm}||l|ll|ll|} \hline
Channel  & No. of   & \multicolumn{2}{c|}{Wait} & \multicolumn{2}{c|}{Idle}  \\ 
	 & Requests &  \multicolumn{2}{c|}{(ns)} &  \multicolumn{2}{c|}{(ns)} \\ \hline \hline
\textit{Pcvalue}  & 1540     & 4805  & 54.0\% & 3850 & 43.2\%  \\ \hline
\textit{PCplus4}  & 1540     & 1227  & 13.8\% & 7545 & 84.7\%  \\ \hline 
\textit{NPC}      & 1540     & 302   & 3.39\% & 8189 & 92.0\%  \\ \hline
\textit{BaseAddID}& 1538     & 5865  & 65.9\% & 2189 & 24.6\%  \\ \hline
\textit{BaseAddEX}& 1538     & 2490  & 28.0\% & 5599 & 62.9\%   \\ \hline
\textit{Offset32} & 980      & 1446  & 16.23\%& 7317 & 82.2\%   \\ \hline
\textit{Sa}       & 65       & 97    & 1.1\%  & 879  & 98.7\%  \\ \hline	 
\textit{RegRead}  & 1538     & 1217  & 13.7\% & 7639 & 85.8\%  \\ \hline
\textit{RegWrite} & 904      & 2704  & 30.4\% & 6374 & 71.6\%  \\ \hline
\textit{EXCtrl}   & 1538     & 2188  & 24.6\% & 5267 & 59.1\%  \\ \hline
\textit{FRACtrl}  & 1538     & 3162  & 3.55\% & 8489 & 95.3\%  \\ \hline
\textit{FWCtrl}   & 1538     & 17    & 0.19\% & 8688 & 97.6\%   \\ \hline
\textit{ReadData0}& 1167     & 1278  & 14.4\% & 7413 & 83.3\%   \\ \hline
\textit{ReadData1}& 1400     & 1411  & 15.9\% & 7237 & 81.3\%   \\ \hline
\textit{FOp0}     & 371      & 95    & 1.07\% & 8693 & 97.6\%    \\ \hline
\textit{FOp1}     & 137      & 41    & 0.46\% & 8819 & 99.0\%   \\ \hline
\textit{MEMCtrl}  & 1114     & 217   & 2.43\% & 8665 & 97.3\%  \\ \hline
\textit{EXRes}    & 1114     & 196   & 2.20\% & 8674 & 97.4\%   \\ \hline
\textit{EXRd}     & 884      & 2557  & 2.87\% & 8621 & 96.8\%   \\ \hline
\textit{MemD}     & 209      & 256   & 2.88\% & 8617 & 96.8\%   \\ \hline
\textit{WBCtrl}   & 884      & 141   & 1.58\% & 8750 & 98.3\%   \\ \hline  
\textit{MEMRes}   & 884      & 85    & 0.95\% & 8806 & 98.9\%  \\ \hline
\textit{MEMRd}    & 884      & 188   & 2.11\% & 8685 & 7.5\%  \\ \hline
\textit{FEXRes}   & 884      & 241   & 2.71\% & 8739 & 98.2\%  \\ \hline
\textit{FMEMRes}  & 884      & 3743  & 42.0\% & 5257 & 59.0\%  \\ \hline
\textit{IDch}	 & 1	    & 1     & 0.01\% & 8901 & 99.97\% \\ \hline
\textit{EXch}	 & 93	    & 72    & 0.8\%  & 8815 & 99.0\%  \\ \hline
\textit{MEMch}	 & 0	    & 0     & 0\%    & 8904 & 100\%   \\  \hline
\end{tabular}
\end{center}
\label{Channel Activity}
\end{table}

\subsection{A Comparative Analysis}

This section presents a comparison between SAMIPS and other Balsa and MIPS implementations, both synchronous and asynchronous. These systems are of course not directly comparable, however the analysis does provide some useful insights about the positioning of SAMIPS in the ecosystem.  Physical level simulation results are used, however it is important to emphasise that even at this level these are still estimated data from simulators and not measured from the fabricated processor. 
\subsubsection{Comparison with another Balsa System}

\begin{table}
\caption{Comparison with SPA}
\begin{center}
\begin{tabular}{|p{1.14cm}|p{0.9cm}||p{0.6cm}|p{1.1cm}|p{0.68cm}|p{1cm}|} \hline                     
 \multicolumn{2}{|c||}{Processor}  & Core  & Transistor & Speed  & Power    \\ 
 \multicolumn{2}{|c||}{}           & size (mm$^2$)    & count       & (MIPS) & (MIPS/W) \\  
 \multicolumn{2}{|c||}{}           &           &             &        &          \\  \hline \hline
 {\em four\_b\_rb}   & SAMIPS  & 0.9    & 168, 906    & 16.6   & 2031.3   \\ \cline{2-6}
	             & SPA     & 1.2    & 318, 054    & 7.1    & 1427	  \\ \hline \hline
 {\em dual\_b}       & SAMIPS  & 2.1    & 441, 935    & 6.7    & 560.4    \\ \cline{2-6}
                     & SPA     & 2.6    & 656, 958    & 3.8    & 509.2    \\ \hline \hline
 {\em one\_of\_2\_4} & SAMIPS  & 2.7	& 582, 237    & 7.6    & 873.6    \\ \cline{2-6}
                     & SPA     & 3.5    & 872, 762    & 4.0    & 610.5    \\ \hline 
\end{tabular}
\begin{footnotesize}
\begin{tabular}{llp{6cm}}
 Benchmarks: & Dhrystone2.1 (SAMIPS); & Arm Test Suit T1 (SPA)\\

 Process: & 0.18$\mu$m@1.8v & \\
\end{tabular}
\end{footnotesize}
\end{center}

\label{Compared to SPA}
\end{table}
 To examine how SAMIPS compares with other BALSA synthesised systems, the SPA processor core has been selected.  SPA is an asynchronous ARM compatible processor core developed at the University of Manchester as  part of the G3Card project \cite{G3Card} and fabricated in the Buxton chip \cite{Plan02} to investigate the applicability of self-timed logic in security-sensitive devices. 
 SPA is particularly suitable for our analysis as it is a RISC processor core, fully specified and synthesised in Balsa.  SAMIPS and SPA have also been contrasted as part of a different exercise to evaluate Balsa \cite{10.1109/ASYNC.2006.29}. 
 
Table \ref{Compared to SPA} presents the comparative simulation results in terms of core size, transistor count, speed and power efficiency. SPA is about 1.2-1.6 times bigger than SAMIPS in area cost and 1.5-1.9 times higher in total number of transistors. The speed and power report of SAMIPS is from the simulation of Dhrystone 2.1 while that of SPA is generated by running test T1 from the ARM Test Suit. The power efficiency of both processor cores in the {\em dual\_b} implementation is similar, but that of SAMIPS is much higher than SPA in {\em four\_b\_rb} and {\em one\_of\_2\_4} implementations. In terms of speed, SAMIPS is about 1.7-2.3 times faster than SPA; besides the ISA difference, the  higher speed of SAMIPS may also be attributed to the fact that SPA is built around a three-stage pipeline architecture with no result forwarding.  

\subsubsection{Comparison with other Asynchronous MIPS}

MiniMIPS, TITAC-2 and ARISC are asynchronous processors based on MIPS architecture. Although  they implement the same ISA as SAMIPS, these processors all use different pipeline architectures, design techniques and implementation technologies. Table \ref{other AsyncMIPS} presents reported data from these processors. 

Both MiniMIPS and ARISC are much larger compared to the single-rail SAMIPS in terms of total transistor count. The results indicate that SAMIPS performs well in terms of power consumption, while speed wise it appears to be slower. In comparison with MiniMIPS, SAMIPS is about nine times slower and two and three times slower than TITAC and ARISC respectively.  The lower speed of SAMIPS can be attributed to the fact that SAMIPS is automatically synthesized and therefore not optimized (e.g. MiniMIPS supports out-of-order completion). 

\begin{table}[t]
\caption{Reported Results from other Asynchronous MIPS Implementations}
\begin{center}
\begin{tabular}{|p{1.1cm}|p{0.8cm}|p{0.8cm}|p{1.1cm}|p{0.8cm}|p{1.1cm}|} \hline
Processor & Process & Core  & Transistor   &  Speed  &  Power     \\ 
          & ($\mu$m)  & size (mm$^2$)   & count        &  (MIPS) &	      \\ \hline \hline
MiniMIPS  & 0.6     & N/A	  & 250, 000   &  180	 & 4W	      \\ \hline
TITAC-2   & 0.5     & N/A       & 496, 367     &  52.3   & 2.11W      \\ \hline
ARISC     & 0.35    & 2.2       & N/A          &  74-123 & 635 MIPS/W \\ \hline
\end{tabular}
\begin{footnotesize}
\begin{tabular}{llp{4cm}}
Conditions: & MiniMIPS, TITAC-2(measured) @3.3v & \\
            & ARISC(simulated) @3.3v & \\
\end{tabular}
\end{footnotesize}
\end{center}

\label{other AsyncMIPS}
\end{table}

\subsubsection{Comparison with Synchronous MIPS}


At the time of writing, and to the best of our knowledge, there are no reported figures for a 0.18$\mu$m MIPS R3000 implementation. In line with the approach taken for the comparative analysis of MiniMIPS  in \cite{Martin01}, figures provided in \cite{Bhan96} can be used as a basis, shown in Table \ref{syncMIPS}. The performance of a hypothetical 0.18$\mu$m can then be extrapolated from Table \ref{syncMIPS}, if we assume that the same scaling factor is applied to the feature size, the thin oxide and the Vdd voltage \cite{Martin01}. In this case,  throughput scales linearly with the feature size. By extrapolation, a 0.6$\mu$m synchronous MIPS R3000 would have a speed of 47MHZ, that is with 7MHZ increase per 0.2$\mu$m. Therefore, the speed of the 0.18$\mu$m synchronous R3000 would be around 62MHZ. The speed of SAMIPS is about one third of that of the synchronous R3000.  

Extrapolating for power efficiency is not trivial \cite{Martin01}. Published figures for a 0.35$\mu$m implementation of MIPS R3000A architecture  report 
power efficiency of 350MIPS/W including the on chip memory (256-512 KB) \cite{Toshiba}. The power efficiency of SAMIPS {\em four\_b\_rb} implementation for 0.18$\mu$m is 2197.8MIPS/W at 1.8v.

\begin{table}
\caption{Performance of Commercial Synchronous MIPS R3000 Implementations \cite{Bhan96}}
\begin{center}
\begin{tabular}{|l|c|c|c|c|c| } \hline
Processor & Process & Transistor & Speed & Power & Condition       \\ 
          & ($\mu$m)  & count     & (MHZ) &	 (W)    &          \\ \hline \hline
R3000    & 1.2      & 110, 000  & 25    &	  4   & cpu only   \\ \hline
R3000A   & 1.0      & 110, 000  & 33    &  N/A   & cpu only        \\ \hline
VR3600   & 0.8      & 190, 000  & 40    &	 N/A   & cpu+fpu   \\ \hline
\end{tabular}
\end{center}

\label{syncMIPS}
\end{table}


\section{Evaluation of the Asynchronous Forwarding Mechanism}
\label{Evaluation of the Asynchronous Forwarding Mechanism}

This section presents a detailed evaluation in terms of the area cost, delay, and power consumptions introduced to implement the asynchronous forwarding mechanism designed for SAMIPS. Both DHDT and DHDQ have been implemented and compared. The simulation results indicate that the extra control cost incurred in the \textsf{RegBank} for the data hazard detection is not significant, compared to the cost of the rest of the \textsf{RegBank}. The scalability of the algorithm for increasing pipeline depth has also been investigated.  

\subsection{The Experimental Frame}

To analyse the forwarding mechanism, a simplified CPU model has been built and used, which consists of two parts namely: (a) A Balsa model, including a \textsf{RegBank}, a \textsf{FWunit} and two Multiplexers, as shown in Figure \ref{Asynchronous Forwarding} and (b) Verilog models for all the other pipeline stages. Verilog has been used for the pipeline modelling for two main reasons: (a) in Verilog, specify or adjust the operation delays can be explicitly defined and adjested; however in Balsa, such latencies are automatically calculated by the Balsa system during simulation;(b) Verilog delivers higher layout and post layout simulation speeds.  The benchmark used (the input data stream of \textit{RegRead}) is extracted from the trace file of SAMIPS simulation of Dhrystone 1.2).

The number \textit{n} of pipeline stages range from 5 to 9, and the assumption is that the execution stage $S_i$ starts from the third stage of the pipeline (i.e. $i=3$ see also  section \ref{Asynchronous Forwarding in SAMIPS}). For the comparison between DHDT and DHDQ, a 5-stage pipeline ($n=5$) is used.

A latency of 20ns has been adopted for the Verilog models of the pipeline stages in the model. This is to reflect the factor of 1.6 between the latencies of the slowest stage (\textsf{EXEUnit}) and the \textsf{RegBank} in SAMIPS. The latency of the Balsa \textsf{RegBank} model for this experiment is approximately 12ns (see Figure \ref{ForMech Delay Analysis}(b)). 

\subsection{Results}
\begin{table}
\caption{Average Idle Time of \textsf{FWunit} (ns)}
\begin{center}
\begin{tabular}{|p{1.3cm}||p{0.8cm}|p{0.8cm}|p{0.8cm}|p{0.8cm}|p{0.8cm}|} \hline
                    & $n=5$ & $n=6$  & $n=7$  & $n=8$ & $n=9$  \\                                        
                    &       &       &       &       &      \\ \hline \hline
{\em four\_b\_rb}   & 13.97 & 13.74 & 13.51 & 13.34 & 13.29 \\ \hline
{\em dual\_b}       & 15.29 & 14.59 & 14.58 & 13.92 & 13.64  \\ \hline
{\em one\_of\_2\_4} & 14.33 & 14.43 & 14.29 & 14.35 & 14.05  \\ \hline
\end{tabular}

\label{Average Idle Time inside the FWunit}
\end{center}
\end{table}


Table \ref{Average Idle Time inside the FWunit} presents the average idle time in \textsf{FWunit}. The idle time is about 14ns, nearly 3/4 of the pipeline stage latencies (20ns assumed). In other words, the \textsf{FWunit} spends most of its time ``idle". This indicates that before the \textsf{RegBank} sends the next pair of \textit{FRACtrl} and \textit{FWCtrl}, the \textsf{FWunit} has already correctly processed all the forwarded results and is ready for the next cycle. 


\begin{table*}[t]
\caption{Latency Analysis of Forwarded Operands (ns)}
\begin{center}
\begin{tabular}{|l|c|p{3.1cm}||p{1.2cm}|p{1.2cm}|p{1.2cm}|p{1.2cm}|p{1.2cm}|} \hline
 \multicolumn{3}{|c||}{}                             & n=5        & n=6       & n=7        & n=8        & n=9        \\
 \multicolumn{3}{|c||}{}                             &            &           &            &            &            \\	        \hline \hline
                    & \textit{Op0} & Forwarding Frequencies & 403 27.0\% & 497 33.4\%& 540 36.2\% & 585 39.3\% & 613 41.1\% \\  \cline{3-8}
{\em four\_b\_rb}   &	  & Average Latency          & 9.93       & 9.58      & 9.65       & 10.01      & 10.07      \\  \cline{2-8}
                    & \textit{Op1} & Forwarding Frequencies & 193 13.0\% & 208 14.0\%& 234 15.7\% & 245 16.4\% & 247 16.6\% \\  \cline{3-8}
                    &     & Average Latency          & 7.71       & 7.66      & 7.64       & 7.56       & 7.60       \\  \hline \hline
                    & \textit{Op0} & Forwarding Frequencies & 283 19.0\% & 401 26.9\%& 492 33.0\% & 545 36.6\% & 598 40.1\% \\  \cline{3-8}
{\em dual\_b}       &     & Average Latency          & 9.45       & 8.28      & 8.32       & 8.36       & 8.61       \\  \cline{2-8}
                    & \textit{Op1} & Forwarding Frequencies & 102 6.85\% & 162 10.9\%& 183 12.3\% & 217 14.6\% & 234 15.7\% \\  \cline{3-8}
                    &     & Average Latency          & 7.58       & 9.45      & 8.86       & 9.62       & 9.72       \\  \hline \hline
                    & \textit{Op0} & Forwarding Frequencies & 403 27.0\% & 440 29.5\%& 460 30.9\% & 507 34.0\% & 552 37.0\% \\  \cline{3-8}
{\em one\_of\_2\_4} &     & Average Latency          & 7.49       & 7.82      & 7.96       & 8.2        & 8.33       \\  \cline{2-8}
                    & \textit{Op1} & Forwarding Frequencies & 193 13.0\% & 180 13.2\%& 196 13.2\% & 204 13.7\% & 226 15.2\% \\  \cline{3-8}
                    &     & Average Latency          & 6.81       & 8.11      & 8.26       & 9.38       & 8.75       \\  \hline	       	       
\end{tabular}

\label{Latency Analysis of Forwarded Operands}
\end{center}
\end{table*}

Table \ref{Average Idle Time inside the FWunit} provides an analysis of the activity of the \textsf{FWUnit} and the associated delays.  Forwarding frequencies refer to the number of times a
forwarded result was used by the \textsf{FWUnit}, or in other words, the number of
times a data hazard occurred. Percentages as calculated out of a total of
1490 instructions (\textit{RegRead} requests). Overall, data hazards increase with
the number of pipeline stages. This is expected as there are more
instructions in the pipeline executing in parallel and therefore the
probability of a data hazard increases. However, it should be emphasised that
because the pipeline is asynchronous the relative sequence of read and write-back requests for a particular register is nondeterministic, something that affects the total number of data hazards.

The latencies in Table \ref{Average Idle Time inside the FWunit} represent the average time it takes for the
forwarded results to reach \textsf{EXEUnit}. Overall it is observed that the average
delay experience small variations. 
This is due to
the fact that the Balsa model for the \textsf{Regbank} has to be modified for
different pipeline stages which result in different Verilog netlists being
produced. This in turn has also an influence in the relative timing of the
events in the system. It should be emphasised that the pipeline stage that
will forward the result depends on the relative sequence of instructions
being executed and therefore remains the same. 



\begin{figure}[t]
\subfigure[]{\includegraphics[width=0.4\textwidth]{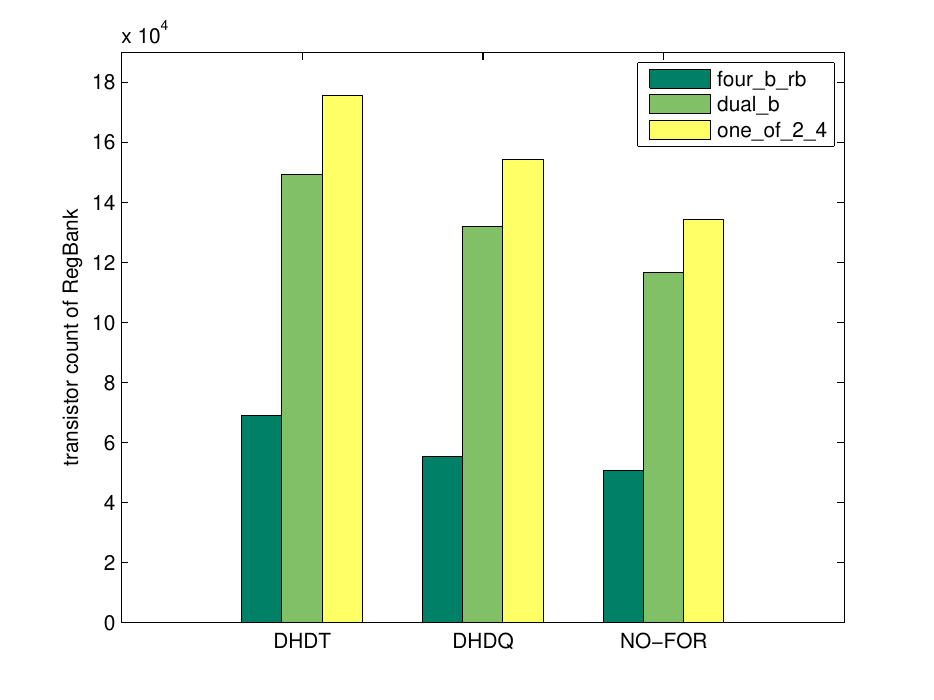}}
\subfigure[]{\includegraphics[width=0.4\textwidth]{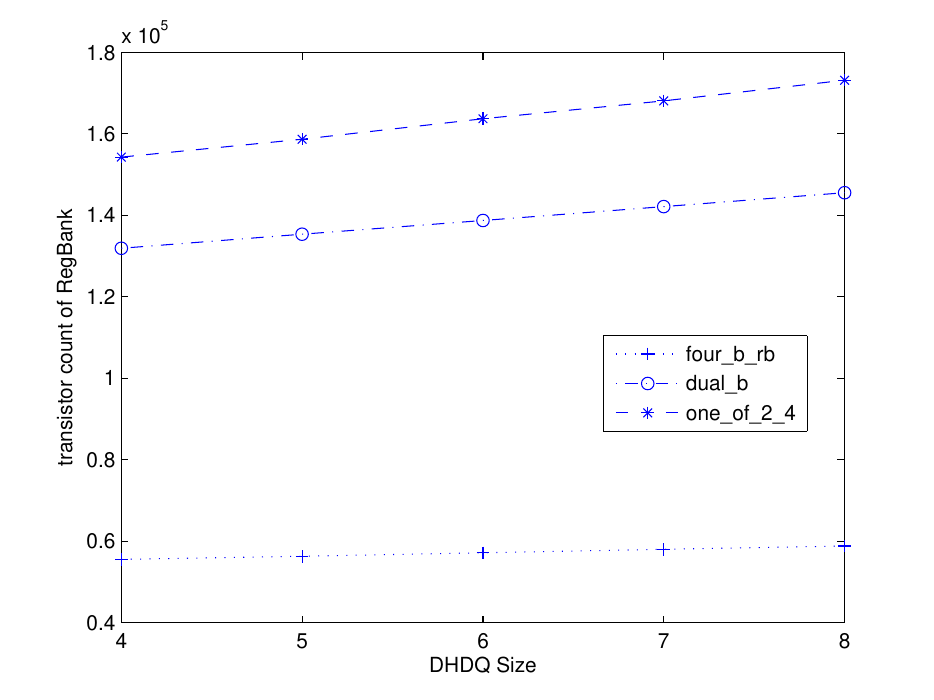}}
\caption{Area Cost Analysis of the Forwarding Mechanism}
\label{ForMech Transistor Count Analysis}
\end{figure}

\begin{figure}
\subfigure[]{\includegraphics[width=0.4\textwidth]{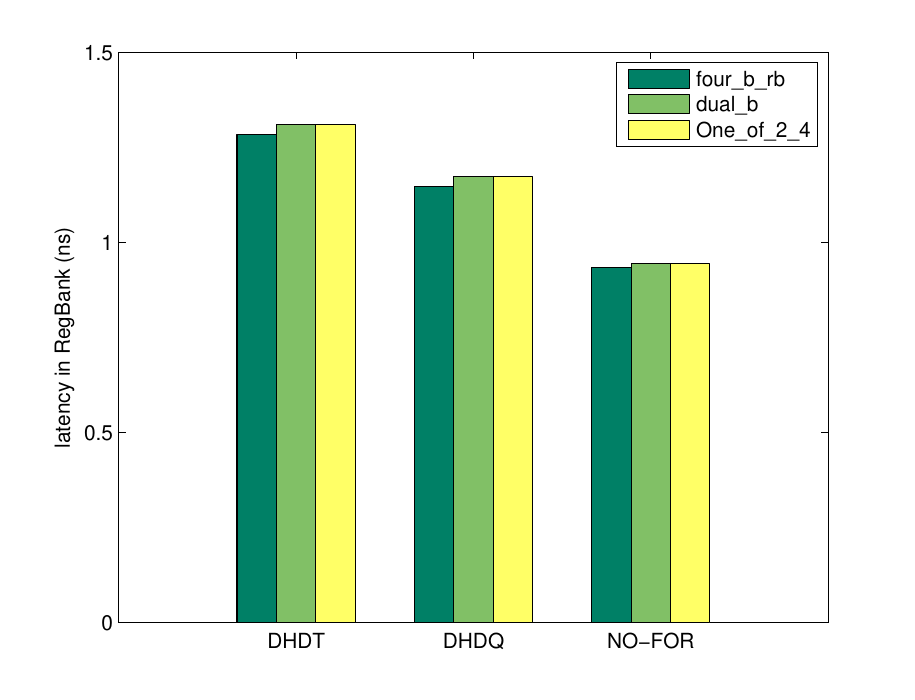}}
\subfigure[]{\includegraphics[width=0.4\textwidth]{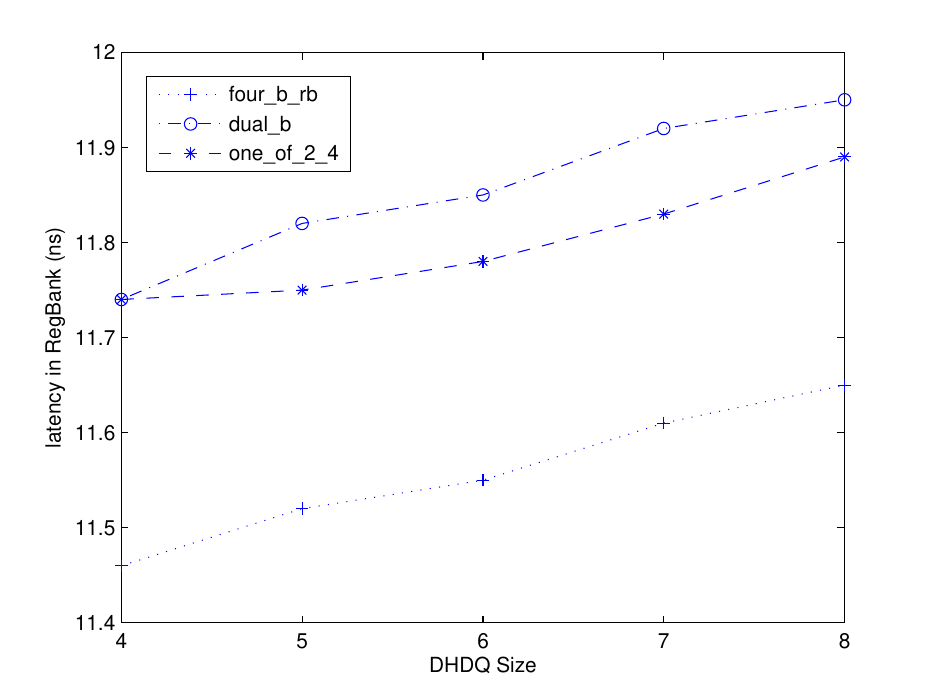}}
\caption{Delay Analysis of the Forwarding Mechanism}
\label{ForMech Delay Analysis}
\end{figure}

\begin{figure}
\subfigure[]{\includegraphics[width=0.4\textwidth]{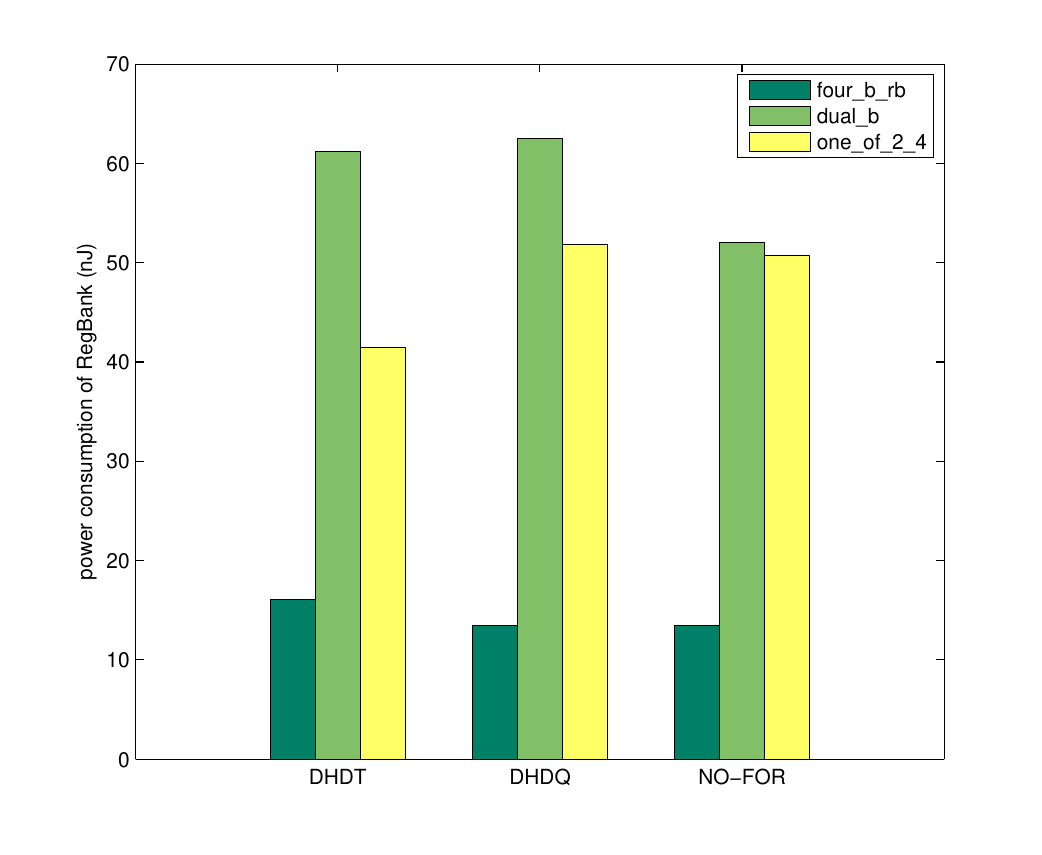}}
\subfigure[]{\includegraphics[width=0.4\textwidth]{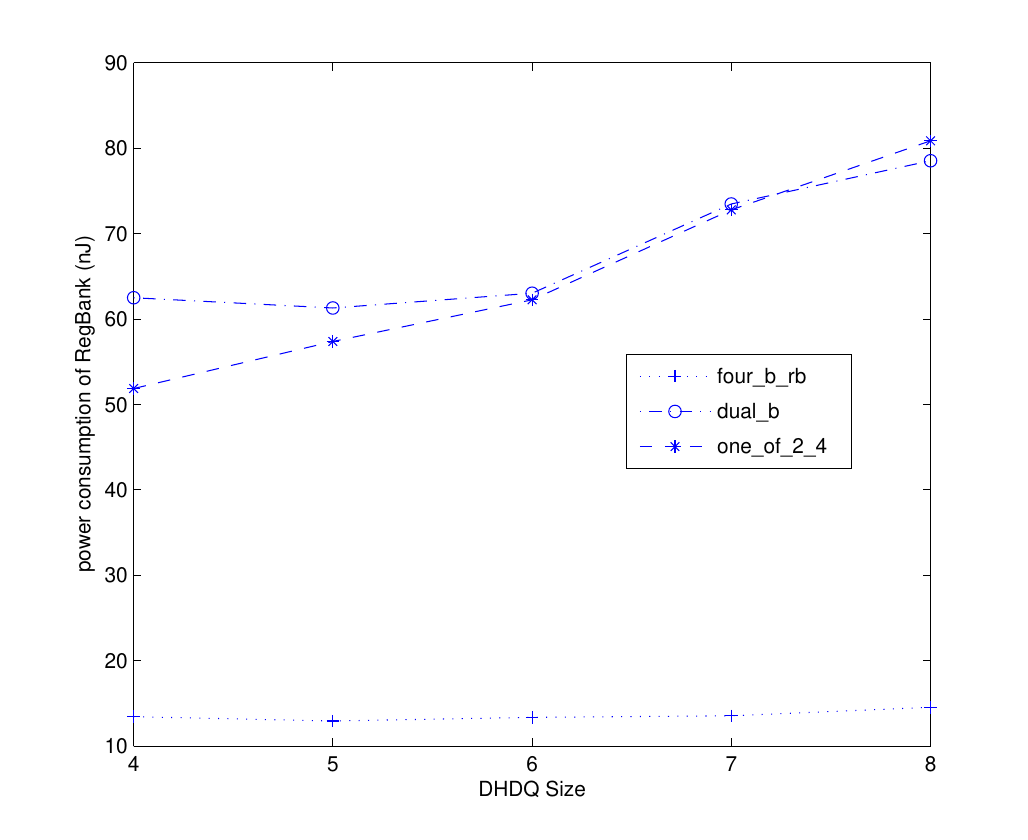}}
\caption{Power Analysis of the Forwarding Mechanism}
\label{ForMech Power Analysis}
\end{figure}

Figures \ref{ForMech Transistor Count Analysis}-\ref{ForMech Power Analysis} present an analysis of the different implementations of the \textsf{RegBank} in terms of area cost, latency and power consumption respectively. Each figure presents (a) a comparison DHDT, DHDQ and No-Forwarding (normal register reading and writing operations without data hazard detection) implementation of the \textsf{RegBank} for a 5-stage pipeline. (b) an illustration of the scalability of the approach, in terms of the overhead introduced as the number of pipeline stages increases; the data used is from the DHDQ implementation for pipeline sizes ranging from 4 to 8 (for a 5-stage pipeline, the size of DHDQ is 4).

\subsubsection{Area Cost Analysis}
Figure \ref{ForMech Transistor Count Analysis} shows area cost in terms of number of transistors, obtained from the Balsa \textit{\textbf{balsa-net-tran-cost}} utility. The comparison between the DHDT and DHDQ implementation clearly shows that the introduced area overhead of DHDT is higher than that of DHDQ. Compared to a \textsf{RegBank} without any forwarding mechanism implemented (NO-FOR column) in {\em four\_b\_rb} encoding, DHDT has introduced 36.10\% extra transistors while DHDQ only 9.28\%. The former is 3.89 times larger than the latter which is very close to what we have calculated in Table \ref{Storage Cost of DHDT and DHDQ} where the cost factor is 3.8 (95/20) when $n=5$. However, in {\em dual\_b} and {\em one\_of\_2\_4} encoding, this cost factor is reduced to 2.14 and 2.06, which is due to the fact that large completion detection circuits are most significant in delay-insensitive data encoding.  DHDQ implementation introduced 9.28\% ({\em four\_b\_rb}), 13.0\% ({\em dual\_b}) and 14.9\% ({\em one\_of\_2\_4}) extra cost to the \textsf{RegBank}, which can be considered acceptable. When DHDQ is adopted, the transistor count of \textsf{RegBank} increases linearly with the total number of the pipeline stages in all three encoding technologies as expected. The increase rates are about 800 transistors per stage in {\em four\_b\_rb}, 3,400 in {\em dual\_b} and 4,700 in {\em one\_of\_2\_4}.  

\subsubsection{Latency Analysis}
As shown in Figure \ref{ForMech Delay Analysis}, for DHDT the data hazard detection has introduced 37.58\% operation latency inside the \textsf{RegBank} while for the DHDQ implementation 22.7\% ({\em four\_b\_rb}), namely,the latency introduced by the DHDT implementation is about 1.12 times higher than that of the DHDQ implementation.  The average latency in \textsf{RegBank} increases as the DHDQ size increases, wih the increase following the data hazard frequency patterns. This is something expected, because the latency inside the \textsf{RegBank} differs when a data hazard occurs and when a normal register file reading is processed, so the overall average latency in the \textsf{RegBank} will change if the data hazard frequency changes. As illustrated in Table \ref{Latency Analysis of Forwarded Operands},the increase of the data hazard is not linear and this will consequently affect the average latency inside the \textsf{RegBank}. 

\subsubsection{Power Analysis}
\label{Forward Power Analysis}
The power consumption of the \textsf{RegBank}, in \textit{nJ}, is presented in Figure \ref{ForMech Power Analysis}. Although the forwarding mechanism seems to have introduced a substantial number of comparison circuits in its Balsa description (see Figure \ref{RegBankcode}), these comparison operations do not really require substantially extra power, especially in {\em four\_b\_rb} implementation. This is because with the use of the forwarding mechanism, there is no output on the channel \textit{ReadData0} or \textit{ReadData0} (illustrated in section \ref{The Forwarding Algorithm}) when a data hazard occurs, which results to power saving. Due to the increased number of transistors required for the {\em dual\_b} and {\em one\_of\_2\_4} implementations the total energy consumed in both of these is  higher than that in the {\em four\_b\_rb}, e.g. for DHDQ and for ($n=5$),  {\em four\_b\_rb} is 13.44, {\em dual\_b} is 62.49 and {\em one\_of\_2\_4} is 51.87. The increase in power consumption as the size of DHDQ increases is small. 


\section{Evaluation of the Multi-Colour Algorithm}
\label{Evaluation of the Multi-Colour Algorithm}
The functional correctness of the multi-colour algorithm  has been validated by incorporating it in SAMIPS and also through formal verification\cite{Wang04, Wang05}. This section provides a evaluation of the algorithm in terms of area cost, speed, and power consumption. The target of the analysis is the overhead introduced by the \textsf{AAU} unit in relation to the rest of SAMIPS. 

\begin{figure}
\subfigure[]{\includegraphics[width=0.4\textwidth]{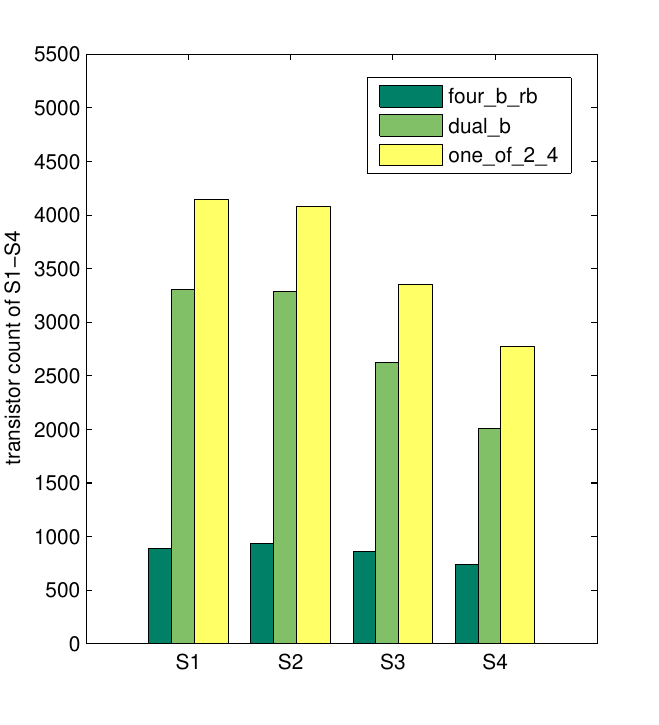}}
\subfigure[]{\includegraphics[width=0.45\textwidth]{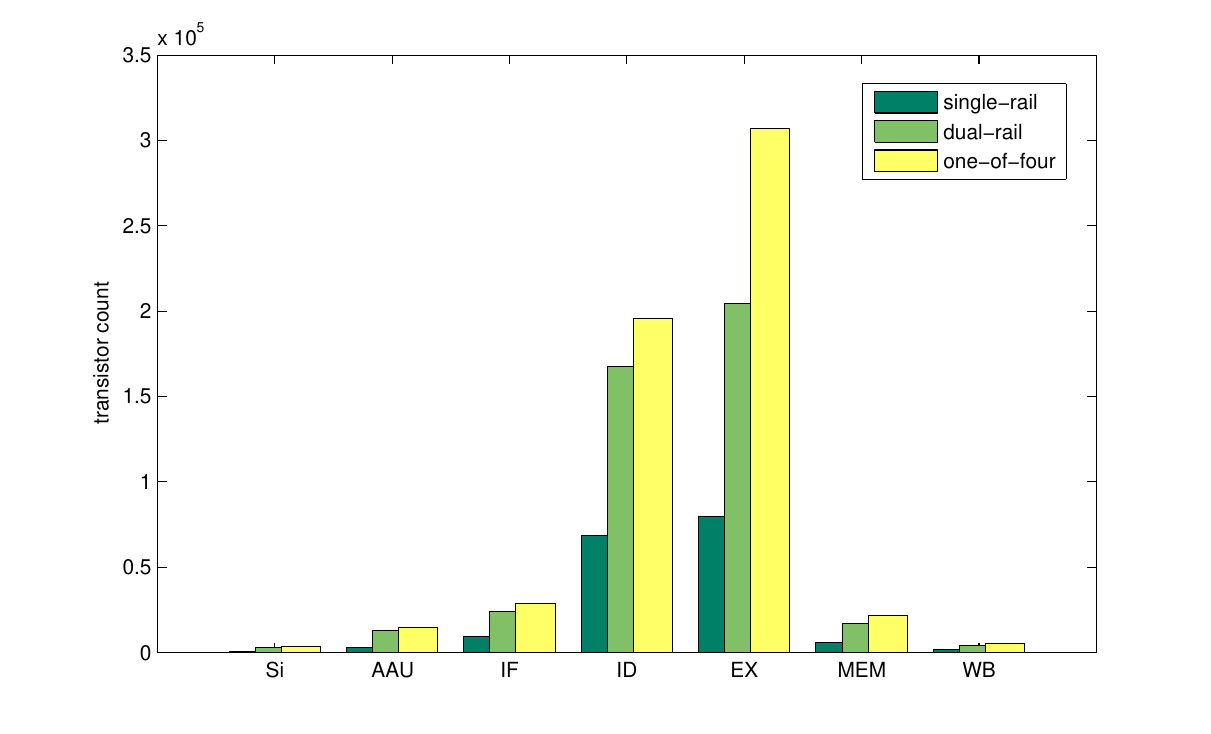}}
\caption{Transistor Count Analysis Compared to SAMIPS}
\label{Transistor Count Analysis}
\end{figure}

\begin{figure}
\subfigure[]{\includegraphics[width=0.4\textwidth]{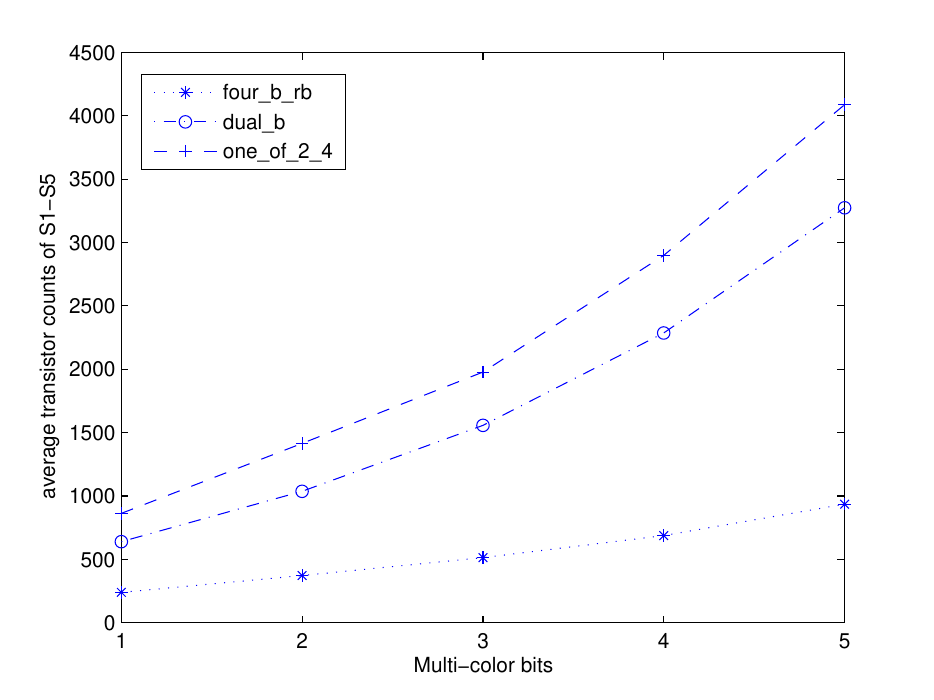}}
\subfigure[]{\includegraphics[width=0.4\textwidth]{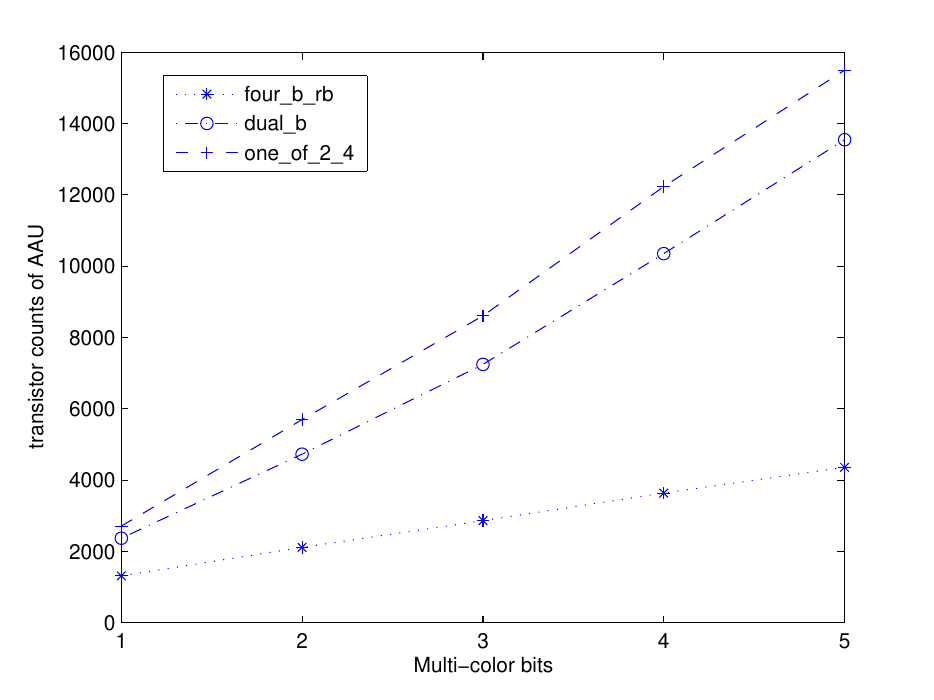}}
\caption{Transistor Count per Colour Vector Size}
\label{Transistor Count Analysis Vector}
\end{figure}

\subsection{The Experimental Frame}

To obtain the actual area cost and delay information introduced by the implementation of the multi-colour algorithm, a simplified Balsa model of a five-stage pipeline is extracted from SAMIPS architecture. The model consists of 7 parallel processes, one for each of the pipeline stages ($S_1$-$S_5$), one for the \textsf{PC} and one for the \textsf{AAU}, in line with Figure \ref{Stages}. 

A Verilog memory model is used for testing, where pseudo instructions are located and fetched sequentially, unless the control flow is changed. Instructions are specially designed, only consisting of two types, namely hazard instructions and NOP instructions. A hazard instruction has 5 bits, which present control hazard generating command for five stages respectively: \{CH1, CH2, CH3, CH4, CH5 \}. The definition of a NOP instruction is the same as MIPS ISA, basically passing through the pipeline with no operations. 

\subsection{Results}
\begin{figure}[t]
\subfigure[]{\includegraphics[width=0.4\textwidth]{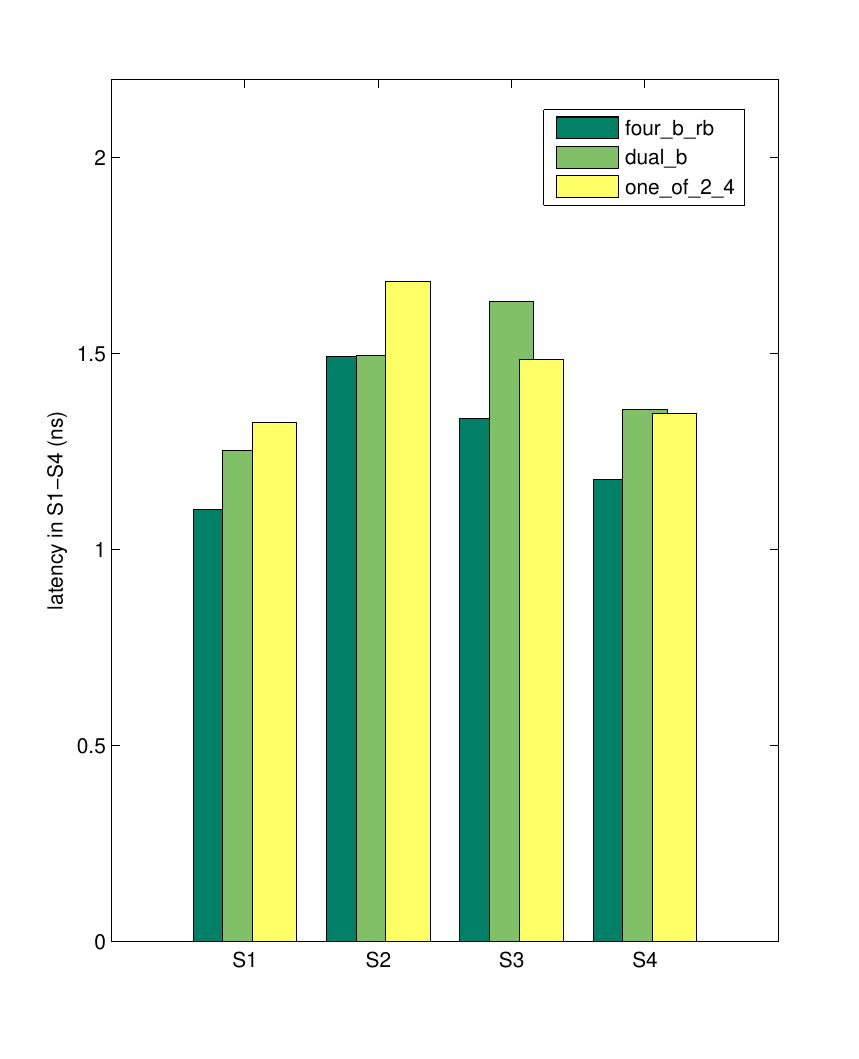}}
\subfigure[]{\includegraphics[width=0.45\textwidth]{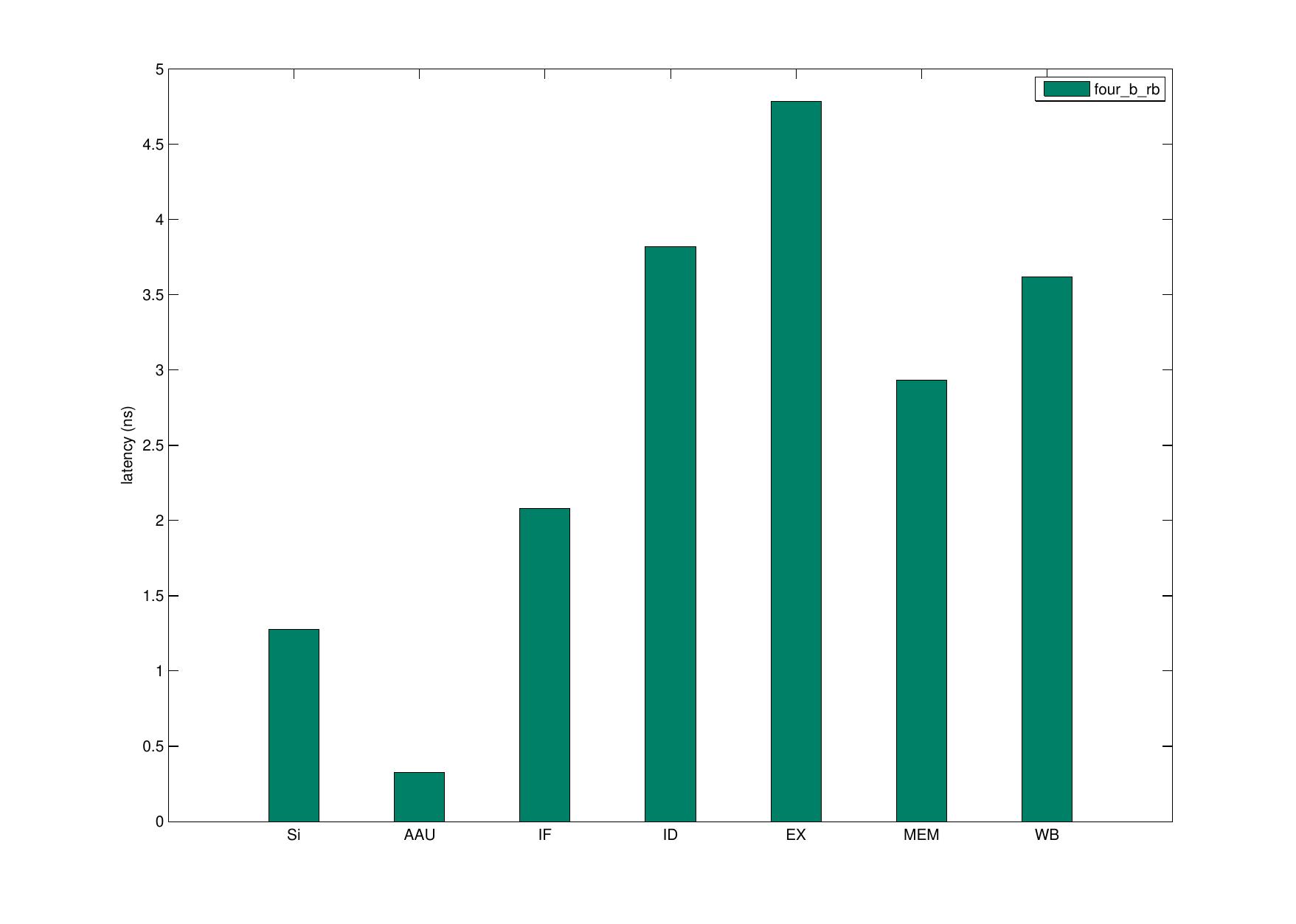}}
\caption{Latency Analysis Compared to SAMIPS}
\label{Operation Delay Analysis}

\end{figure}  

\begin{figure}[t]
\subfigure[]{\includegraphics[width=0.4\textwidth]{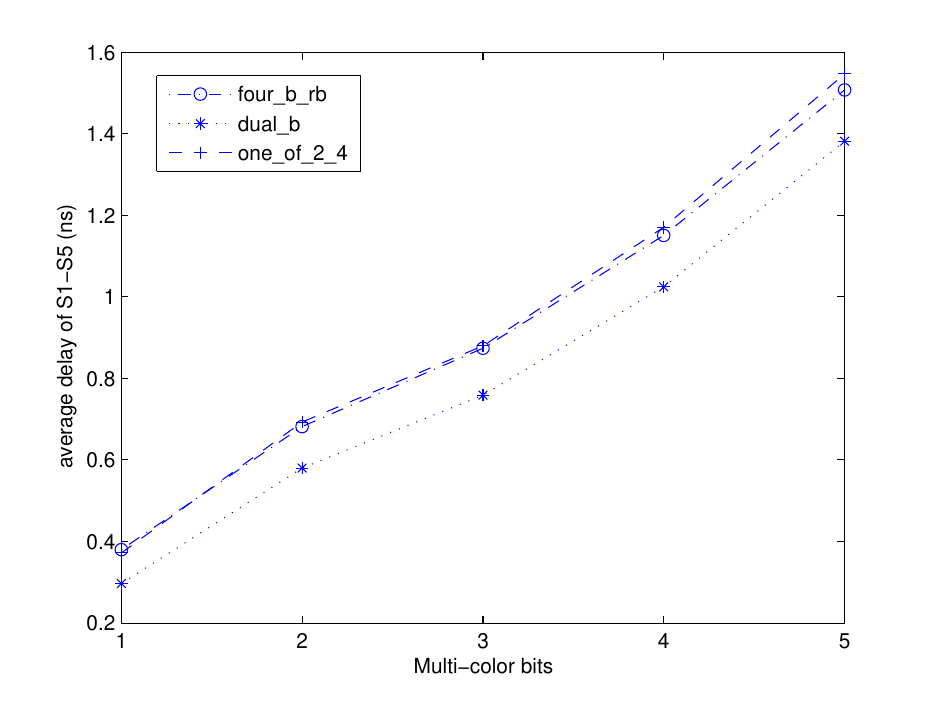}}
\subfigure[]{\includegraphics[width=0.4\textwidth]{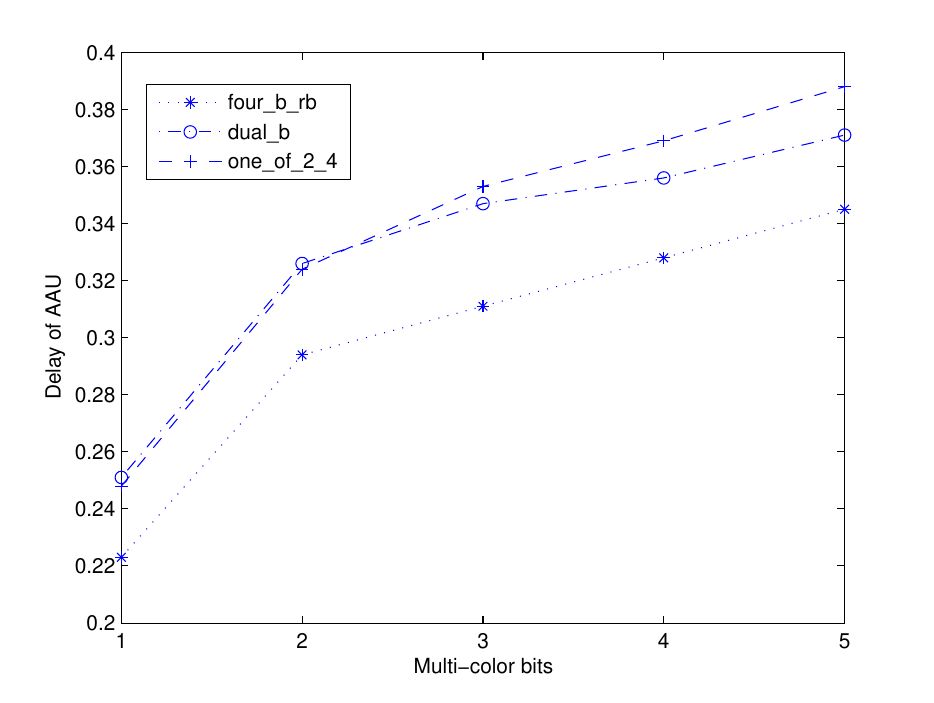}}
\caption{Latency Analysis on Colour Vector Size}
\label{Delay Analysis}
\end{figure}  

Figures \ref{Transistor Count Analysis}-\ref{Power Analysis} present the obtained results for the area cost, latency and power consumption respectively.  Two sets of results are presented for each case. The first set evaluates the algorithm by comparing the obtained results from $S_1$-$S_5$ and \textsf{AAU} with SAMIPS (in SAMIPS, the colour vector size is 4); the second set illustrates the scalability, in terms of overhead introduced with increasing colour vector size from 1 to 5.    
\subsubsection{Area Cost Analysis}

Figures \ref{Transistor Count Analysis} and \ref{Transistor Count Analysis Vector} illustrate the transistor count analysis, obtained from the Balsa (\textit{\textbf{balsa-net-tran-cost}}) utility count.  Figure \ref{Transistor Count Analysis}(a) shows the transistor count for different pipeline stages. The difference in costs is due to the different colour comparison circuits in the different pipeline stages. It is obvious that {\em four\_b\_rb} implementation introduces the minimum extra cost. Figure \ref{Transistor Count Analysis}(b) shows the average cost of pipeline stages ($S_1$-$S_5$, shown as $S_i$ in the figure) and the cost of the \textsf{AAU} in relation to the rest of functional blocks of SAMIPS. Clearly the overheads introduced are insignificant.

 Figure \ref{Transistor Count Analysis Vector}(a) and (b) present the increase in transistor count for pipeline stages (average) and the \textsf{AAU} respectively with an increasing number of color bits. The overhead introduced by {\em four\_b\_rb} implementation is the smallest. 

\subsubsection{Latency Analysis}

Figure \ref{Operation Delay Analysis} and \ref{Delay Analysis} present latency results.  In Figure \ref{Operation Delay Analysis}(a) the latencies in each pipeline stage is illustrated, with {\em four\_b\_rb} delivering the best result. Figure \ref{Operation Delay Analysis} the latency in the \textsf{AAU} for the {\em four\_b\_rb} implementation in relation to the rest of the SAMIPS components is indeed small. 

In Figure \ref{Delay Analysis}(a), the operation latency for {\em four\_b\_rb} implementation is still the smallest, albeit the difference with the other two implementations is smaller than that in area cost and power analysis. The steep slop between the first two points in Figure \ref{Delay Analysis}(b) is attributed to the fact that the comparison circuits are much simpler for the 1-bit colour algorithm than for the n-bit (n$>$1) algorithm, as there is no need for the priority checking.

\subsubsection{Power Analysis}

\begin{figure}
\subfigure[]{\includegraphics[width=0.4\textwidth]{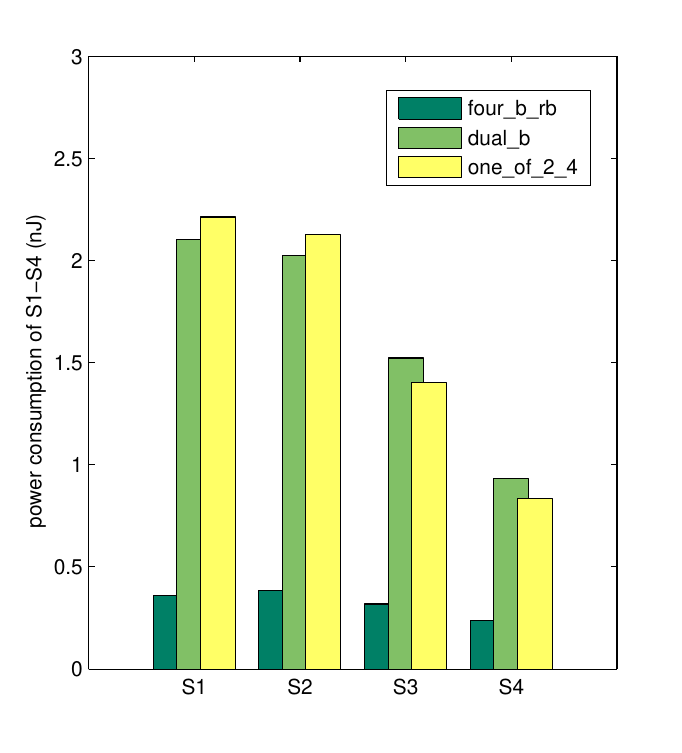}}
\subfigure[]{\includegraphics[width=0.45\textwidth]{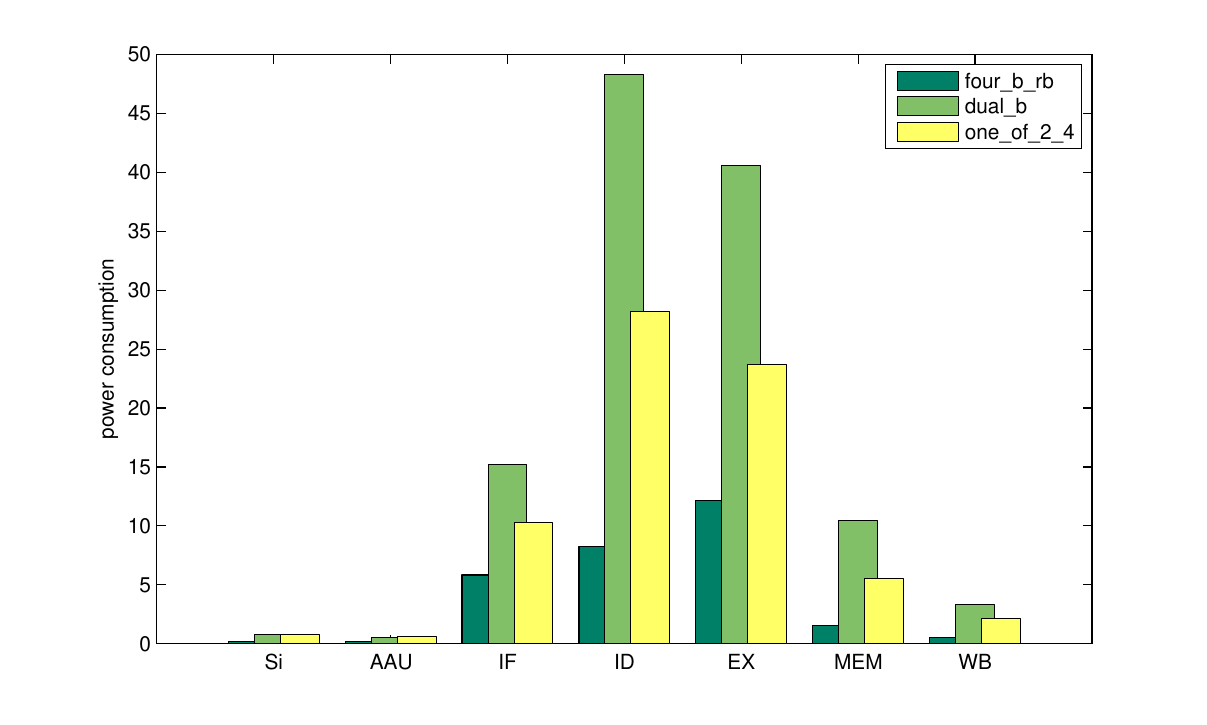}}
\caption{Power Analysis Compared to SAMIPS}
\label{Power Consumption Analysis}
\end{figure}

\begin{figure}
\subfigure[]{\includegraphics[width=0.4\textwidth]{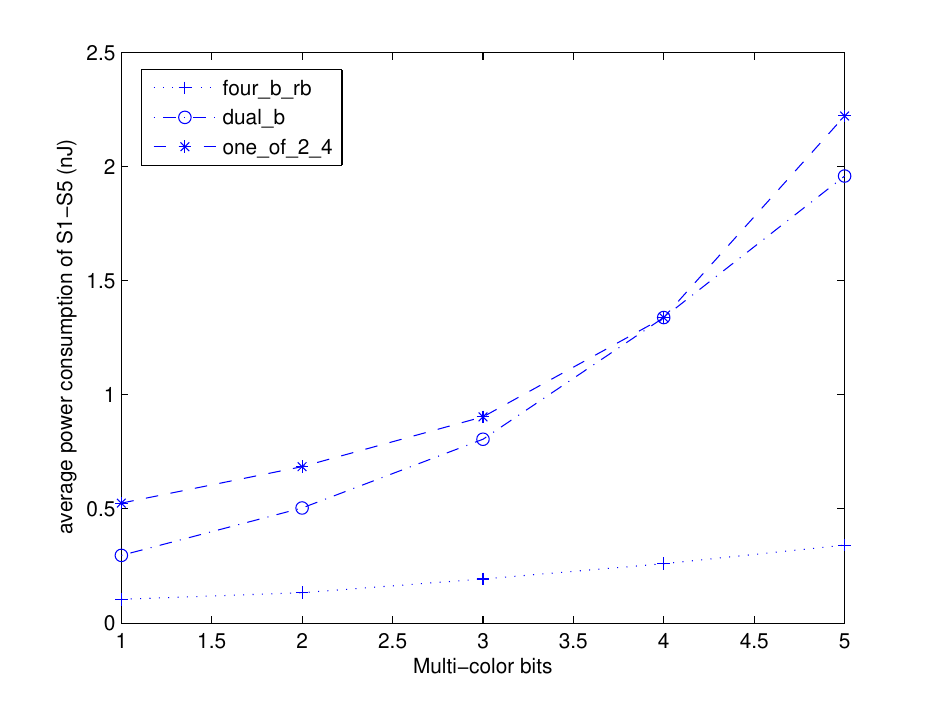}}
\subfigure[]{\includegraphics[width=0.4\textwidth]{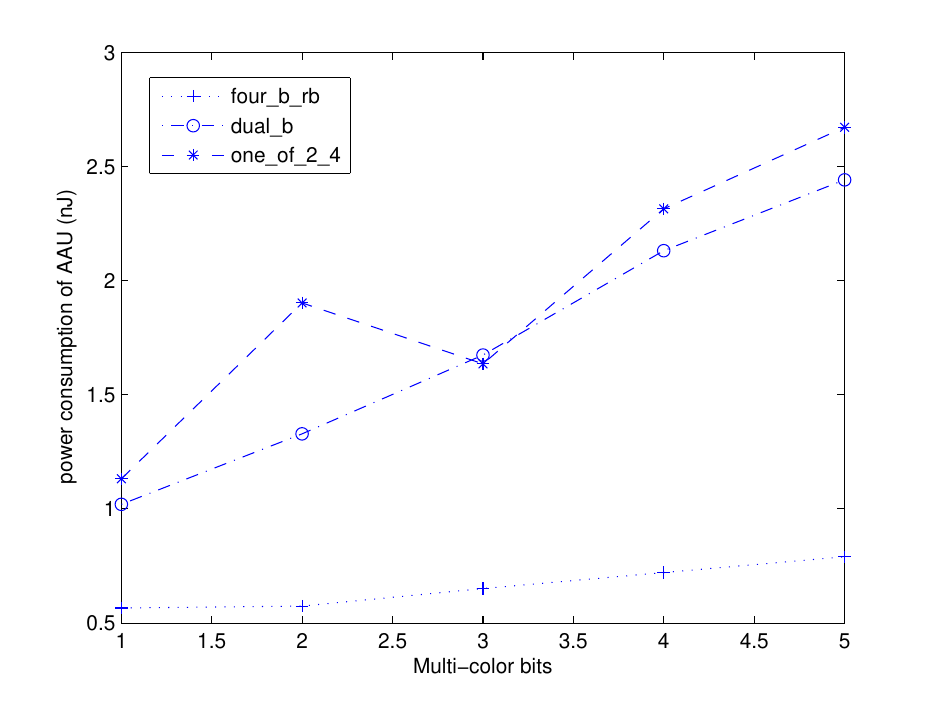}}
\caption{Power Analysis on Colour Vector Size}
\label{Power Analysis}
\end{figure}

Figure \ref{Power Consumption Analysis} provides results regarding  the power consumption. Again {\em four\_rb\_b} achieved the best power efficiency while the power consumed to implement the multi-colour algorithm is very small compared to that in SAMIPS. 
The increase of average power consumption of pipeline stages in both {\em four\_rb\_b} and {\em dual\_b} \ref{Power Analysis} implementation is linear. The increase rate for {\em four\_rb\_b} is very small. As mentioned in section \ref{Forward Power Analysis}, the irregular change of the {\em one\_of\_2\_4} implementation in Figure \ref{Power Analysis}(b) is due to the encoding of {\em one\_of\_2\_4} datapath where the most-significant bit is implemented in {\em dual\_b} and hence the communication energy required to transmit 2n-1 and 2n bits remains the same \cite{Will}.

\section{Enhancing Performance}
\label{Speed Based Refinement} 

The evaluation of SAMIPS presented in the previous section has provided useful insights about its behaviour and efficiency. SAMIPS power consumption and area cost have been shown to be satisfactory, while there seems to be scope for improvement of its performance. 
This is expected as the Balsa system aims for ease of development rather than optimisation of the synthesised products. Nevertheless, the distribution analysis presented in the previous section has revealed parts of SAMIPS that could be potential targets of performance enhancement. The rest of the section provides an outline of this endeavour, which may act as a guideline approach for other synthesised systems.

\subsection{SAMIPS Critical Path}
\label{The Speed Analysis of An Asynchronous Pipeline}

The performance of a pipeline can be characterised by means of the latency and throughput. 
Due to clock synchronisation, for a fix delay pipeline, where each stage has the fixed operation delay, 
In the case of a synchronous pipeline, where the clock frequency is determined by the slowest stage, the total latency through the pipeline is is the maximum stage latency multiplied by the number of stages (equation \ref{6}). In contrast, the latency of an asynchronous pipeline is the sum of the latencies in all stages   (equation \ref{7}). The throughput of both the synchronous and the asynchronous pipeline is limited by the slowest stage   (equation \ref{8}).

\begin{eqnarray}
Latency\_s & =  & n * MAX_{i=1}^{n}(T_{Si}) \label{6} \\
Latency\_a & =  & \sum_{i=1}^n(T_{Si}) \label{7}\\
Throughput & =  & \frac{1}{MAX_{i=1}^{n}(T_{Si})}\label{8}
\end{eqnarray}
{\em where n is the total number of pipeline stages and $T_{Si}$ is the latency of $i_{th}$ stage.}  
For a variable delay pipeline, where each stage may perform different operations and thus have various internal latencies, the throughput of a synchronous pipeline is limited by the longest latency of the slowest stage   (equation \ref{9}), while that of an asynchronous pipeline is limited by the average latency of the slowest stage  (equation \ref{10}).    
\begin{eqnarray}
Throughput\_s & =  & \frac{1}{MAX_{i=1}^{n}\{MAX_{j=1}^{m}(T_{Sij})\}} \label{9} \\
Throughput\_a & =  & \frac{1}{MAX_{i=1}^{n}\{AVG_{j=1}^{m}(T_{Sij})\}} \label{10} 
\end{eqnarray}
{\em where m is the aggregate latency in one stage and $T_{Sij}$ is the $j_{th}$ latency in the $i_{th}$ stage.} 


\begin{figure*}[t]
\begin{center}
\includegraphics[width=0.8\textwidth]{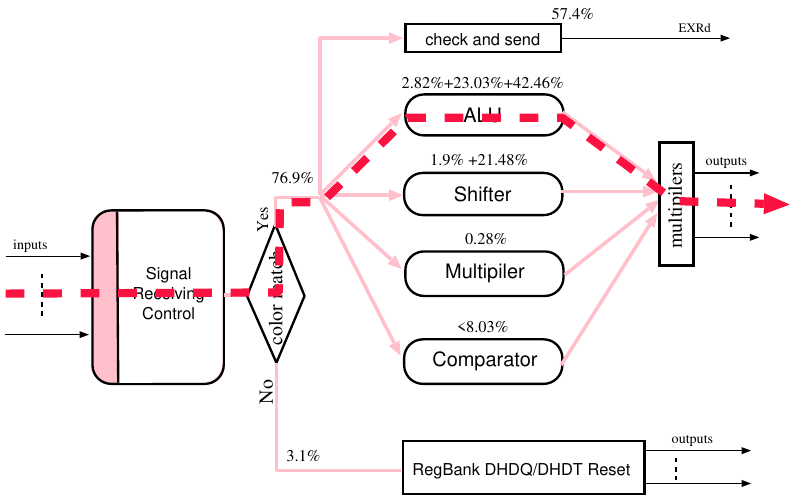} 
\caption{\textsf{EXEunit} Organisation and Critical Path Calculation}
\label{EXEunit Organisation}
\end{center}
\end{figure*}

The critical path is the slowest logic path in the circuit, or, in other words, the latency on that path constrains the overall speed of the processor.  The latency distribution results presented in section \ref{Analysis of Latency Distribution} clearly show that the \textsf{EXEunit} is the busiest block of SAMIPS. 


\begin{figure*}
\begin{center}
\includegraphics[width=0.8\textwidth]{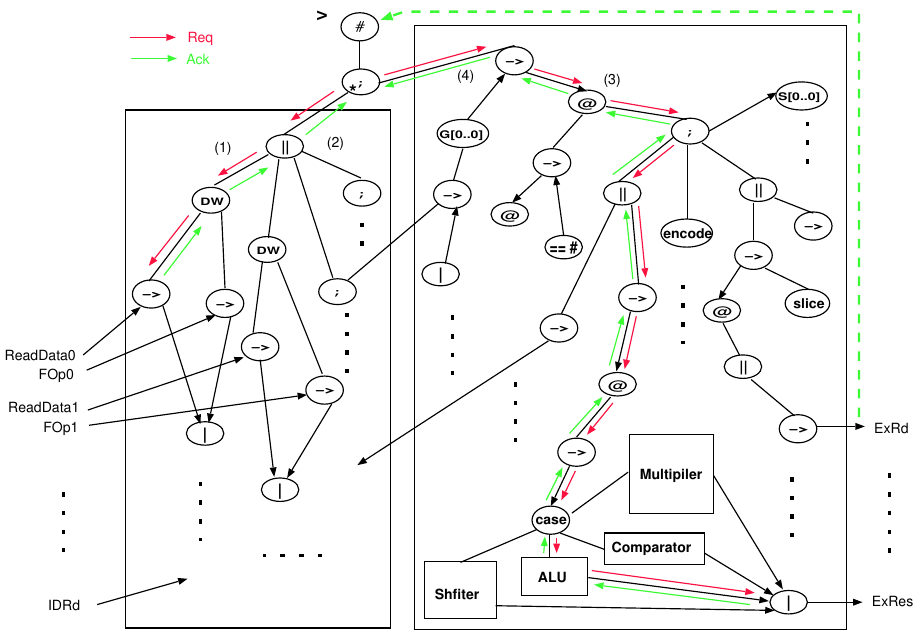}
\end{center}
\caption{Handshake Circuits of the \textsf{EXEunit}}
\label{Handshake Circuits of EXEunit}
\end{figure*}


The \textsf{EXEunit} incorporates four processing elements, namely the \textsf{ALU}, the \textsf{Shifter}, the \textsf{Multiplier} and the \textsf{Comparator}. Because of the different instruction usages, their contributions to the overall performance also vary. An alternative view of  \textsf{EXEunit} is provided in Figure \ref{EXEunit Organisation}.  
Before an instruction is processed in the \textsf{EXEunit}, all the input signals need to be correctly acknowledged and buffered in the \textsf{Signal Receiving Control} unit, followed by the colour vector checking. This applies to every instruction and thus any improvement in these two parts should change the average latency of the \textsf{EXEunit}. Examining execution traces from Dhrystone1.2, 76.9\% instructions are colour matched, which means the upper path in Figure \ref{EXEunit Organisation} (the "Yes" branch) contributes much more than the lower one (the "No" branch). Around 57\% of the instructions need to check and send the \textit{EXRd}, which is processed in parallel with the operations of the four processing units. However, due to the very short internal latency, this path is not crucial. Dynamic instructions usage in Table \ref{Dynamic Instructions Usage} illustrates that about 2.82\%(logic) +  23.03\%(arithmetic) + 42.46\%(memory) = 68.2\% instructions require ALU operations. Based on this observations, the thick dotted arrow in the figure marks the predicted \textit{critical path} in \textsf{EXEunit}, which contributes the most to the overall latency of SAMIPS.

\subsection{Levels of Optimisation}
\label{Levels of Optimisation}
Within the Balsa synthesis system, there are four different levels that could be targeted for performance optimization, namely architecture level, Balsa modelling level, handshake circuits level and handshake component level. 

Architecture level optimisations would imply substantial changes in the architecture of SAMIPS. These could include changes in the pipeline depth and structure, out-of-order instruction execution, etc. Such changes would change the entire philosophy of SAMIPS and therefore are considered outside the scope of our analysis.  Similarly, handshake optimisation at the handshake component level requires new, more efficient implementations of handshake components. 

\begin{figure*}
\begin{center}
\includegraphics[width=0.8\textwidth]{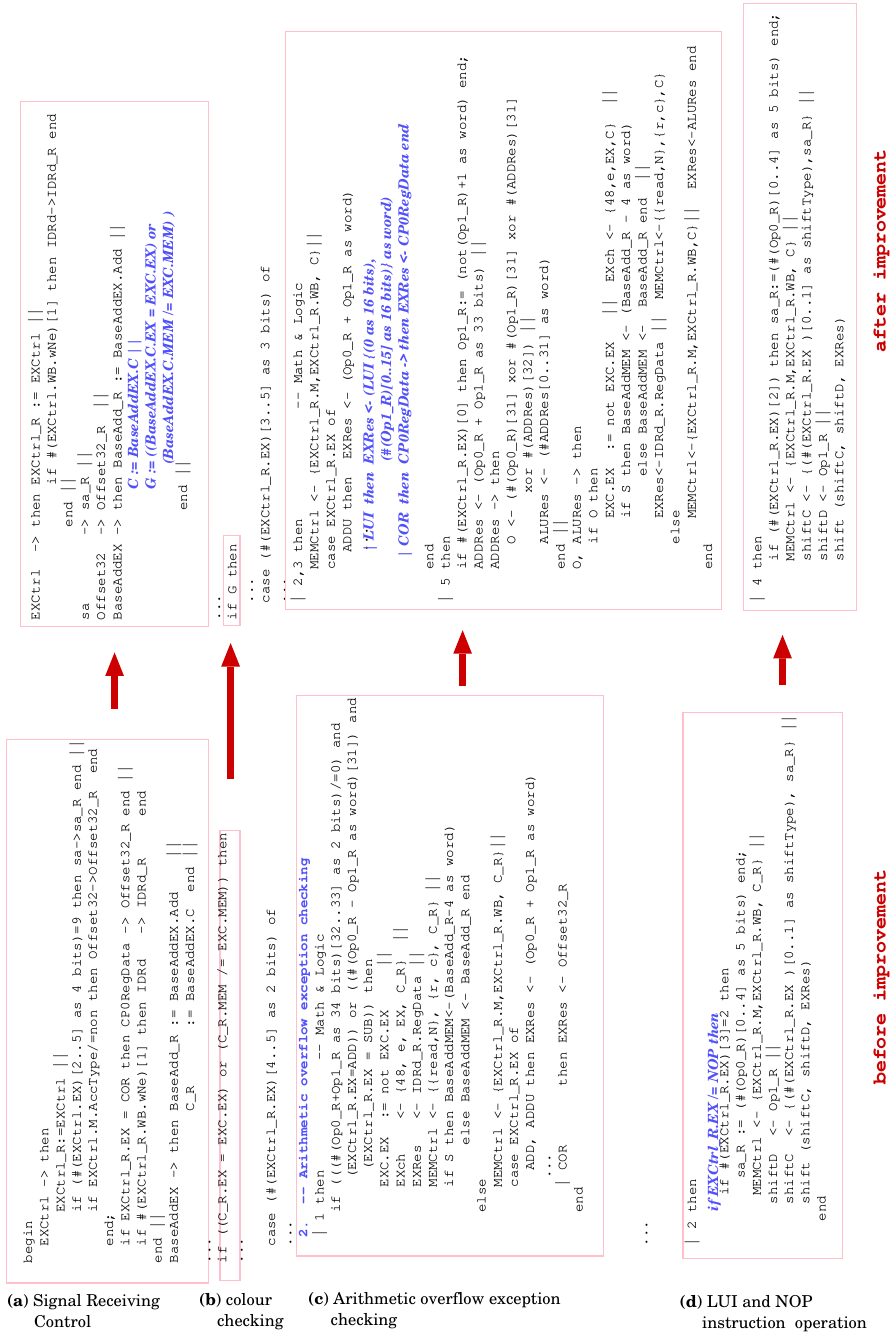}
\end{center}
\caption{\textsf{EXEunit} Modifications}
\label{EXEunit Modification}
\end{figure*}



\subsubsection{Handshake Circuits Level Optimisation}
As discussed in section \ref{Balsa}, given the same Balsa specification, there is a range of matchable handshake circuits networks. The  Balsa compiler is not an optimising one, the syntax-driven translation used by the compiler often produces inefficient circuits and results in performance overheads.  

The Balsa system includes a visualization tool which generates a colour-based graph animation to highlight the handshake circuits control flows \cite{Janin04}. The graph derived for the SAMIPS \textsf{EXEunit} is illustrated in Figure \ref{Handshake Circuits of EXEunit}.  The main loop has been` compiled into two handshake trees, highlighted in two rectangles in the figure, for inputs, and outputs and processing respectively.  In such a tree structure, a leaf node starts to work after a REQ handshake chain finishes from the root (path (1) in Figure \ref{Handshake Circuits of EXEunit}) and reports the completion in an ACK handshake chain (path (2)) back to the root. These handshake chains could be very long in a complex design and thus the time consumed is not negligible. In our \textsf{EXEunit}, after all the output channels are acknowledged by the \textsf{MemInt}, the new operation cycle cannot start until path (4) completes all the ACK handshakes, which take about 10\% of total execution time (65 out of 565 time units, measured in the Balsa behaviour simulator). The circuits would be more efficient if that could be hidden in other operations, e.g. if a control signal (the dotted line in the figure) links directly to the top repeat component after all the output channels finish operation to enable a new cycle in the left rectangle without waiting for path (4)'s completion. Time could thus be saved from the overlapped operation between the two handshake trees.  

Such manual optimisations are clearly not practical. Several automated optimisation approaches have been proposed for Balsa handshake circuits  \cite{1402060,6598349,Alekseyev2009OptimisationOB,10.1007/978-3-540-95948-9_19} and \cite{7584899}; the latter also reports results from applying their approach (which is based on STGs and incorporates a logic synthesizer for Balsa)  onto SAMIPS.  



\subsubsection{Balsa Modelling Level Optimisation}

\begin{table*}
\caption{New Control Signal Encode for the EX Stage} 
\begin{small}
\begin{center}
\begin{tabular}{p{12.7cm}r} 
\end{tabular} \\ 
\begin{small}
\begin{tabular}{|c||c|c|c|c|c|c|c|c|}    \hline 
\diagbox{bits 2..0}{bits 5..3}  & 000   & 001  & 010  & 011  & 100  & 101    & 110  & 111     \\ \hline \hline 
 000    & BEQ   & BNE  & BGTZ & BLEZ & BLTZ & BLTZAL & BGEZ & BGEZAL  \\ \hline
 001    & *     & JR   & JALR & JAL  & *    & *      & *    & *       \\ \hline
 010    & *     & *    & ADDU & SUBU & AND  & OR     & XOR  & NOR     \\ \hline
 011    & EXC   & EXCS & MA   & COR  & SLTU & SLT    & \textbf{\textit{LUI}}  
        & *     \\ \hline
 100    & SLLV  & SRLV & SRAV & *    & SLL  & SRL    & SRA  & *       \\ \hline
 101    & \textbf{\textit{ADD}} & \textbf{\textit{SUB}}  & *    & *   
        & *     & *    & *    & *	 \\ \hline
 110    & MULTU & MULT & DIVU & DIV  & MTHI & MTLO   & MFHI & MFLO    \\ \hline
 111    & \textbf{\textit{NOP}} & *    & *    & *    & *    & *      & *    &         \\ \hline 
\end{tabular}
\end{small}
\begin{tabular}{lp{12cm}}
 
\end{tabular} 
\label{NewContrlEX} 
\end{center}
\end{small}
\end{table*}

To illustrate the effect that Balsa modelling level modifications can have on the performance of the system, and after careful consideration of the automatically produced handshake circuits, two elements of the critical path have been identified as indicative candidates for improvement in the Balsa code: 

\begin{enumerate}
\item In the \textsf{EXEunit}, all input channels are first acknowledged and buffered by the \textsf{Signal Receiving Control} and then a ``colour" comparison starts before an instruction is executed. These two steps originally operate sequentially and both have a long internal delay.  
\item In the ALU unit, there is a check for arithmetic overflow exception each time an arithmetic and logic instruction enters into this unit, even though only four instructions out of fourteen may potentially cause an overflow exception. 
\end{enumerate}

A performance improvement plan was then devised targeting these two aforementioned elements. Figure \ref{EXEunit Modification} highlights the modified  Balsa specification while Table \ref{NewContrlEX} presents the necessary adjustments made to implement the new design in the control signal encoding for the EX stage. The main modifications are listed below.  

\begin{enumerate}
\item Check the ``colour" vector in parallel with the inputs receive step and save the comparison result (Figure \ref{EXEunit Modification} (a) and (b)).
\item Minimise the usage of the conditional checking instruction in the \textsf{Signal Receiving Control} reduce the internal delay (Figure \ref{EXEunit Modification} (a)).
\item Move the receiving of \textit{CP0RegData} from the \textsf{Signal Receiving Control} to the ALU unit, since it is not frequently used (Figure \ref{EXEunit Modification} (a) and (c)).
\item In the ALU, separate the operation of those exception raising instructions from other ordinary ones (Figure \ref{EXEunit Modification} (c)).
\item Modify the exception checking part to make it more efficient by using a different arithmetic logic. The addition operation does not need to be re-performed during the exception checking (Figure \ref{EXEunit Modification} (c)).
\item Remove the LUI and NOP operations from the shifter unit and treat them separately since they are more frequently used (Figure \ref{EXEunit Modification} (c) and (d)).
\end{enumerate}


\subsection{Evaluation}

\begin{table}
\caption{Performance Gained after Modifications}
\begin{center}
\begin{tabular}{|l||c||p{1.5cm}|c|c|} \hline                     
SAMIPS     & Transistor & Benchmarks   & Speed  & Power    \\ 
 	   &  count	&	       & (MIPS) & (MIPS/W)  \\  \hline \hline
           & 	    	& Dhrystone1.2 & 16.9   & 2197.8   \\  \cline{3-5}
Original & 168,906	& Quicksort    & 16.8   & 2120.7   \\  \cline{3-5}  
           & 	    	& Heapsort     & 17.8   & 2206.0   \\  \cline{3-5}
	   & 	    	& ExcTest      & 17.4	& 2132.1   \\  \hline  \hline 
	   & 	    	& Dhrystone1.2 & 21.8	& 2227.2   \\  \cline{3-5}		  
Modified	   & 	    	& Dhrystone2.1 & 21.2	& 1983.7   \\  \cline{3-5}		  
 & 169,634	& Quicksort    & 22.1	& 2168.7   \\  \cline{3-5}		
           & 	    	& Heapsort     & 21.2	& 2185.1   \\  \cline{3-5}
	   & 	    	& ExcTest      & 21.1	& 2146.9   \\  \hline 
\end{tabular}

\label{Performance Gained after Refinement}
\end{center}
\end{table}


To evaluate the performance improvements achieved from the Balsa modelling level improvements, the modified SAMIPS model has been re-synthesised into its final layout. 

Table \ref{Performance Gained after Refinement} presents a set of comparative results obtained from  
physical level simulation of the original and modified BALSAMIPS specification respectfully.  As it is evident from the table, the modifications result in over 30\% performance improvement. The speed increase of Heapsort is relatively lower (20\%) and this may be attributed to the high usage of the branch and shift operations, which have not been included in the modification.  
The price paid for this performance improvement in terms of power consumption and transistor count is very small. The modifications result to a slight decrease of power efficiency by  1\%-2\%  for Dhrystone2.1 and Heapsort, and 1\%-2\% increase for the other benchmarks. Nearly no extra area cost has been introduced.  This suggests that a careful specification of the system at the Balsa modelling phase can have a significant impact on the performance of the produced system.

\section{Summary and Conclusions}
\label{Summary and Conclusions}
Interest in asynchronous circuit design continues  and asynchronous design techniques are increasingly finding their way in the mainstream VLSI design, e.g. in the form of GALS, Globally Asynchronous Locally Synchronous systems, which take advantage of both synchronous and asynchronous methodologies. This paper aspires to provide  insights about the design and automatic systhesis of an asynchronous processor that the developers of asynchronous systems might find useful as a roadmap for their designs. MIPS forms the basis of the increasingly popular RISC-V architecture, as well as other important systems such as the Loongson/Godson processor series. It is anticipated that the work presented in this paper may offer guidance for asynchronous implementations of such systems. 

The paper has provided a holistic description of the system and its components, the innovative solutions that have been developed to address hazard challenges and a quantitative analysis that provides insights to both: the performance characteristics of SAMIPS and the effectiveness of Balsa as a  hardware description language (HDL) and synthesis system. The main advantage of automated synthesis and silicon compilation is its simplicity and short development period while the major drawback is the performance. Determining factors for the performance achieved are the efficiency of the compiler, the implementation of the cell library and the HDL specification of the model by the designer. This paper has demonstrated that only minor changes in the model can have significant impact on the final performance of the system.







\ifCLASSOPTIONcaptionsoff
  \newpage
\fi



%

\bibliographystyle{plain}
\bibliography{bibliography}




%
\begin{IEEEbiography}[{\includegraphics[width=1in,height=1.25in,clip,keepaspectratio]{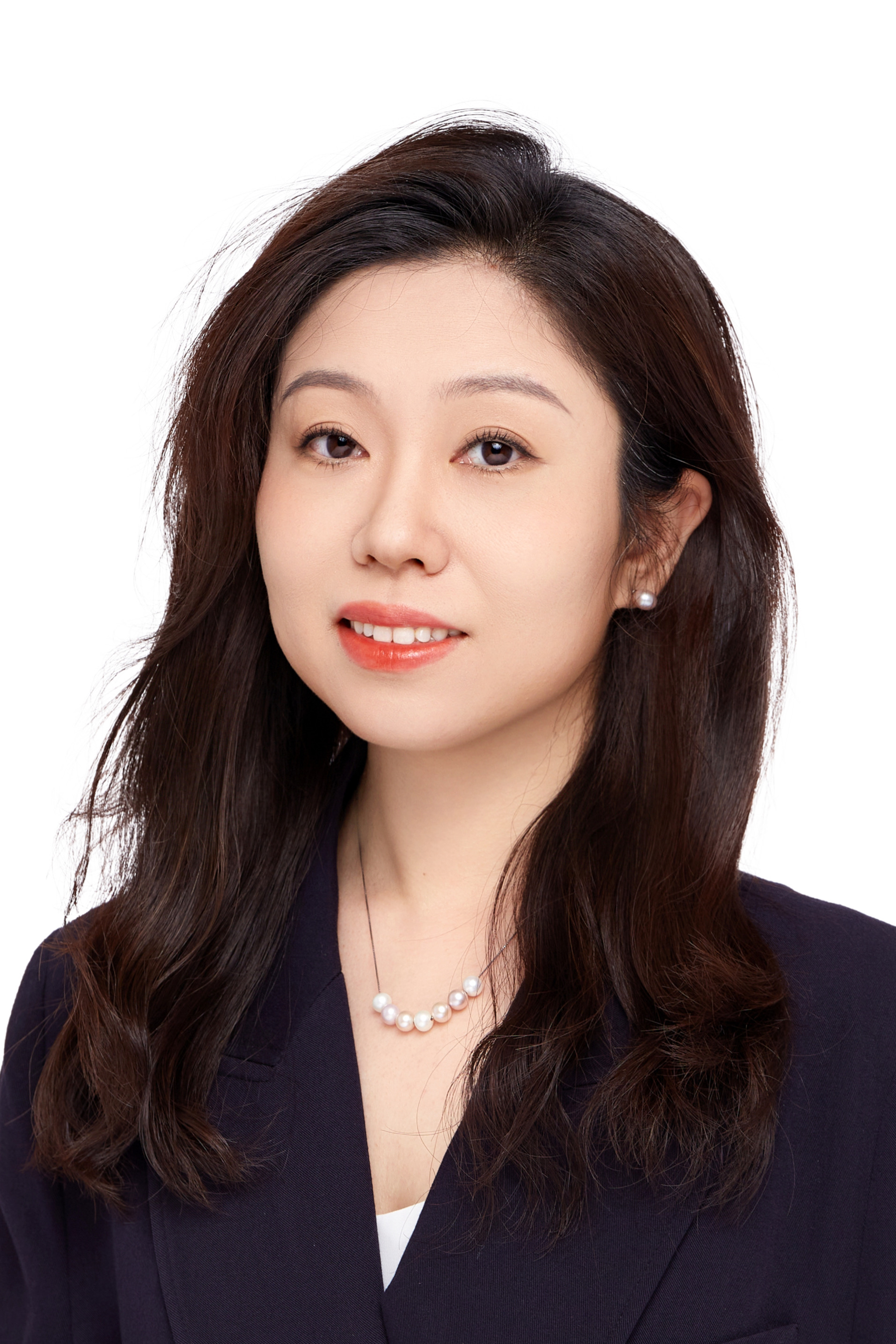}}]
{Qianyi Zhang} is currently Product Engineering Director at Cadence Design Systems in Shanghai, China. She is an expert in parasitic extraction, library characterization, static timing analysis, chip level power and rail signoff with years of hands on tapeout experience using lower power and smart hierarchical design flows. Her design experience extends from ARM Cortex CPU cores to the latest 8-CPU smart phone chips and 64-CPU server chips. She holds a PhD in Computer Science from the University of Birmingham, UK. 
\end{IEEEbiography}
\vspace{-5cm}
\begin{IEEEbiography}[{\includegraphics[width=1in,height=1.25in,clip,keepaspectratio]{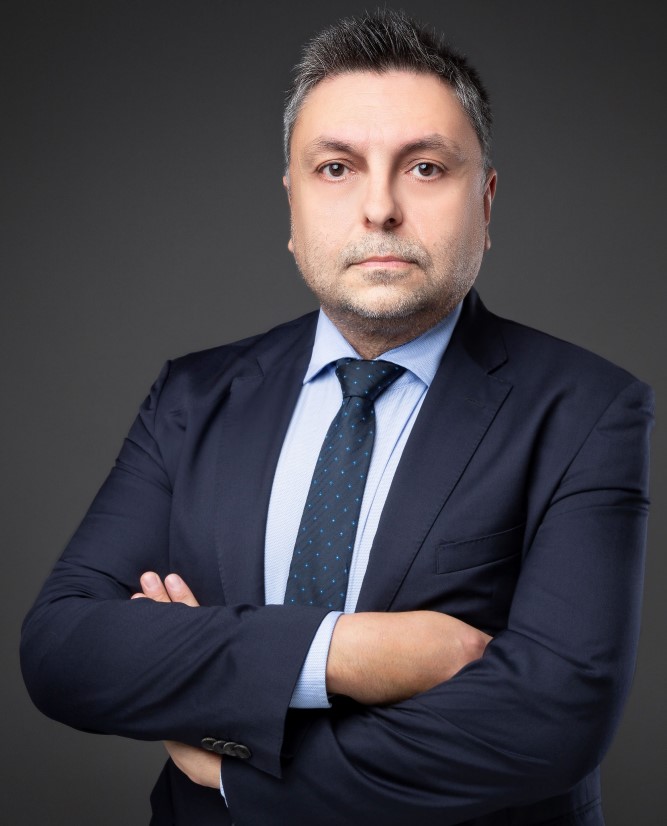}}]
{Georgios Theodoropoulos} is currently a Chair Professor at the Department of Computer Science and Engineering at SUSTech in Shenzhen, China.  He was previously the inaugural Executive Director of the Institute of Advanced Research Computing, a Chair Professor in Computer Engineering and the Head of the Innovative Computing Group at the School of Engineering and Computing Sciences at the University of Durham, UK. He has been a Senior Research Scientist with IBM Research and senior faculty at the University of Birmingham, UK, where he was also founding Director of one of UK's  e-Science Centres of Excellence. He has held an Adjunct Chair at the Trinity College Dublin and visiting appointments at the Nanyang Technological University and National University in Singapore. His research interests and contributions are in Info-Symbiotic Systems, Distributed Simulation and Distributed Virtual Environments, Parallel,  Distributed and Intelligent Computer Systems, Complex and Multi-agent systems, Interdisciplinarity and Sustainability. 
He is Ordinary Member of the European Academy of Sciences and Arts,  a Fellow of the World Academy of Art and Science, an Accredited Board Director of the Singapore Institute of Directors, a Chartered Engineer, and holds a PhD from the University of Manchester, UK.

\end{IEEEbiography}




\end{document}